\documentclass[epsfig,useAMS,usenatbib]{mn2e}

\input{epsf}

\usepackage{graphicx}

\def\gsim{ \lower .75ex \hbox{$\sim$} \llap{\raise .27ex \hbox{$>$}} }
\def\lsim{ \lower .75ex \hbox{$\sim$} \llap{\raise .27ex \hbox{$<$}} }

\title[Simulating the Bullet Cluster]{Simulating the Bullet Cluster}

\author[Mastropietro \& Burkert]
{Chiara Mastropietro $^{1}$ \thanks{E-mail: chiara@usm.uni-muenchen.de} \& 
Andreas Burkert $^{1}$\\  
$^{1}$Universit\"ats Sternwarte M\"unchen, Scheinerstr.1, D-81679 M\"unchen,Germany}
\begin{document}


\pagerange{\pageref{firstpage}--\pageref{lastpage}} \pubyear{00} 

\maketitle  

\label{firstpage}

\begin{abstract}
We present high resolution N-body/SPH simulations of the interacting cluster
1E0657$-$56. 
The main and the sub-cluster are modeled using extended cuspy $\Lambda CDM$ 
dark matter halos and isothermal $\beta$-profiles for the collisional 
component.
The hot gas is initially in hydrostatic equilibrium inside the global potential of 
the clusters.
We investigate the X-ray morphology and derive the most likely impact parameters, mass ratios and initial relative velocities.
We find that the observed displacement between the X-ray peaks and the associated mass distribution, the morphology of the bow shock, the surface brightness and projected temperature profiles across the shock discontinuity can be well reproduced by offset 1:6 encounters where the sub-cluster has initial velocity (in the rest frame of the main cluster) close to 2 times the virial velocity of the main cluster dark matter halo.
A model with the same mass ratio and lower velocity ($1.5 $ times the main cluster virial velocity) matches quite well most of the observations. However, it does not reproduce the morphology of the main cluster peak.
Dynamical friction strongly affects the kinematics of the sub-cluster so that the low velocity bullet is actually bound to the main system at the end of the simulation.
We find that a relatively high concentration  (c=6) of the main cluster dark matter halo is necessary in order to prevent the disruption of the associated X-ray peak. 
For a selected sub-sample of runs we perform a detailed three dimensional analysis following the past, present and future evolution of the interacting systems.
In particular, we investigate the kinematics of the gas and dark matter components as well as the changes in the density profiles and  the motion of the system in the $L_X-T$ diagram.

\end{abstract}


\section{Introduction}

The ``bullet'' cluster 1E0657-56 represents one of the most complex and
unusual large-scale structures ever observed.
Located at a redshift $z = 0.296$ it has the highest X-ray luminosity and 
temperature of all known clusters as a result of overheating due to a recent 
supersonic Mach $M \sim 3$ \citep{Markevitch02, Markevitch06} central encounter of a sub-cluster 
(the bullet) with its main cluster.
The 500 ks Chandra ACIS-I image of 1E0657-56 (Fig. 1 in Markevitch 2006) 
shows  two plasma concentrations
with the bullet sub-cluster on the right of the image being deformed in a classical bow shock on
the  western side as a result of its motion through the hot gas of the main
cluster. The analysis of the shock structure leads to the conclusion that
the bullet is now moving away from the main cluster
with a velocity of $\sim 4700$ km s$^{-1}$ \citep{Markevitch06}.
The line-of-sight velocity difference between the two systems is only  600 km s$^{-1}$ 
suggesting that the encounter is seen nearly in the plane of the sky \citep{Barrena02}. 
As the core passage must have occurred $\sim 0.15$ Gyr ago we have the unique
opportunity to study this interaction in a very special short-lived stage,
far away from thermal and dynamical equilibrium. 
As a result of the encounter, the collision-dominated hot plasma and the
collisionless stellar and dark matter components have been separated.
The galaxy components of both clusters are clearly offset from the associated X-ray
emitting cluster gas \citep{Liang00, Barrena02}.
In addition, weak and strong lensing maps \citep{Clowe04, Clowe06, Bradac06} show that
the gravitational potential does not trace the distribution of the hot
cluster gas that dominates the baryonic mass but follows approximately
the galaxy distribution as expected for a collisionless dark matter
component.
The likelihood to find such a high velocity cluster encounter in a
$\Lambda$CDM cosmology has recently been
investigated by \citet{Hayashi06} using the Millenium Run simulation .
According to the newest estimates from X-ray and gravitational lensing results the \citet{Hayashi06} 
likelihood becomes $\sim 0.8 \times 10^{-7}$ \citep{Farrar07} which means that 1E0657-56 represents an 
extremely rare system in a $\Lambda$CDM universe.
Recent numerical works \citep{Milosavljevic07, Springel07} 
have demonstrated that the relative velocity of the dark matter components associated 
to the main and the sub-cluster are not necessarily coincident with the bullet velocity inferred from the shock analysis.
In details, \citet{Milosavljevic07} using a 2-D Eulerian code well reproduced the observed increase 
in temperature across the shock front with a dark halo velocity $\sim 16\%$ lower than that of the shock, while \citet{Springel07} 
found even a larger difference between the shock velocity ($\sim 4500$ km s$^{-1}$ in their best model) and the speed of the halo (only $\sim 2600$ km s$^{-1}$). 
Moreover, according to \citet{Milosavljevic07}, due to a drop in ram-pressure after the cores' interaction 
the gas component of the sub-cluster can eventually be larger than that of its dark matter counterpart.

\begin{table*} 
\caption{Initial conditions of the simulations. For each run dark matter virial mass 
  ($M_{vir}$) and concentration $c$ of the main and the sub-cluster are indicated. $f_g$, $b$, $v_i$ and $v$ are the adopted gas mass fraction, impact parameter and initial velocity of the sub-cluster in the system of reference where the main-cluster is at rest and in the center of mass rest-frame, respectively. The last column indicates if radiative cooling is included.   }
\begin{tabular}{l|c|c|c|c|c|c|c|c|c|}
\hline
Run & $M_{vir}$(main)&  $c$ (main)& $M_{vir}$(bullet)  & $c$ (bullet) &$f_g$&  $b$ & $v_i$ & $v$ & Model\\
& [$10^{14}$ $M_{\odot}$]& & [$10^{14}$ $M_{\odot}$] & & & [kpc]& [km/s]& [km/s]& \\
\hline
1:6b0 &  7.13  & 6 & $1.14$ & 8& 0.17 & 0& 5000 &4286 &adiabatic\\
1:6v3000b0 &  7.13  & 6& 1.14 & 8 &  0.17& 0 & 3000&2571 &  adiabatic \\
1:6 &  7.13  & 6 & $1.14$ & 8& 0.17 & 150 & 5000&4286 &adiabatic\\
1:6v3000 &  7.13  & 6& 1.14 & 8 &  0.17& 150 &3000&2571 & adiabatic\\
1:6v2000 &  7.13  & 6& 1.14 & 8 &  0.17& 150 &2000&1714 & adiabatic\\
1:3 &  7.13  & 6& 2.4 & 7 &  0.17& 150 &5000&3750 &adiabatic \\
1:8 &  7.13  & 6& 0.91 & 8 &  0.17& 150 &5000&4445 & adiabatic\\
1:6v3000big & 14.2& 6 & 2.4 & 7 & 0.17 & 150& 3000& 2571 & adiabatic\\
1:6c4 &  7.13  & 4 & $1.14$ & 8& 0.17 & 150 &5000&4286 &adiabatic\\
1:6lfg &  7.13  & 6 & $1.14$ & 8& 0.12 & 150 &5000&4286 &adiabatic\\
1:3lfg &  7.13  & 6& 2.4 & 7 &  0.12& 150 &5000&3750 & adiabatic\\
1:6c &  7.13  & 6 & $1.14$ & 8& 0.12 & 150&5000&4286 & cooling\\
1:3clfg &  7.13  & 6 & $2.4$ & 7& 0.12 & 150 &5000&3750 &cooling\\
\hline
\label{runs}
\end{tabular}
\end{table*} 

The simulations of \citet{Springel07} represent the most complete three dimensional numerical modeling of the 1E0657-56 system so far.
Nevertheless they focus preferentially on the speed of the bullet but fail in reproducing the observed displacement of the X-ray peaks which represent an important indicator of the nature of the interaction. 
In particular, they do not obtain any displacement in the X-ray distribution associated with the main cluster suggesting that the baryonic component is suffering too little ram-pressure. 
Moreover, the concentrations used for the main cluster (and obtained by modeling the lensing data) are much smaller than those suggested by $\Lambda$CDM \citep{Maccio07} for halos of similar masses.

The aim of this paper is to investigate the  evolution of the bullet cluster in details
using high resolution SPH simulations. 
We quantify the initial conditions that
are required in order to better reproduce its observed state and predict its subsequent
evolution.

Our model allows us to determine in details the spatial, thermal and
dynamical
state of the dark matter and hot gas distribution in the bullet cluster.

The paper is organized as follows.
Section 2 describes the adopted cluster models and orbital parameters.
In Section 3 we perform a projected analysis of our simulations comparing it
with the latest X-ray and gravitational lensing results. 
In Section 4 we select some significant models and study in details the
three dimensional kinematics and morphology of the interacting systems and their
past and future evolution with time, as well as the motion of the main cluster along
the $L_X-T$ diagram.

\section{Models}

Both the main and the sub-cluster are two components spherical systems
modeled assuming a cuspy dark matter halo and a distribution of hot gas
in hydrostatic equilibrium within the global potential of the cluster.
The dark halo has a NFW \citep{Navarro97} profile:
\begin{equation}
\rho(r) =\rho_{crit} \frac{\delta_c}{(r/r_s)(1+r/r_s)^2},
\label{NFW}
\end{equation}
where $\rho_{crit}$ is the critical density of the universe at the time of the
halo formation, $r_s$ is a scale radius and $\delta_c$ the characteristic halo
overdensity. The virial mass $M_{vir}$ and radius $r_{vir}$ are related by:
$M_{vir} = \Delta_{vir}\rho_{crit}(4\pi/3)r_{vir}^3$, where the density
contrast $\Delta_{vir}$ is set equal to 200. The concentration
parameter $c=r_{vir}/r_s$ is assumed to be dependent on the halo mass \citep{Maccio07}. 
The velocity distribution at a given point in space is approximated by a
Gaussian, whose velocity dispersion is given by the solution of the Jeans
equation at this point \citep{Hernquist93}. 
The distribution of hot gas follows an isothermal $\beta$-model \citep{Cavaliere76} of the form:
\begin{equation}
\rho(r) = \rho_0(1+(r/r_c)^2)^{-3/2\beta},
\label{betamodel}
\end{equation}
We take the asymptotic slope parameter $\beta=2/3$ \citep{Jones84}
and $r_c = 1/2 r_s$ \citep{Ricker01}. The adopted gas fraction ranges from a minimum value of $8\%$ to $17\%$, consistent with the recent WMAP results \citep{Spergel07}, with an intermediate value of $12\%$, comparable with the gas mass fraction provided by X-ray observations of galaxy clusters \citep{Vikhlinin06, McCarthy07}.

Assuming a spherically symmetric model, the temperature profile is determined by
the condition of hydrostatic equilibrium by the cumulative total mass
distribution and the density profile of the gas \citep{Mastropietro05}:
\begin{equation}
T(r) = \frac{\mu m_p}{k_B} \frac{1}{\rho(r)} \int_{r}^{\infty} \rho(t)\frac{GM(t)}{t^2} \, dt \, , 
\label{hydrostaticeq}
\end{equation} 
where $M(r)$ is the total mass within the radius $r$, $m_p$ is the proton mass and $\mu$ the mean molecular weight. We assume $\mu=0.6$ for a gas of primordial composition, which appears to be a reasonable approximation since the mean temperature of 1E0657-56 is $T\sim 14$ keV according to \citet{Markevitch06} and cooling is dominated by bremsstrahlung and almost independent of the metallicity. $G$ and $k_B$ are the gravitational and Boltzmann constants.

\begin{figure}
\includegraphics[%
  height=50mm]{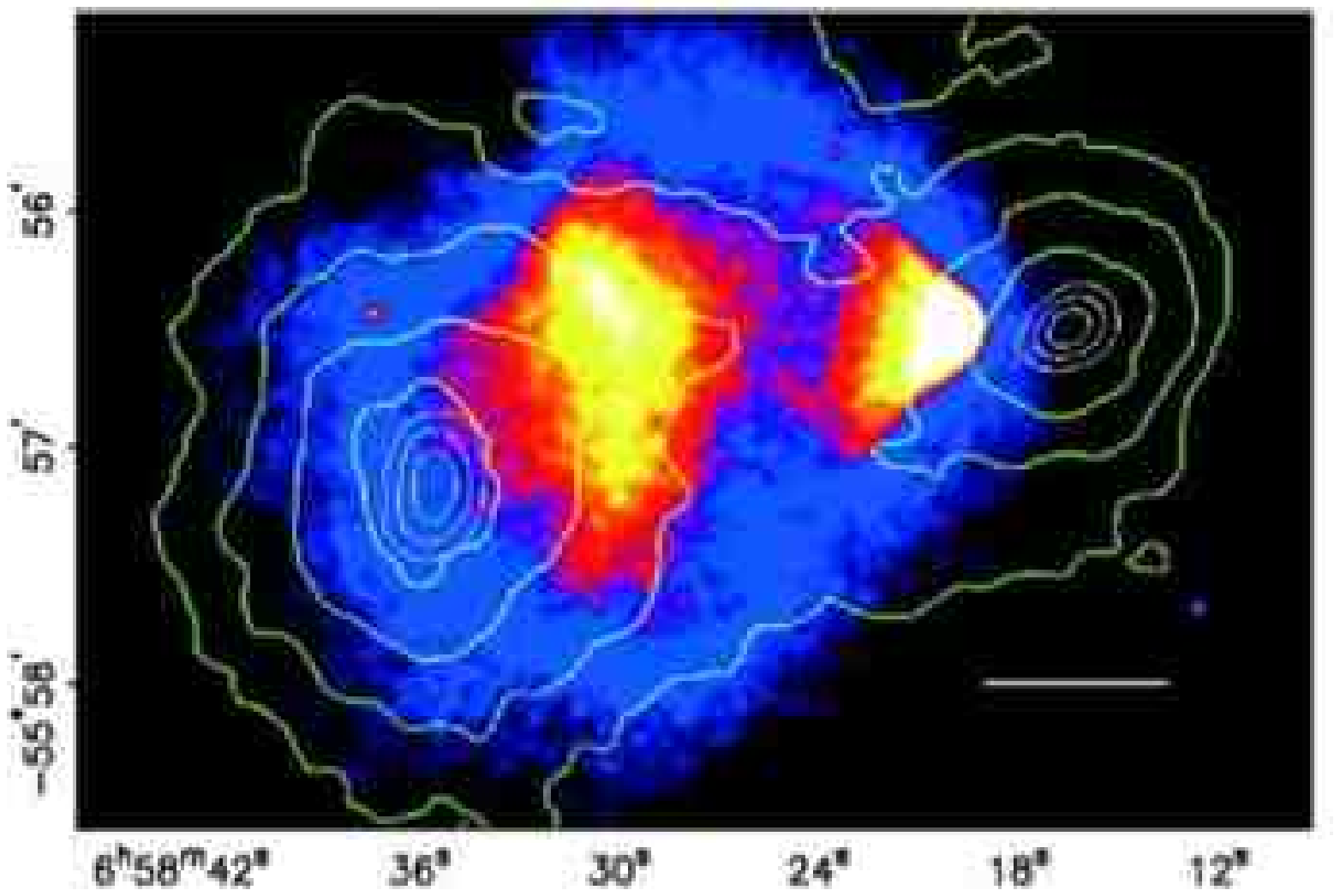}
\includegraphics[%
  height=75mm]{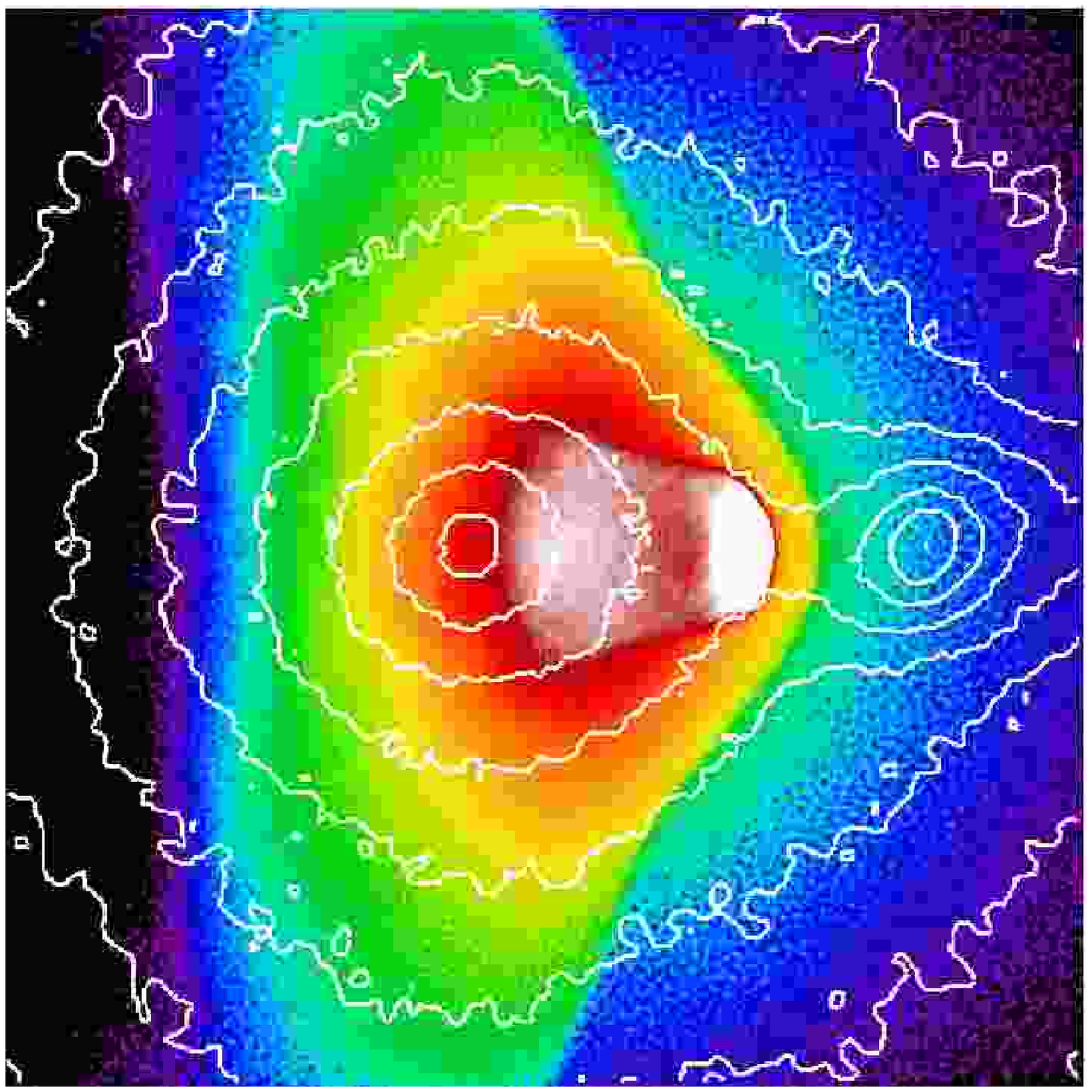}
\includegraphics[%
  height=75mm]{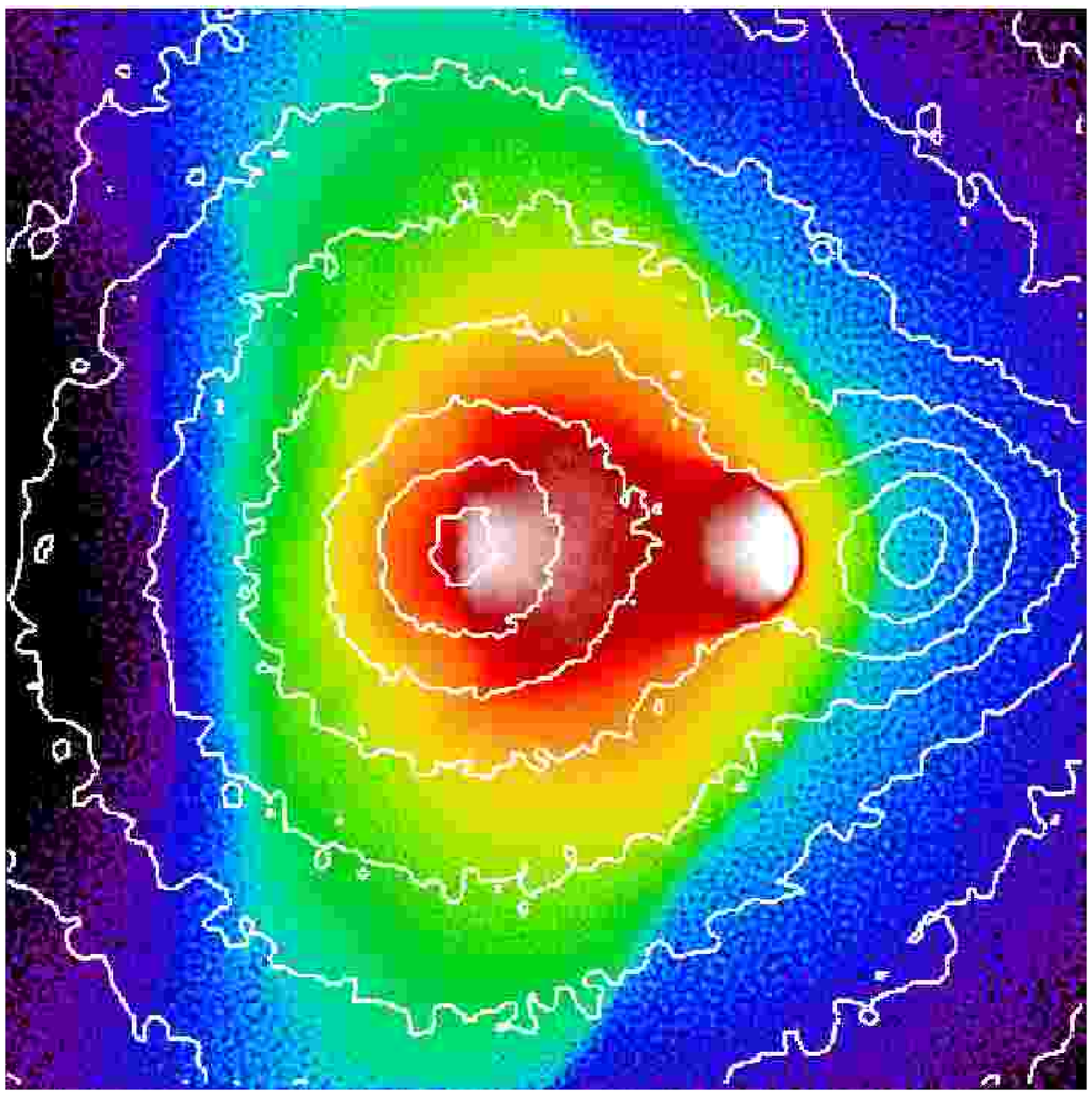}
\includegraphics[%
  height=4mm]{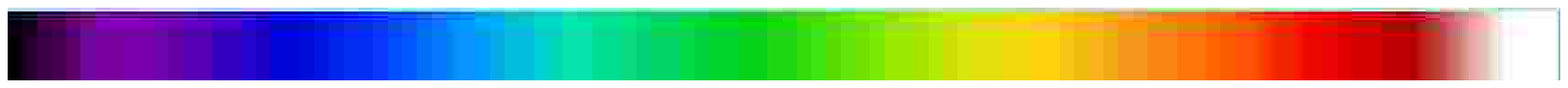}
\caption{Upper panel. 500 ks Chandra image of the system with weak lensing $k$ reconstruction shown in green (courtesy of D. Clowe). Central and bottom panel. $0.8-4$ keV surface brightness maps of runs 1:6vb0 and 1:6v3000b0. Logarithmic colour scaling is indicated by the key at the bottom of the figure with violet corresponding to $10^{38}$ erg s$^{-1}$ kpc $^{-2}$ and white to $2 \times 10^{41}$ erg s$^{-1}$ kpc $^{-2}$. White contours trace the total surface mass density of the system within $2.3 \times 10^{3}$ and $2.3 \times 10^{8}$ M$_{\odot}$ kpc$^{-2}$. The box size is 1800 kpc. }
\label{b=0}
\end{figure}

Masses are assigned to the models according to the weak and strong lensing
mass reconstruction of \citet{Bradac06}. 
In particular we assume that the inferred mass enclosed within the field of the 
HST Advanced Camera for Surveys  (ACS) \citep{Bradac06} is comparable with the total projected mass of our simulated system (calculated when the two centers of the mass distribution are at a distance similar to the observed one) within the same area.
Since the ACS field represents only the central fraction of the area covered by the entire system, this mass constraint is strongly influenced by the concentration of the dark matter halos \citep{Nusser07}. 
With a cosmologically motivated choice of $c =6$ \citep{Maccio07} for the main cluster initial halo model, we can reproduce the lensing mass reasonably well adopting a main cluster total mass (within the virial radius) of $\sim 8.34 \times 10^{14} M_{\odot}$ (Table \ref{runs}), almost a factor 1.8 smaller than the mass obtained by fitting lensing data with extremely low concentrated ($c<2$) NFW halos where the inner density profile is much flatter than the one suggested by $\Lambda$CDM simulations.
One of the simulations presented in this paper adopts a main halo with a lower ($c=4$) concentration value. We will see that in this case the X-ray intensity peak associated with the main cluster is easily destroyed during the interaction.

We model encounters with mass ratios 1:3, 1:6 and 1:8 between the sub and the main cluster in order to investigate the effects of tidal and ram-pressure stripping, which  significantly reduce the mass associated to the sub-cluster and lead  to values closer to the 1:10 ratio inferred from lensing observations \citep{Clowe04, Clowe06, Bradac06}. 
\begin{table*}
\caption{Present time. $\Delta$ is the projected (perpendicular to the plane of the encounter) distance between the peaks of the total mass distributions, associated with the two clusters. 
The third
  and fourth columns represent the projected offset between each X-ray peak and the
  associated mass density peak. $v_{gas}$ and $v_{dark}$ are the sub-cluster gas and
  dark matter velocity calculated in the center of mass system of reference. }
\begin{tabular}{l|c|c|c|c|c|c|c|c|}
\hline
Run & $\Delta$ & offset (bullet)& offset (main) & $v_{gas}$ & $v_{dark}$ \\
\hline
1:6b0 & 753 & 278 & 188  & 3215& 4715\\
1:6v3000b0 & 741 & 213  & 66  & 3131 & 3134\\
1:6 & 742& 237&  128&  3609& 4756\\
1:6v3000 & 729&  185& 172& 2893& 3137\\
1:6v2000 &721& 126  &  230 & 2849& 2425\\ 
1:3 &  784&  162& 223&  3908& 4076\\ 
1:8 & 737&  228& 117& 3647& 4858\\
1:6v3000big & 725 &139 & 92&  3927 & 3528\\
1:6c4 & 735& 151& 192&  4168& 4799\\
1:6lfg & 718& 200& 189& 3811& 4804\\ 
1:3lfg & 779& 197& 242&  3746 & 4145\\
1:6c &  736&  234& 127& 3497& 4806\\
1:3clfg &780& 228 & 272&  3595& 4205\\ 
\hline
\end{tabular}
\label{runsresults}
\end{table*} 

The main cluster is initially at rest and the sub-cluster moves in the $x$ direction with a velocity which ranges from  2000 to 5000 km s$^{-1}$.
The initial conditions of the different runs are summarized in Table \ref{runs}.  
The velocities of the sub-cluster relative to the center of mass of the system
are listed in the last but one column of the Table.

All the simulations  were 
carried out using GASOLINE, a parallel SPH tree-code with multi-stepping
\citep{Wadsley04}. 
Most of the runs are adiabatic, with $\gamma = 5/3$. Radiative cooling for a primordial mixture of hydrogen and helium in collisional equilibrium is implemented in 1:6c and 1:3lfgc.
The consequences of assuming  a lower concentration for the main dark halo are investigated in run 1:6c4, where $c=4$.

The main cluster is modeled with $1.8 \times 10^6$ particles, $10^6$ SPH and $8 \times 10^5$ collisionless. The sub-cluster, with the exception of run 1:8 (where the number of gas particles in the bullet is $4 \times 10^5$), has $9\times 10^5$ particles, $5\times 10^5$ collisional and the remainder dark matter particles. The gravitational spline softening is set equal to 5 kpc for the gaseous and
dark component. 

\section{Projected analysis}

Each numerical work which aims to simulate the bullet cluster should be able to reproduce simultaneously the main features observed in  X-ray maps (the bow shock, the relative surface brightness of the bullet and the main cluster),  and the observed surface brightness and temperature profiles across the shock discontinuity. An additional constraint is provided by the observed displacement between X-ray and lensing maps, which is not negligible in both the main and the sub-cluster ($\sim 110$ and $\sim 270$ kpc respectively, according to Clowe et al. 2006).

A first indication about the validity of a model arises from the qualitative comparison of our simulated X-ray surface brightness maps with the X-ray 500 ks \emph{Chandra} ACIS-I images provided by \citet{Markevitch06}.

The impact parameter $b$ is not strictly constrained by observations. 
Nevertheless a head-on merger, with $b=0$, seems to be excluded by comparing deep X-ray observations and weak lensing maps. In particular in the top panel of Fig. 1 -- which is Fig. 1b  of \citet{Clowe06} -- the brightest gas associated with the main cluster is not located along the line which connects the centers  of the two total mass distributions.
Moreover, the X-ray emission from the main cluster is asymmetric, with a peak in the north of the image and an extended tail of less bright material pointing south.
These features are hardly associable with a zero impact parameter interaction, as shown in the middle and bottom panels of Fig. \ref{b=0} which illustrate two simulated 1:6 head-on
encounters where the sub-cluster moves from the left to the right of the image
($x$ axis of the simulation) with decreasing initial velocities (the middle panel
corresponds to run 1:6vb0 of Table \ref{runs}, the bottom one to 1:6v3000b0).
Images are projected along an axis perpendicular to the collision plane (the
encounter is seen face-on) and the selected snapshot is the one which most
closely matches the observed distance between the centers of the total mass distributions,
associated with the two clusters -- about 720 kpc from \citet{Bradac06} -- once
the sub-cluster has passed through the core of the main system
(hereafter the present time).

In the lower two panels of Fig. \ref{b=0} colours represent X-ray maps in the \emph{Chandra} energy band (0.8-4 keV) generated using the Theoretical Image Processing System (TIPSY), which produces projected X-ray surface brightness maps with the appropriate variable SPH kernel applied individually to the flux represented by each particle. 
Assuming complete ionization and zero metallicity  (metal lines are expected to provide a significant contribution to emissivity only at relatively low temperatures, less than 2 keV ) the X-ray luminosity in a given energy band is defined as \citep{Borgani04}
\begin{equation}
L_X = (\mu m_p)^{-2}\sum_{i}^{N_{gas}} m_i \rho_i \Lambda  (T_i), 
\end{equation}
where $\Lambda (T_i)$ is the cooling function in the specific band,  $T_i$, $\rho_{i}$ and $m_i$ are temperature, density and mass associated with the $i-th$ hot ($T_i>10^5 $K ) gas particle, respectively, $m_p$ is the proton mass and  $\mu$= 0.6 the mean molecular weight.
The sum runs over all the $N_{gas}$ particles within an oblong of base equal to the pixel size and major axis oriented along the line of sight.
When the X-ray luminosity of the entire cluster is calculated  $N_{gas}$ is the number of hot particles within the virial radius $r_{vir}$.  
The cooling function is computed using a Raymond-Smith code \citep{Raymond77} for a gas of primordial composition. The energy band (0.8-4 keV) is chosen in such a way to reproduce the 500 ks \emph{Chandra} ACIS-I image of the bullet cluster \citep{Markevitch06}.
The entire energy band, used to calculate bolometric X-ray luminosities in the following of the paper, goes from 5 eV to $5 \times 10^4$ keV.

The lower two panels of Fig. \ref{b=0} show that after an encounter with zero impact parameter the displacement of the main cluster's X-ray peak is aligned with the $x$ axis. A large relative velocity (central panel) induces a significant offset between the dark and baryonic component of the main cluster (see Table \ref{runsresults} for details) but it also leads to substantial disruption of the main cluster gaseous core.
Moreover, the displacement of the bullet from its dark halo (278 kpc) is much larger than observed.
Decreasing the relative velocity between the two clusters (bottom panel) two X-ray peaks
are clearly visible but the displacement of the main cluster gas is now negligible due to
the lower ram-pressure experienced by the main cluster core.
\citet{Springel07} (Fig. 7 in their article) provide further examples of head-on encounters with even lower mass ratios and relative velocities ($v=2600$ km s$^{-1}$ in the center of mass rest frame). Even assuming extremely low concentrations ($c=2$) for the main halo, the authors never reproduce the displacement observed in the two systems. 
Increasing the concentration strongly increases the luminosity of the main
cluster, which appears much brighter than the bullet, contrary to what  is
observed.

A bow shock is clearly visible on the right of each image. The shape of the
shock front is only marginally dependent on the kinematics  of the model while the
distance between the edge of the bullet (the so called contact discontinuity)
and the shock front becomes larger for decreasing bullet velocities. 
The contact discontinuity itself is much flatter in the case of 1:6b0 than in
the low velocity encounter 1:6v3000b0 and clearly not comparable with observations,
which show a more narrow structure. 
In general, a more efficient ram-pressure during the phase of core-core
interaction is associated with a larger opening angle of the contact
discontinuity at the present time \citep{Quilis01}.  

The rest of the runs listed in Table \ref{runs} have an impact parameter $b$ equal to 150 kpc, comparable with the core radius $r_c$ of the main cluster gas distribution for most of the models.
For a plasma distributed according to a $\beta$ profile like the one adopted in this paper the assumption $b=150$ kpc implies that the maximum external density crossed by the core of the sub-cluster is 1.5 times smaller than the one 
it would pass through for $b=0$ kpc. 
For the same initial velocity, the sub-cluster also sees a weaker potential with respect to the case of a head-on encounter and experiences a smaller maximum orbital velocity. 
A choice of a much larger value of $b$ would further decrease the mutual ram-pressure between the two systems and require much higher relative velocities in order to explain the observed offset between the gas and dark matter.

\begin{figure*}
\includegraphics[%
  height=70mm]{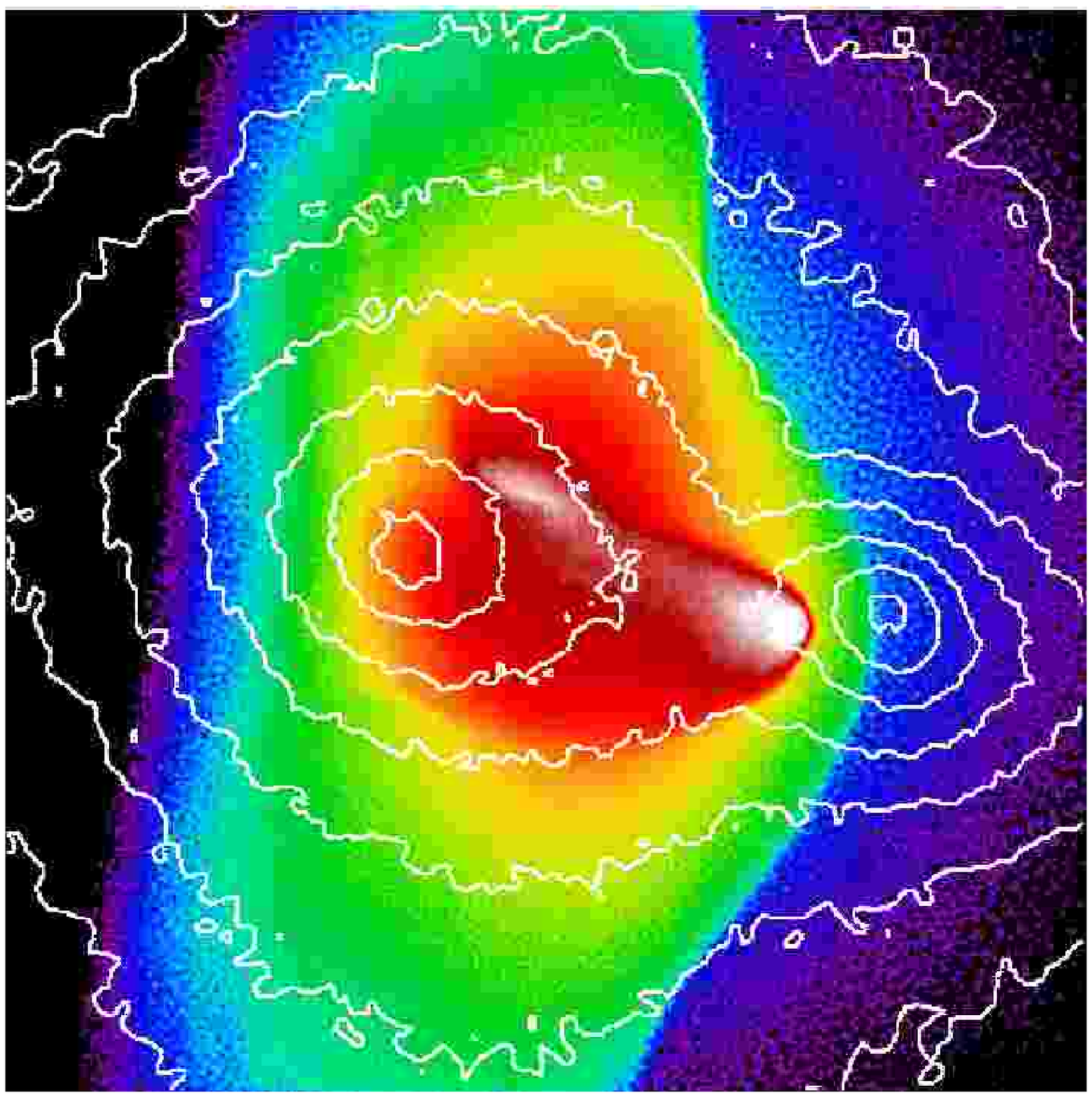}
\includegraphics[%
  height=70mm]{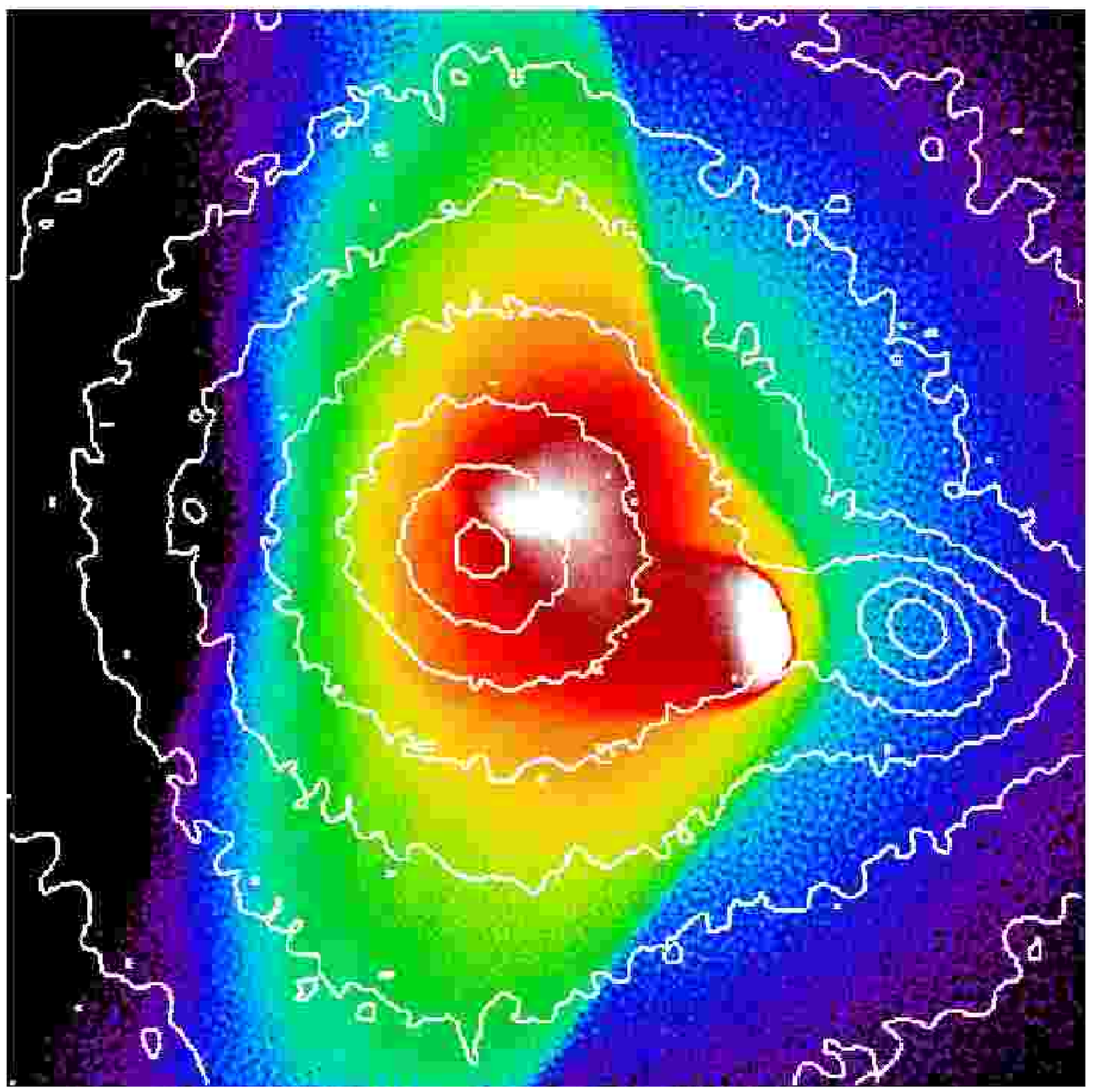}
\includegraphics[%
  height=70mm]{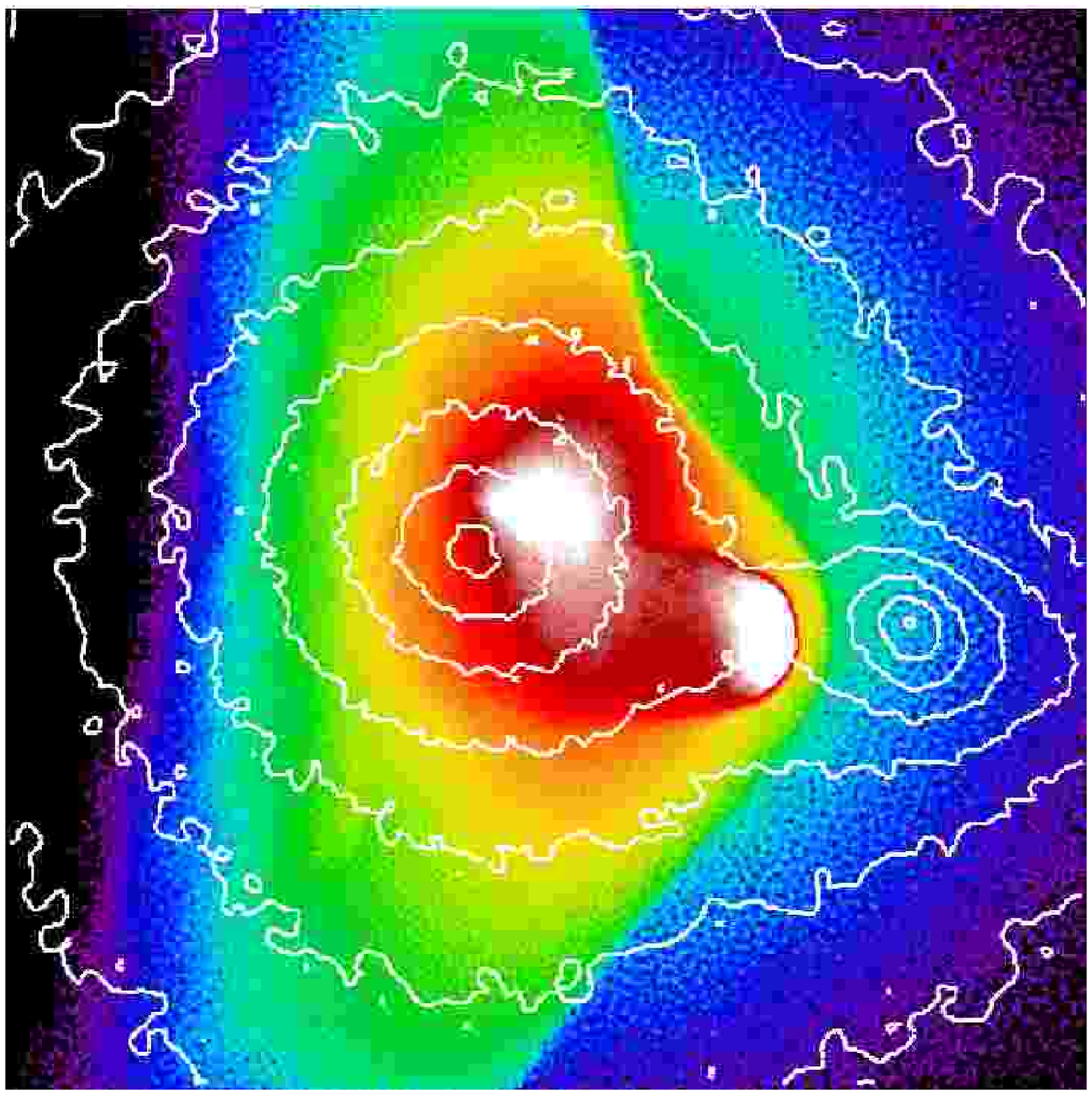}
\includegraphics[%
  height=70mm]{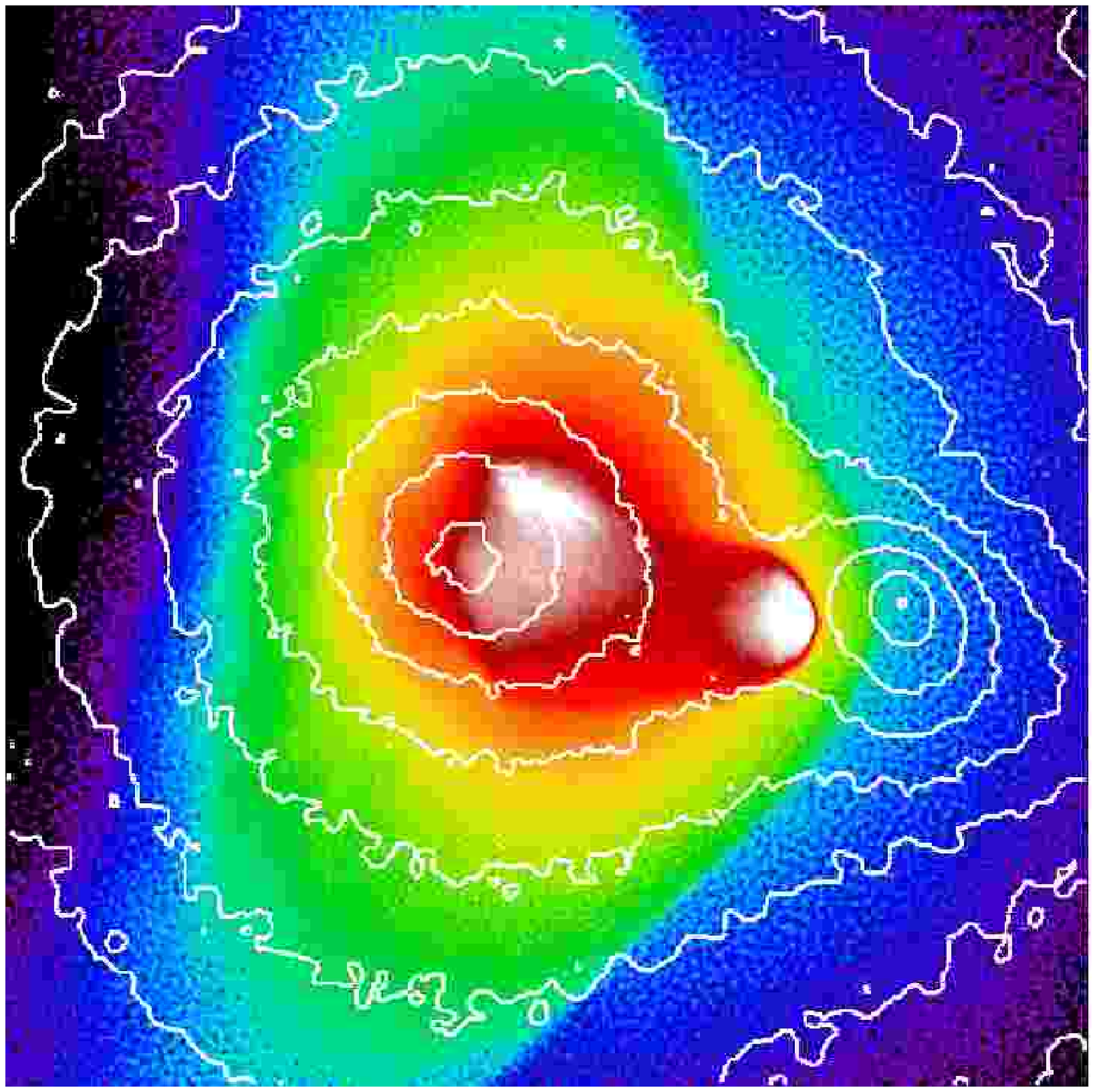}
\includegraphics[%
  height=70mm]{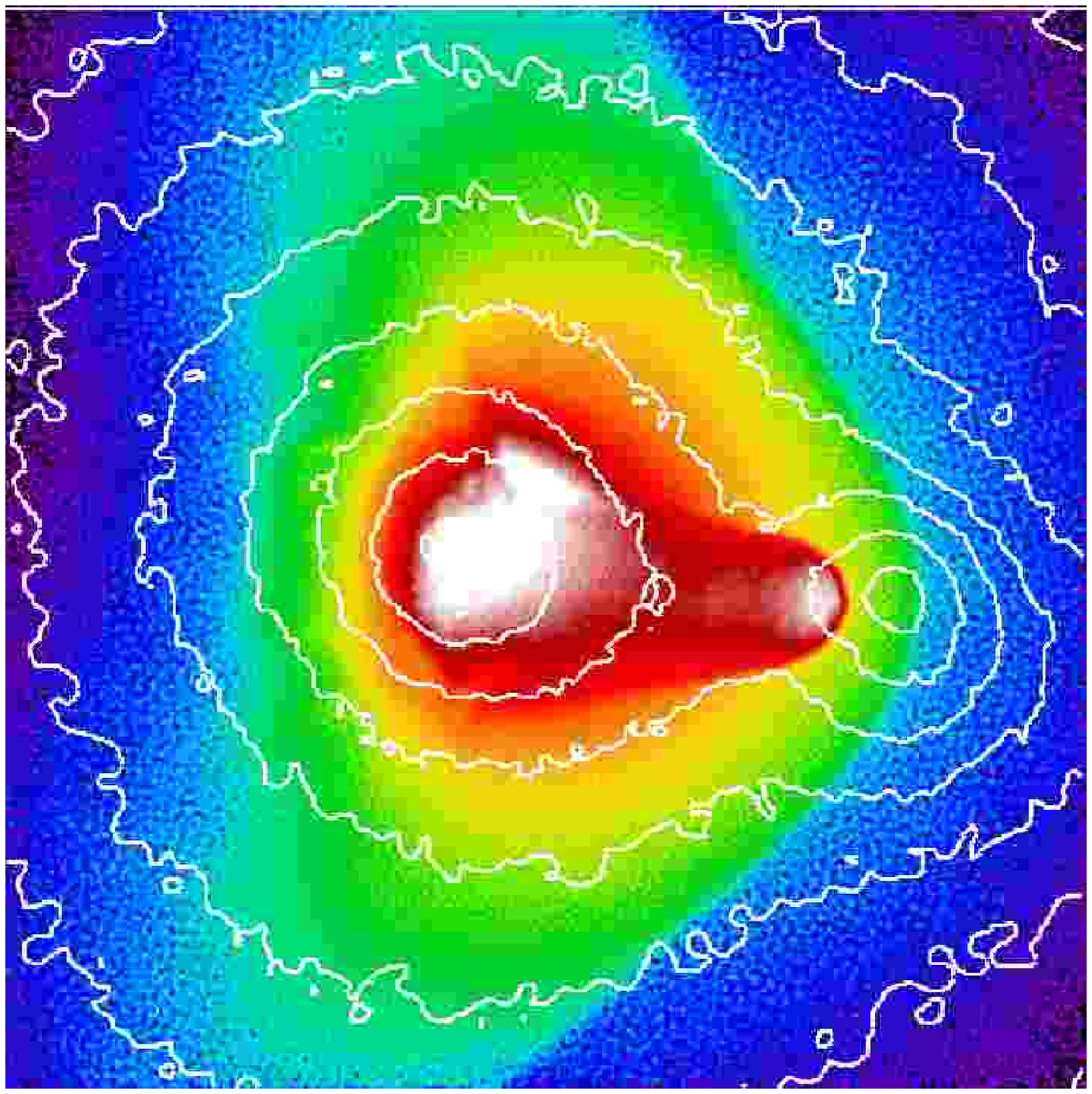}
\includegraphics[%
  height=70mm]{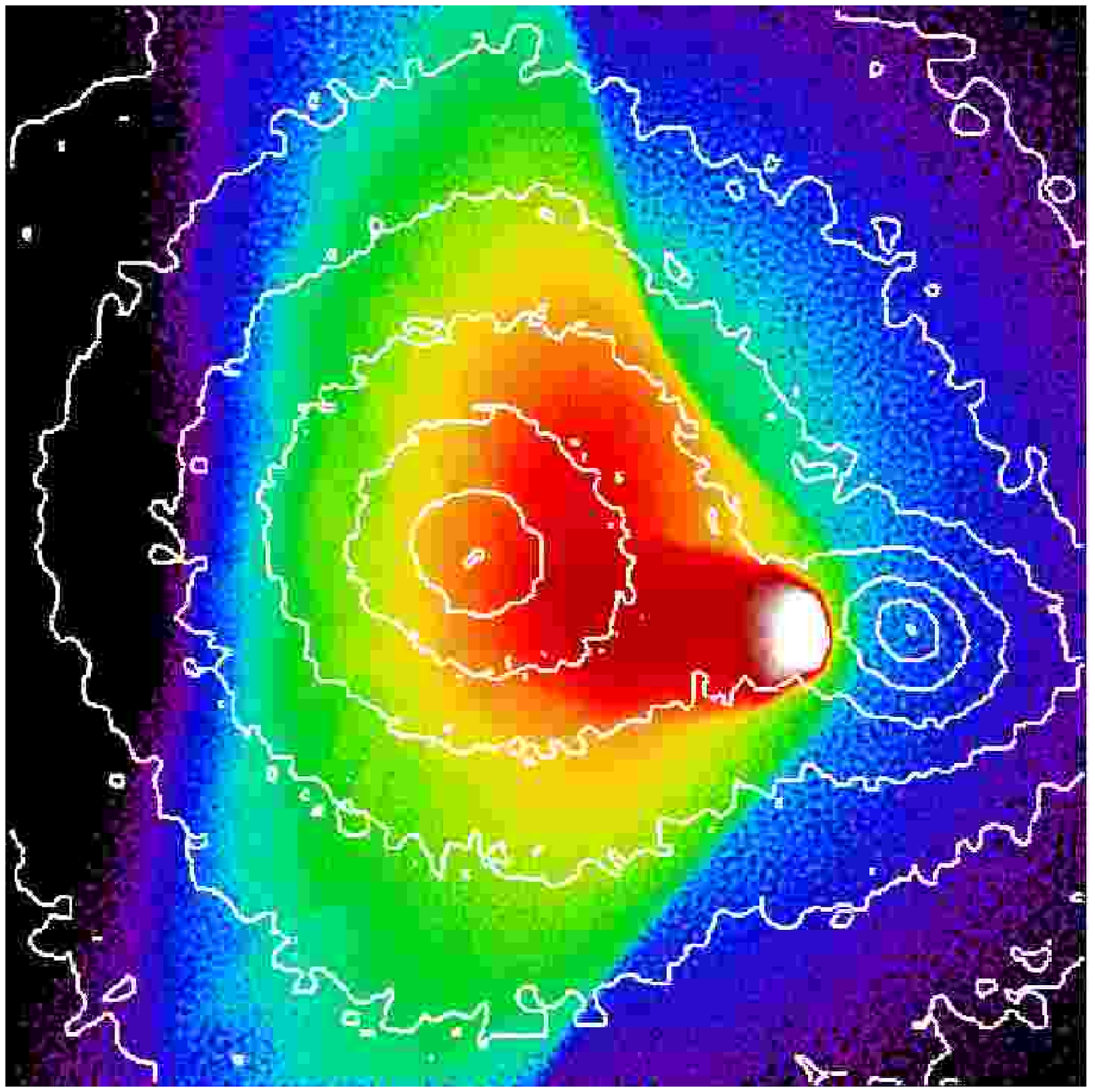}
\includegraphics[%
  height=5mm]{colorbar.ps}
\caption{0.8-4 keV surface brightness maps of runs (from the top left to the bottom right) 1:3, 1:8, 1:6, 1:6v3000, 1:6v2000 and 1:6c4. Logarithmic colour scaling is indicated by the key at the bottom of the figure, with violet corresponding to $10^{38}$ ergs$^{-1}$ kpc$^{-2}$ and white to $1.8 \times 10^{41}$ erg s$^{-1}$ kpc$^{-2}$ in runs 1:8 and 1:6c4, to $2.34 \times 10^{41}$ erg s$^{-1}$ kpc$^{-2}$ in run 1:3 and to $2 \times 10^{41}$ ergs$^{-1}$ kpc$^{-2}$ in the remaining cases. Projected isodensity contours of the total mass distribution are shown. Limits are  $2.3 \times 10^{3}$ and $2.3 \times 10^{9}$ M$_{\odot}$ kpc$^{-2}$. Each box size is 1800 kpc.}
\label{b>0}
\end{figure*}

\begin{figure*}
\includegraphics[%
  height=70mm]{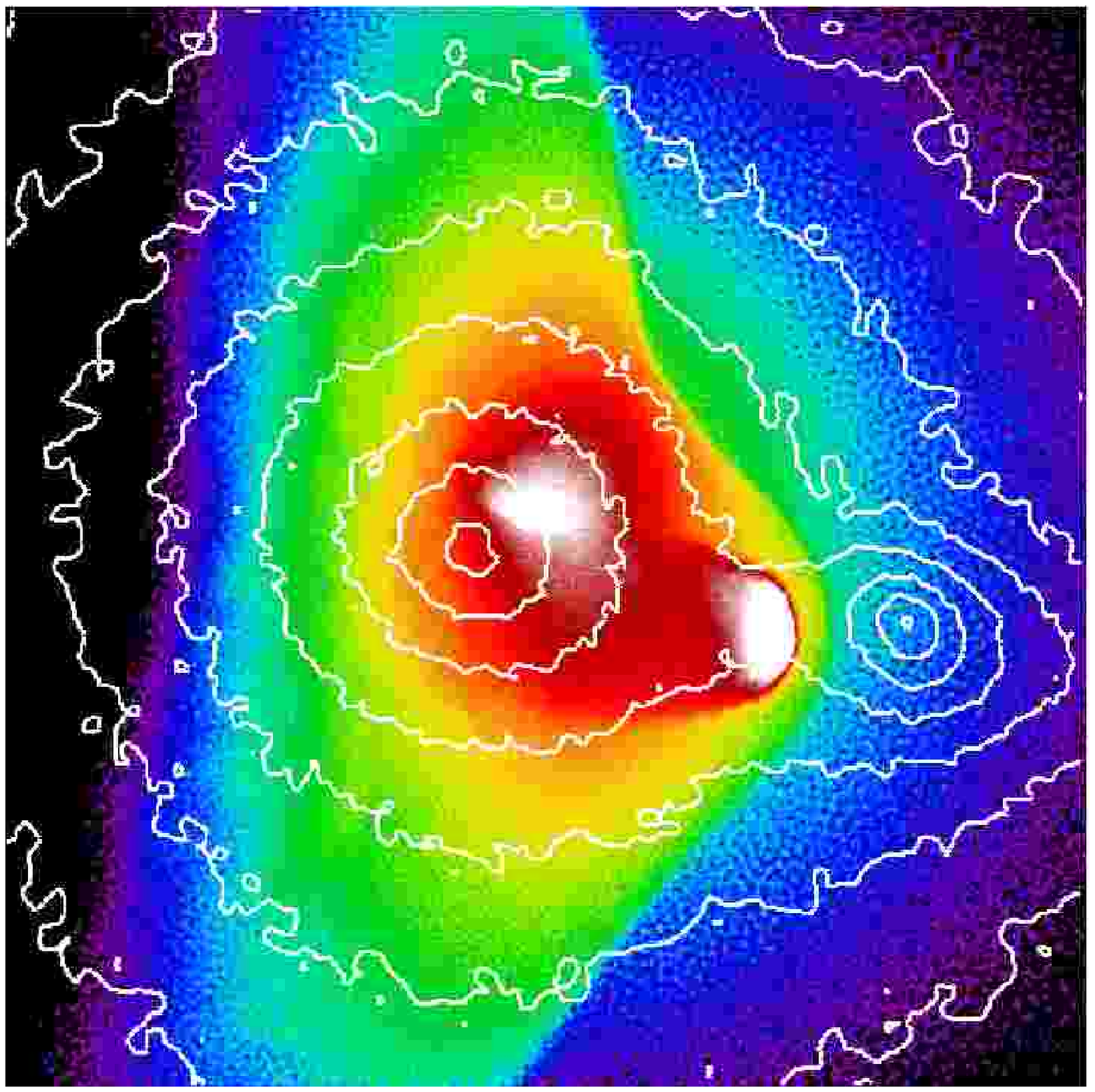}
\includegraphics[%
  height=70mm]{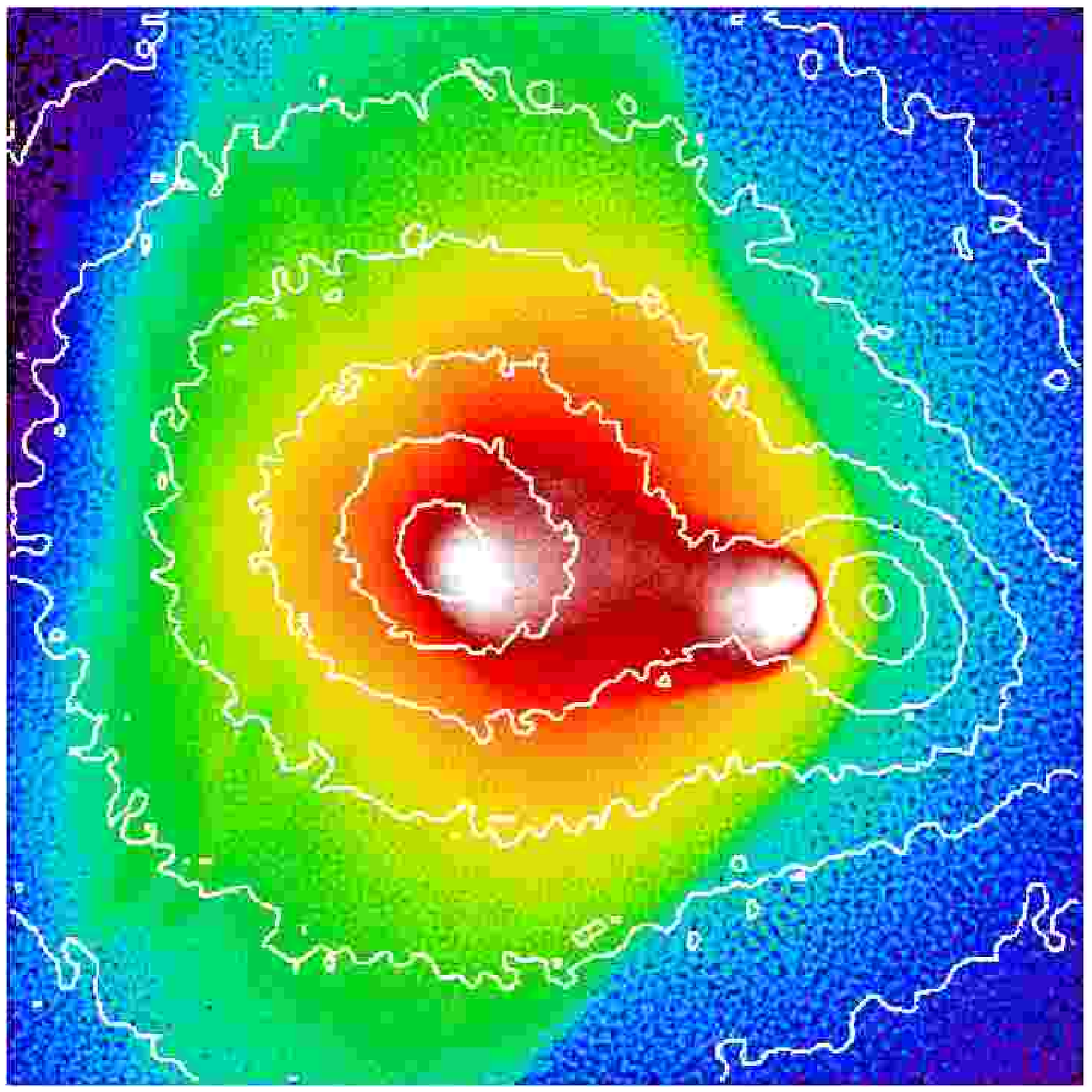}
\includegraphics[%
  height=5mm]{colorbar.ps}
\caption{Same as in Fig. \ref{b>0} for the cooling run 1:6c (the X-ray upper limit is $1.8 \times 10^{41}$ erg s$^{-1}$kpc$^{-2}$) and run 1:6v3000big ($3.9 \times 10^{41}$ erg s$^{-1}$kpc$^{-2}$). }
\label{b>02}
\end{figure*}

In Fig. \ref{b>0} and \ref{b>02} we illustrate the projected X-ray surface
brightness maps of some interesting models with impact parameter $b= 150$ kpc. 
The encounters are shown face-on. 
The box size and the surface mass density contours are the same as in
the last two panels of Fig. \ref{b=0}. 
In order to underline the morphological details of the high emission regions
the  upper limit of the surface brightness scale varies in the
different images. 
Individual values are indicated in the captions.

Among this sub-sample of runs, 1:3 produces the largest displacement of the X-ray peak associated with
the main cluster, but the X-ray map differs from the observations. 
In particular a large strip of strongly emitting gas still connects the two
X-ray peaks while the morphology of the main cluster peak is much more
elongated than observed.
Decreasing the mass ratio between the two interacting systems (runs 1:6 and
1:8) the displacement in the bullet becomes larger than that in the main
cluster.
The run 1:6 is characterized by an initial sub-cluster velocity of 5000 km
s$^{-1}$ (as well as runs 1:3 and 1:8) in the system of reference where the main cluster is at rest, that
corresponds to a present time velocity of $\sim 4300$ km s$^{-1}$ in the
center of mass rest frame. The same model is simulated assuming lower
relative velocities (1:6v3000 and 1:6v2000). 
With decreasing velocities the offset in the sub-cluster becomes smaller
and the X-ray emission from the bullet less pronounced
with respect to the bright X-ray emitting region at the center of the main cluster.
At the same time the shape of the contact discontinuity changes, getting progressively more narrow while the
distance between the contact discontinuity and the shock front increases as 
will be shown more quantitatively in the next section.
 The displacement associated with the main-cluster is determined by the distance of closest approach between the centers of the two clusters, which becomes smaller  -- assuming the same initial impact parameter $b$ -- with decreasing bullet velocities. 
Indeed, the separation between the main cluster X-ray emission peak and its
dark matter counterpart is maximum in the case of the low velocity run 1:6v2000.

\begin{figure}
\includegraphics[%
  height=75mm]{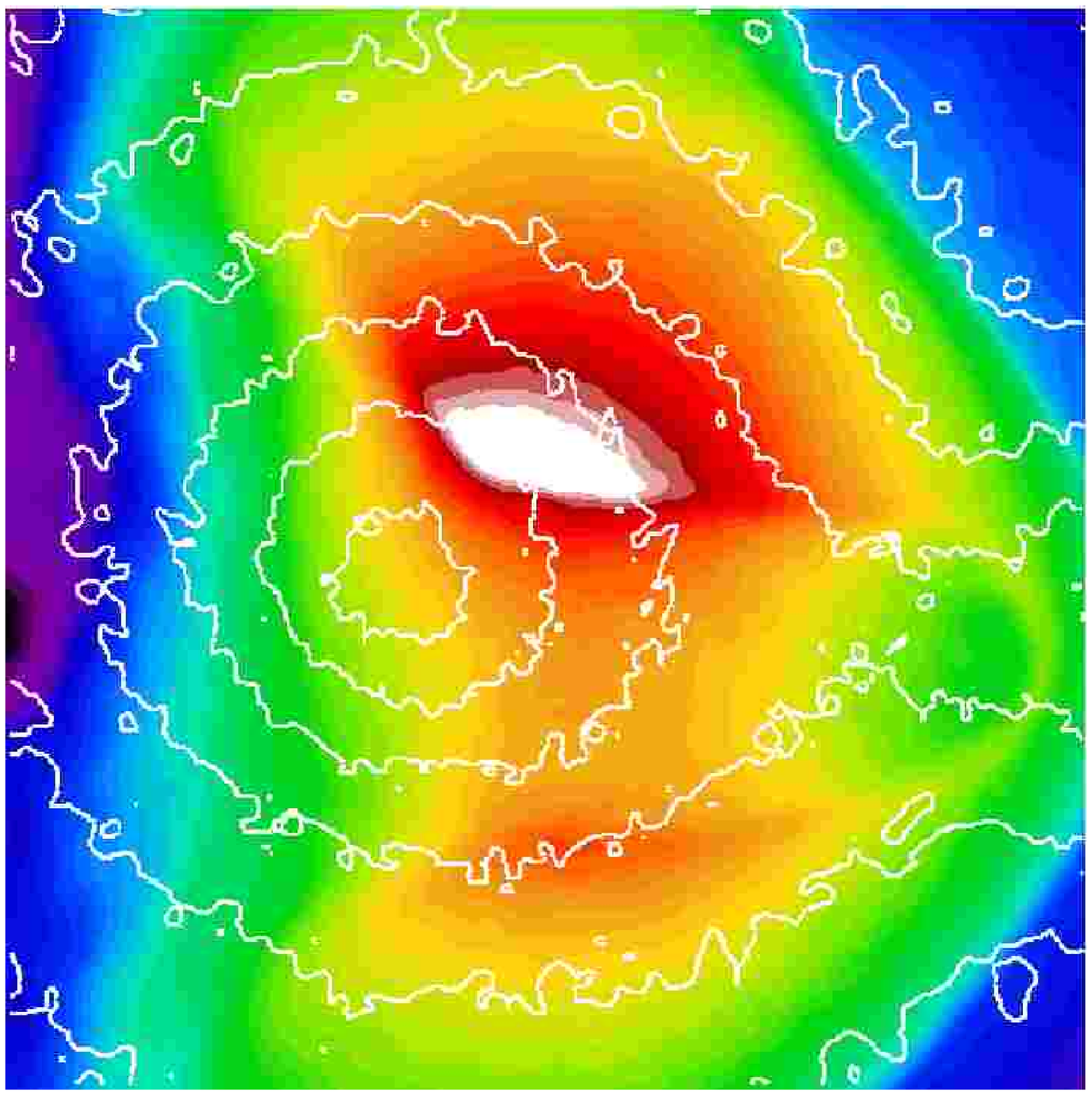}
\includegraphics[%
  height=75mm]{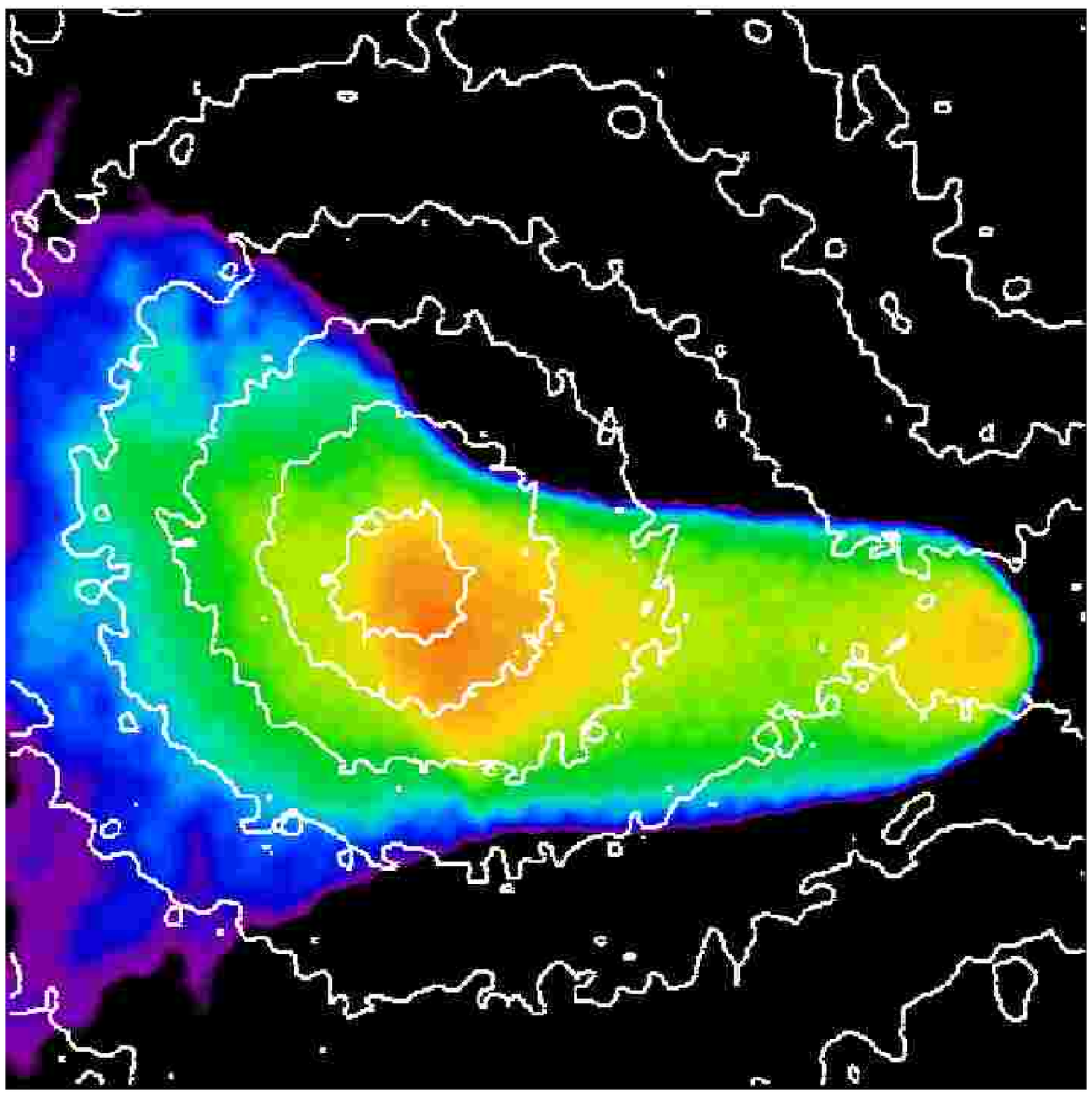}
 \label{future}
\includegraphics[%
  height=4mm]{colorbar.ps}
\caption{Run 1:6v3000. Gas originating from the main (top panel) and the sub-cluster (bottom) is projected individually along the $z$-axis perpendicular to the plane of the encounter. Violet corresponds to a surface density of $2.3 \times 10^{5} M_{\odot}$ kpc$^{-2}$ and white to $2.3 \times 10^{8} M_{\odot}$ kpc$^{-2}$. Projected isodensity contours of the total mass distribution are drawn on  top of the image. The box size is 1 Mpc. }
\label{mixing}
\end{figure}

A not negligible fraction of the X-ray emission visible at the present time near the center of the main-cluster is actually associated with hot gas stripped from the external regions of the bullet. 
Fig. \ref{mixing} refers to run 1:6v3000. It shows the individual distribution of gas originating from the main (top) and sub-cluster (bottom) and lying at the present time within 1 Mpc from the center of the system.
Comparing these images with the middle right panel of Fig. \ref{b>0} it appears evident that the bright elongated X-ray feature crossing the second innermost isodensity contour is associated with the displaced gaseous center of the main cluster while the surrounding more diffuse region hosts a significant amount of sub-cluster gas.  
Indeed, as it shown in the upper panel of Fig. \ref{mixing}, the motion of the bullet  across the inner regions of the main system creates a low density ``tunnel''  in the main cluster gas distribution. At the same time the sub-cluster  looses a large amount of baryonic material during the phase of core-core interaction. This material, which fills the tunnel, falls back into the gravitational center of the main cluster and resides now at more than 500 kpc distance from the X-ray bullet. As we will show in the next section, the amount of gas deposited by the sub-cluster in the central regions of the main system increases with decreasing relative velocities. This trend explains the relative increase in luminosity of the diffuse strongly emitting component if compared to the peak associated to the main cluster core gas (always in the upper right region with respect to the center of the mass density distribution), when we compare run 1:6 with the low velocity run 1:6v2000 where it becomes the primary peak of X-ray emission.  

As shown in the bottom right panel of Fig. \ref{b>0} a main halo with low concentration ($c=4$) does not survive a 1:6 sub-cluster encounter with velocity $v=5000$ km s$^{-1}$ and its X-ray peak is destroyed.

As previously noticed by \citet{Springel07} the choice of a lower gas fraction
($f_g = 0.12$ in three of the last four runs of Table \ref{runs}) does not affect
significantly the X-ray map morphology although the displacement of the two
luminosity peaks with respect to their dark matter counterparts slightly changes.

Finally, we tested the consequences caused by including radiative cooling and choosing a larger main halo model (Fig. \ref{b>02}).
Cooling makes the contact discontinuity more narrow and the amount of diffuse X-ray gas around the peak associated with the main cluster smaller. The offset in the main and sub-cluster remains unaltered.
Run 1:6v3000big is characterized by a main cluster total mass of $\sim 1.64
\times 10^{15} M_{\odot}$, which is closer to the value adopted by
\citep{Springel07} and predicted by  fitting the large field weak
lensing data with extremely low concentrated ($c \sim 2$) NFW
halos. Although the initial relative velocity is only 3000 km s$^{-1}$, due to
the large mass of the host halo the present time velocity of the bullet in the
center of mass rest frame is much higher than in the corresponding 1:6v3000
run. Consequently the offset of the X-ray peak associated with the
main cluster is less than 100 kpc and the amount of gas lost by the bullet in
the core of the main system closer to that observed in 1:6 than in 1:6v3000. 

Summarizing, run 1:6v3000, with mass ratio 1:6, initial relative velocity $v=3000$ km s$^{-1}$ and present time sub-cluster velocity $v \sim 3100$ km s$^{-1}$ in the center of mass rest frame, best reproduces the main features observed in X-ray maps, in particular the peculiar morphology of the X-ray emission associated with the main cluster, the relative surface brightness between the main and the sub-cluster, the shape of the shock front and of the contact discontinuity. Although the low velocity run 1:6v2000 leads to a X-ray displacement closer to that observed by \citet{Clowe06} both in the main and in the sub-cluster, this model seems to be excluded on the basis of a pure morphological comparison with the observational data. Indeed, the bullet seems to be much less bright than the center of the main cluster.

\begin{figure}
\epsfxsize=8truecm \epsfbox{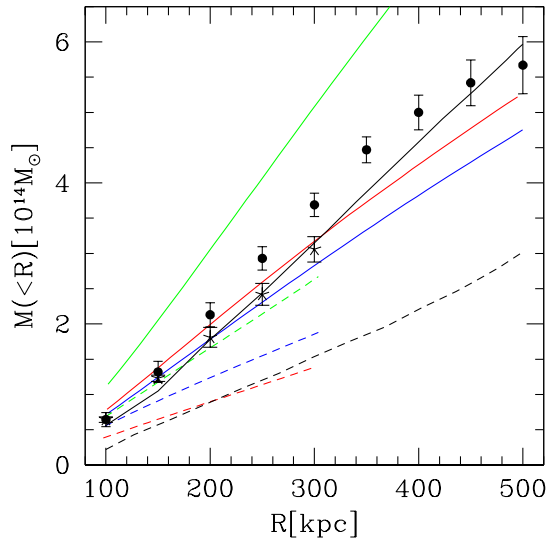} 
\caption{Cumulative projected mass profiles. Points represent the observational results
  \citep{Bradac06} for the main (dots) and the sub-cluster (stars). Black
  curves refer to the model of \citet{Springel07} (solid and dashed black line
  indicates the main and sub-cluster mass, respectively). Coloured lines
  refer to the models in this work, always with solid curves indicating the main cluster and dashed ones  the bullet.  In details the red lines are associated to our favourite model 1:6v3000, blue to 1:3lfg and green to 1:6v3000big.}
\label{masses}
\end{figure}

Fig. \ref{masses} compares the integrated mass profiles of the main and the
sub-cluster for different mass models with the gravitational lensing
results from \citet{Bradac06} and previous simulations by \citet{Springel07}.
The profiles inferred from observations are obtained by measuring the enclosed
mass in cylinders centered on the southern cD of the main cluster and on the
BCG of the sub-cluster.
Here we calculated the present time projected mass within cylindrical bins centered on 
the center of mass of the two clusters.  
The massive cluster run 1:6v3000big (green lines), overestimates the cumulative 
profile of the main cluster, while an initial main cluster mass of
$\sim 8.3 \times 10^{14} M_{\odot}$  (1:6v3000, in red) fits the observational points better than
the large-mass-low-concentration model of \citet{Springel07} in the central
regions and underestimates the projected mass by less than $ 20\%$ between 250
and 450 kpc from the center.
Blue lines refer to run 1:3lfg where the main halo has the same dark matter mass than 1:6v3000 but lower baryonic fraction.
It is interesting to notice that using almost the same initial sub-cluster mass  (the sub-cluster model in run 1:6v3000big and 1:3lfg only differs in the baryonic fraction) we can nearly reproduce the observed projected mass in the bullet only in the case of a very massive host halo (green dashed line, 1:6v3000big run) which overpredicts the main cluster profile (green solid line).
On the  other hand a smaller main halo -- which better fits the observational data (blue solid line) -- associated with a similar sub-cluster mass (1:3lfg)  underpredicts the lensing results (blue dashed line). Our favourite run 1:6v3000 gives even lower projected sub-cluster mass estimates (red dashed line) and  lies close to the model proposed by \citet{Springel07} (their sub-cluster mass is similar to the one used in our 1:6 models).  
These results  seem to indicate that the mass of the sub-cluster is a
significant part of the main cluster  mass, with initial mass ratio between the sub and the main cluster much larger than 1:3. 
\citet{Nusser07} successfully reproduces the projected mass associated with the sub-cluster only  assuming very massive halos and high mass ratios.
In particular the smallest mass ratio allowed by their models is 1:2.7, with a main cluster  and bullet mass of $3.2 \times 10^{15} M_{\odot}$ and $1.2 \times 10^{15} M_{\odot}$, respectively.
Such high masses are however  clearly incompatible with the values inferred by galaxy kinematics.
In particular the virial mass derived by \citet{Barrena02} for the main cluster is $1.24 \times 10^{15}$ (comparable with the value obtained by Girardi \& Mezzetti 2001). They also measure the main cluster total R-band luminosity and find $L_R = 10^{12} L_{\odot}.$ Using $L_{R}$ to estimate the cluster mass \citep{Miller05} we obtain $M_{200} \sim 10^{15} M_{\odot}$.
The bullet mass determination of \citet{Barrena02} is less reliable since it is based on only seven galaxies and on the assumption of equilibrium while the sub-cluster seems to be tidally perturbed beyond $\sim 200$ kpc from its center.  
They find $M_{200} = 1.2 \times 10^{13} M_{\odot}$. From the total R-band luminosity ($L_R = 0.2 \times 10^{12} L_{\odot}$) we get a value almost five times larger ($M_{200} \sim 5 \times 10^{13} M_{\odot}$). 
The bullet mass proposed by \citet{Nusser07} is even one order of magnitude larger than that obtained from weak lensing analysis ($1.5 \times 10^{14} M_{\odot}$) of a new larger lensing field which covers most of the area occupied by the system  (Clowe et al. in prep.).
With such low mass ratios we cannot reproduce the morphology of the X-ray maps (upper left panel of Fig. \ref{b>0}) since the luminosity peak associated with the main cluster core is almost completely destroyed by the interaction.
    
Actually any attempt to derive the mass of the main and sub-cluster simply by projecting the mass of two isolated NFW halos and fitting the observed lensing data  -- like the analysis of  \citet{Nusser07} -- is going to provide wrong results. 
Indeed, as we will show in Section 4, the dark matter profiles of the main and sub-cluster  change significantly during the central phases of the interaction, with a density increase in the inner regions and a decrement beyond 0.3 $r_{vir}$. This effect, which characterizes both the main and the sub-cluster, is not related to stripping processes  but can be explained  by the fact that the two halos are strongly perturbed and cannot be idealized as the sum of two isolated equilibrium  models.
Fitting the projected cumulative mass around the centers of the two interacting clusters with NFW models we would overestimate either the virial mass or the concentration of the original clusters.

\begin{figure}
\includegraphics[%
  height=75mm]{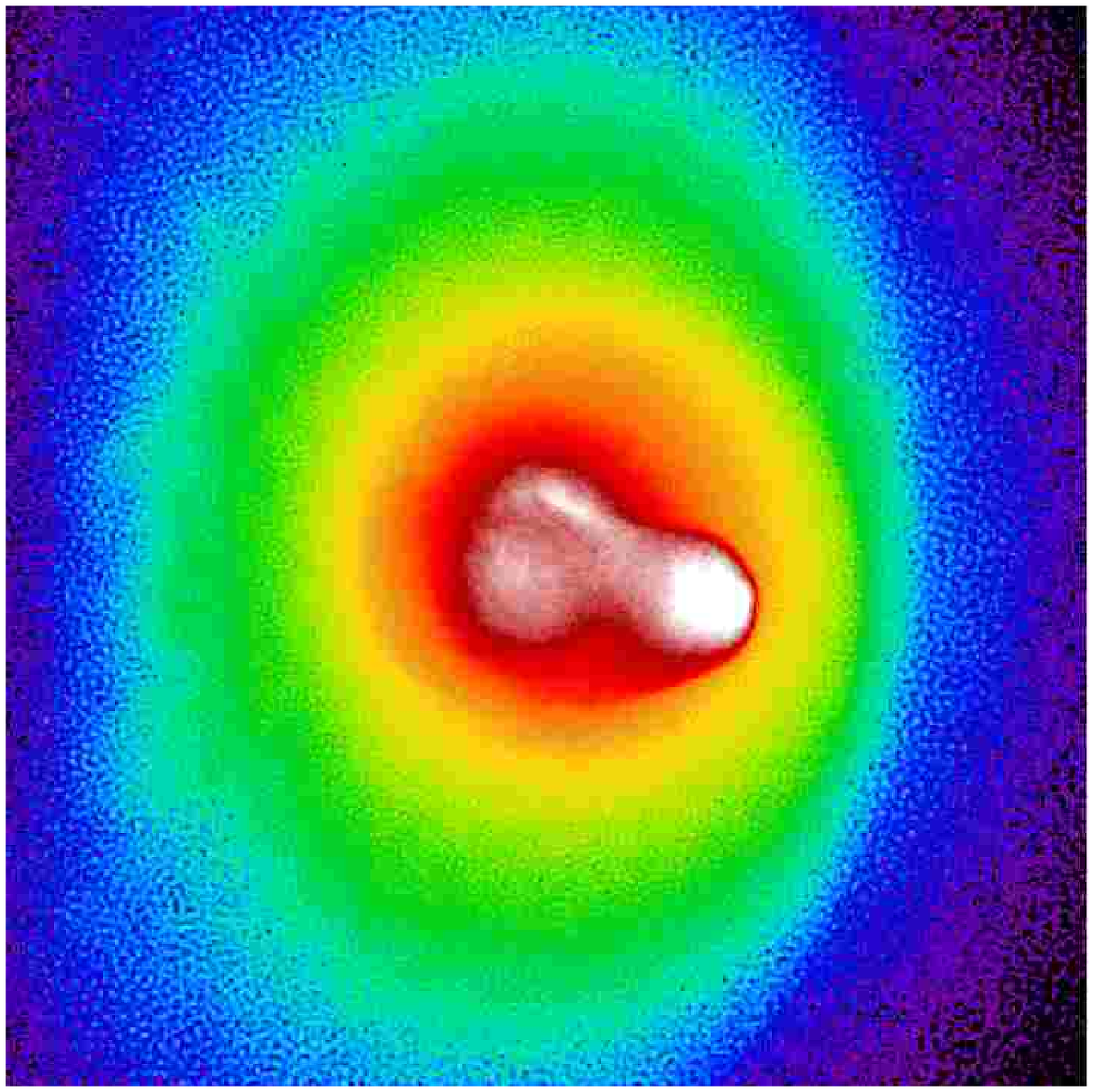}
\includegraphics[%
  height=75mm]{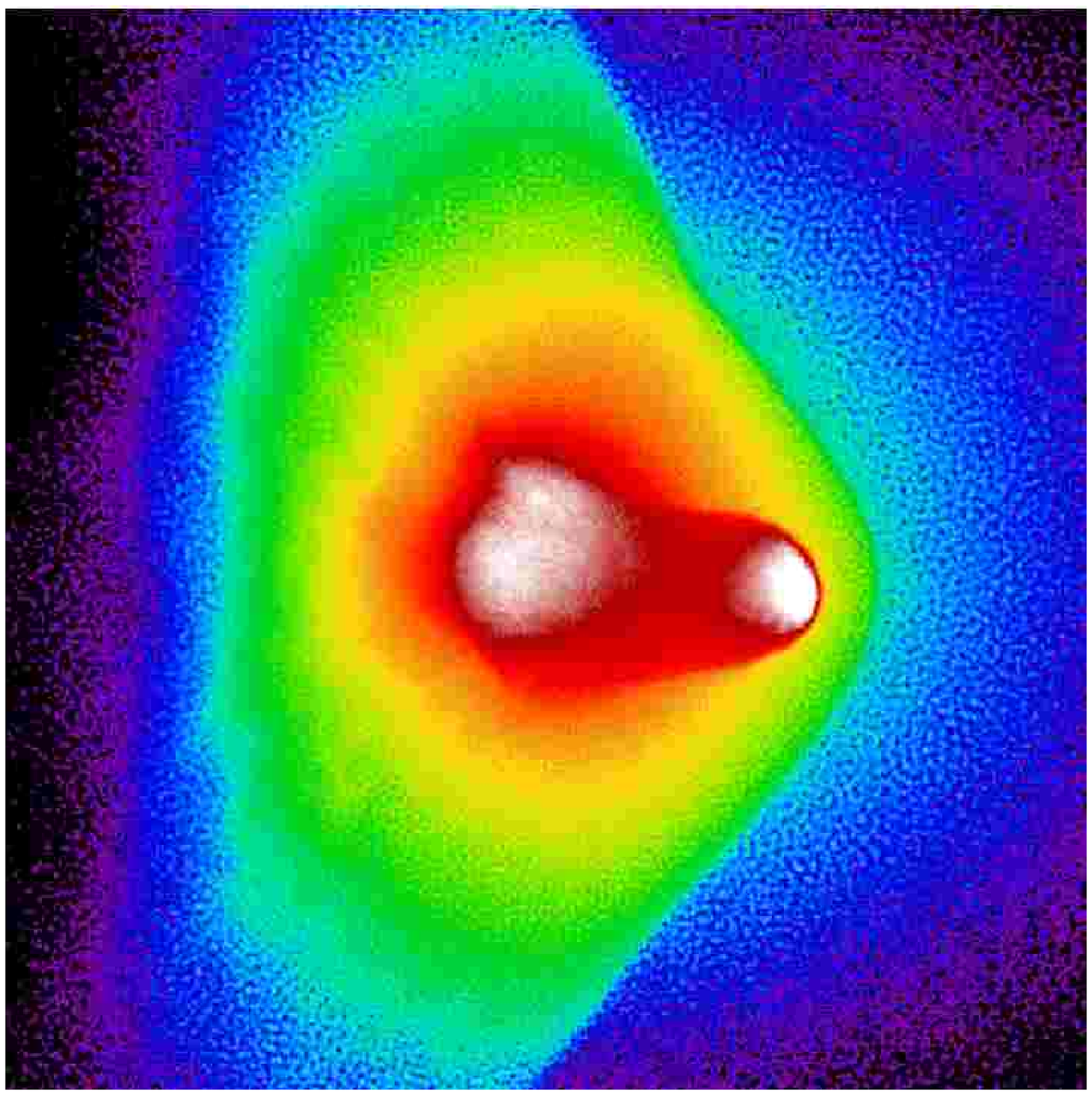}
 \label{future}
\includegraphics[%
  height=4mm]{colorbar.ps}
\caption{X-ray surface brightness maps for different inclinations of run 1:6v3000. In the top panel the orbital plane is left rotated by 45 degrees with respect to the projection plane while in the panel on the bottom the encounter is seen with a top-down inclination of 45 degrees. The colour scaling is the same as in the corresponding face-on map (Fig. \ref{b>0}, middle right panel). The box size is 1.8 Mpc. }
\label{inclination}
\end{figure}

According to \citet{Barrena02} the line of sight velocity difference between the main and the sub-cluster is relatively small, about 600 km s$^{-1}$, which implies that the encounter must be occurring nearly in the plane of the sky.
In Fig. \ref{inclination} we show the X-ray surface brightness map of run 1:6v3000 when the orbital plane is viewed inclined by 45 degrees left (top panel) and down (bottom panel).
In the upper image the bullet (which is closer to the observer) appears to be much brighter than the main cluster while the contact discontinuity and the shock front are much rounder than in the case where the encounter is seen perpendicular to the orbital plane. Such a large inclination has to be excluded by a comparison with observations. Moreover, the line of sight velocity difference between the two clusters (we assume the line of sight kinematics of the dark matter to be coincident with that of the stellar component observed by Barrena et al. 2002) is much larger than the observed one ($\sim 2700$ km s$^{-1}$) due to the fact that the main component ($\sim 3200$ km s$^{-1}$ in the center of mass rest frame) of the sub-cluster velocity is oriented in the direction of the positive $x$ axis in the original face-on encounter. A velocity difference comparable to the observed one is reached for a right to left inclination of $\sim 10$ degrees. 
On the opposite, a top-down inclination of the plane of the encounter with respect to the line of sight (bottom panel) does not produce drastic changes in the morphology of the bow shock and of the edge of the bullet. Indeed the main and the sub-cluster still lie close to the plane of the sky, but the relative surface brightness between the dense and the diffuse strongly emitting components at the center of the main system changes. In particular the peak of X-ray emission due to the presence of hot gas deposited by the sub-cluster within the core of the main cluster is now comparable with the main cluster X-ray peak. The difference in line of sight velocity in this case is only $\sim 100$ km s$^{-1}$ and would probably not be distinguishable with respect to a pure face-on encounter.

\begin{figure}
\epsfxsize=8truecm \epsfbox{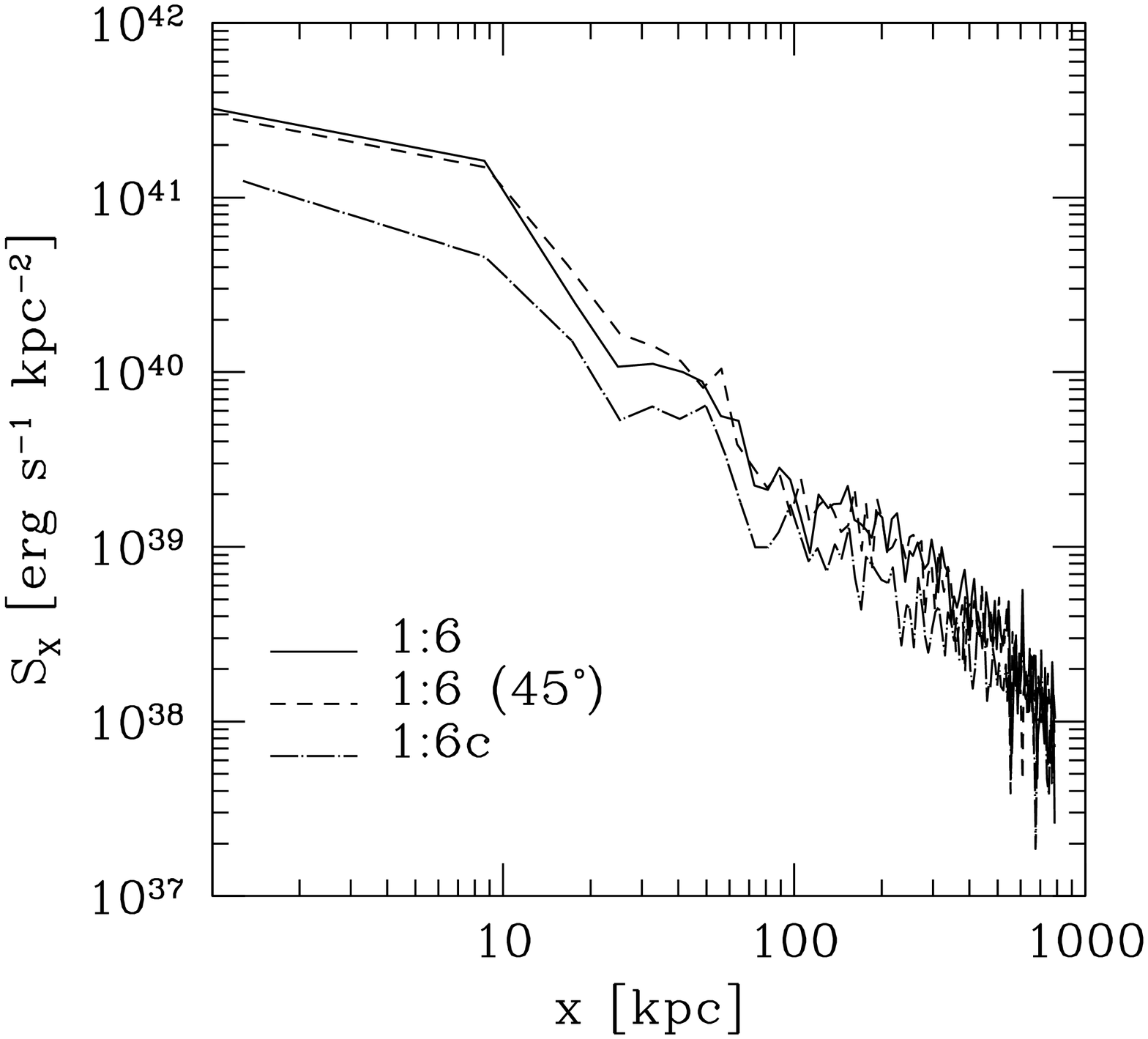} 
\caption{Runs 1:6 and 1:6c: X-ray surface brightness profiles across the shock discontinuity. The bullet is located at $x=0$ kpc and the shock front at $\sim 100$ kpc.  }
\label{sb}
\end{figure}

Fig. \ref{sb} represents the X-ray brightness profile measured in a narrow slit (with thickness 20 kpc) parallel to the $x$-axis across the shock front. The bullet is located at $x=0$ with $x$ increasing towards the pre-shock region. 
The general trend does not depend on the model and is comparable with the
profile suggested by observations \citep{Markevitch06}: all the simulations
show an inner bump associated with the bullet and an outer one (between $x=20$ and $x=100$ kpc) caused by
the shock. The abrupt jump at $x \sim 100$ kpc is the shock front while the
pre-shock region is well fitted by a two dimensional $\beta$-profile $S_X(x) = S_{X0}[1+(x/x_c)^2]^{-3\beta + 1/2}$. In the case of run 1:6  $x_c = 150$ kpc, $\beta=2/3$ and $S_{X0} = 8 \times 10^{40}$ erg s$^{-1}$ kpc$^{-2}$. 
A different orientation ($45^{\circ}$, bottom-up inclination of the
plane of the encounter) of the line of sight does not affect substantially the surface brightness profile. If cooling is activated the surface brightness of the pre and post-shock region is smaller, but the thickness of the shock-front and the jump in surface brightness are similar.

In order to calculate projected temperatures we need to define a weighting function.
The emission weighted temperature $T_{ew}$ was originally introduced to provide a better comparison between simulations and observations with respect to a simple mass weighted temperature definition and has been commonly used in the analysis of simulations \citep{Borgani04}. It assumes a weighting function proportional to the emissivity of each hot gas particle and is defined as
\begin{equation}
T_{ew} = \frac{{{\sum}_i}^{N_{gas}} m_i {\rho}_i \Lambda(T_i) T_i}{{{\sum}_i}^{N_{gas}} m_i  \rho_i \Lambda(T_i)}.
\end{equation}
Recently \citet{Mazzotta04} have demonstrated that for clusters with a complex thermal structure the emission weighted temperature always overestimates the spectroscopic temperature obtained from X-ray observations due to fact that the source is not a single temperature plasma.  
For clusters with a temperature $T > 3$ keV the discrepancy is  $\sim 20-30 \%$. This difference becomes particularly large in the presence of strong temperature gradients like shocks which appear to be much weaker in observations than what is predicted by emission-weighted temperature maps (Mathiesen \& Evrard 2001, Gardini et al. 2004, Rasia et al. 2005). 

\begin{figure*}
\includegraphics[%
  scale=0.4]{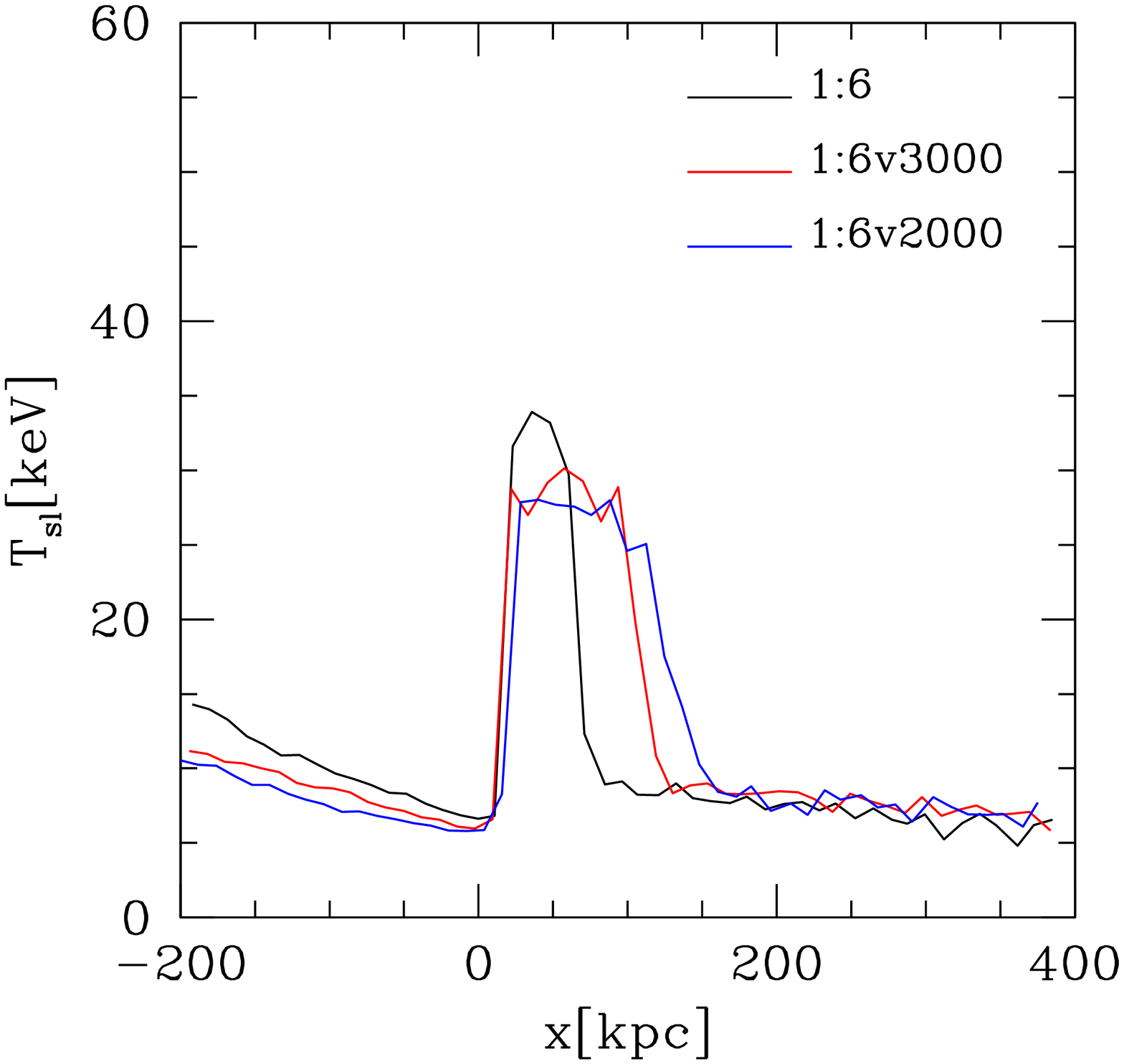}
\includegraphics[%
  scale=0.4]{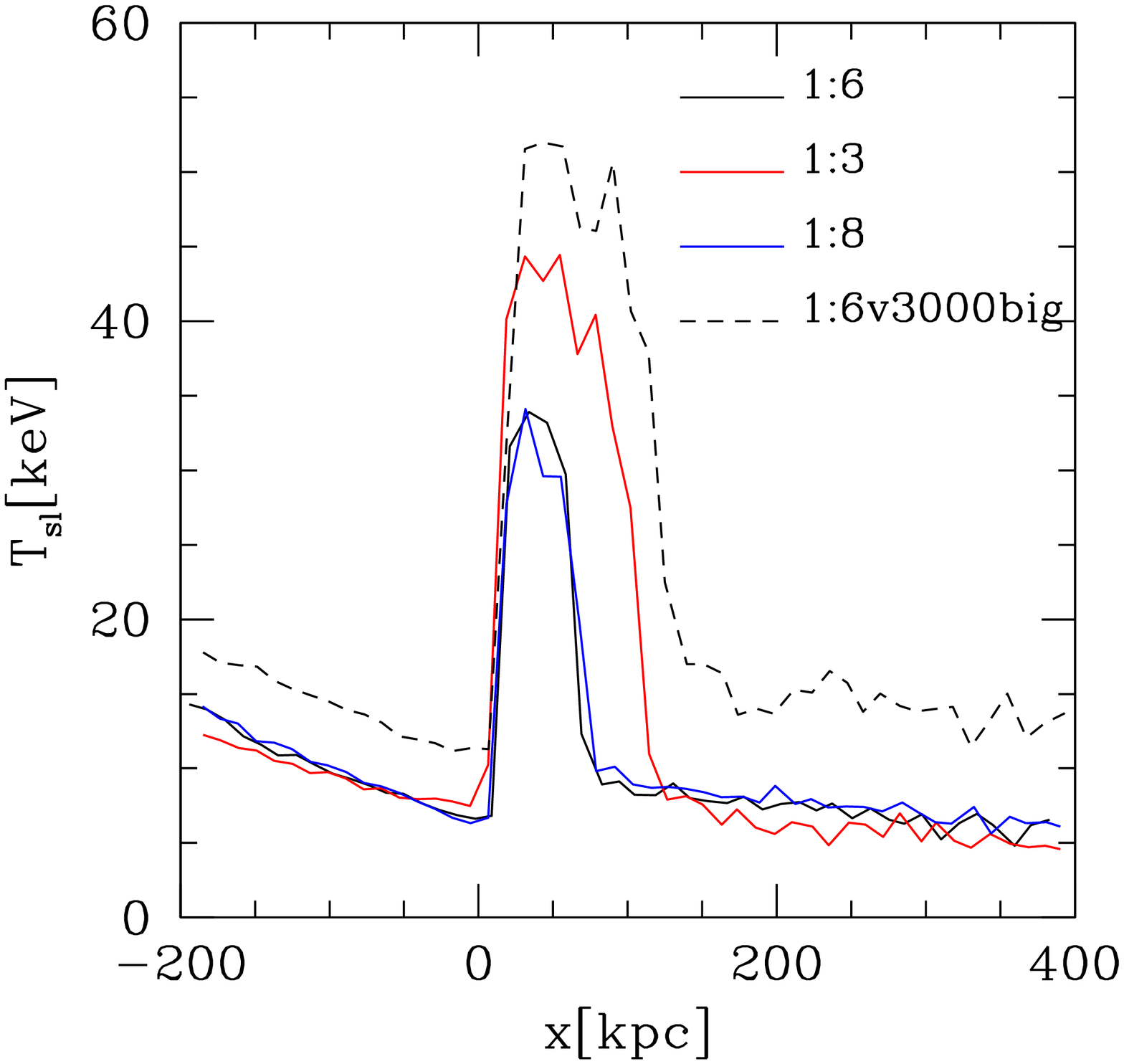}
\includegraphics[%
  scale=0.4]{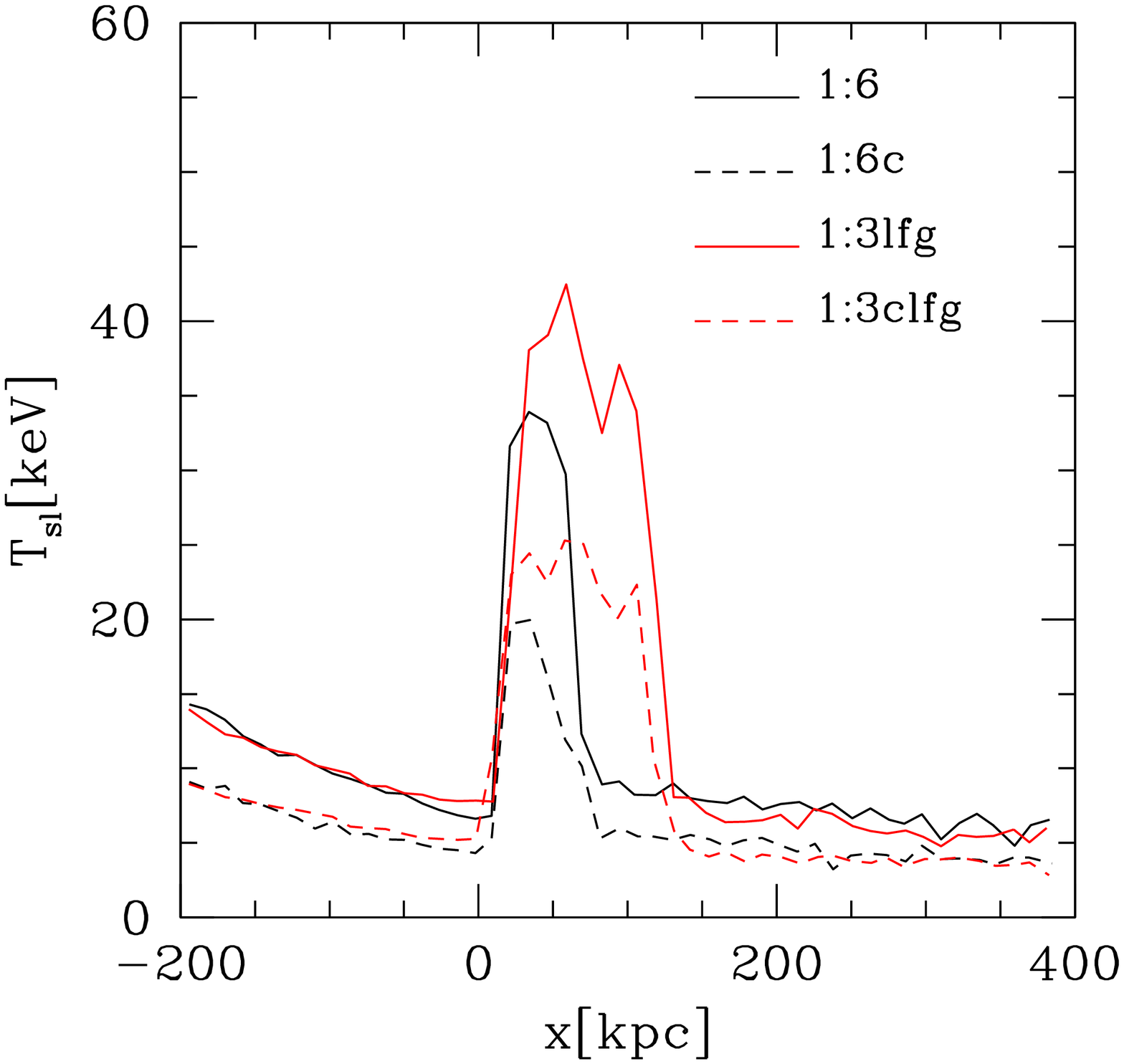}
\includegraphics[%
  scale=0.4]{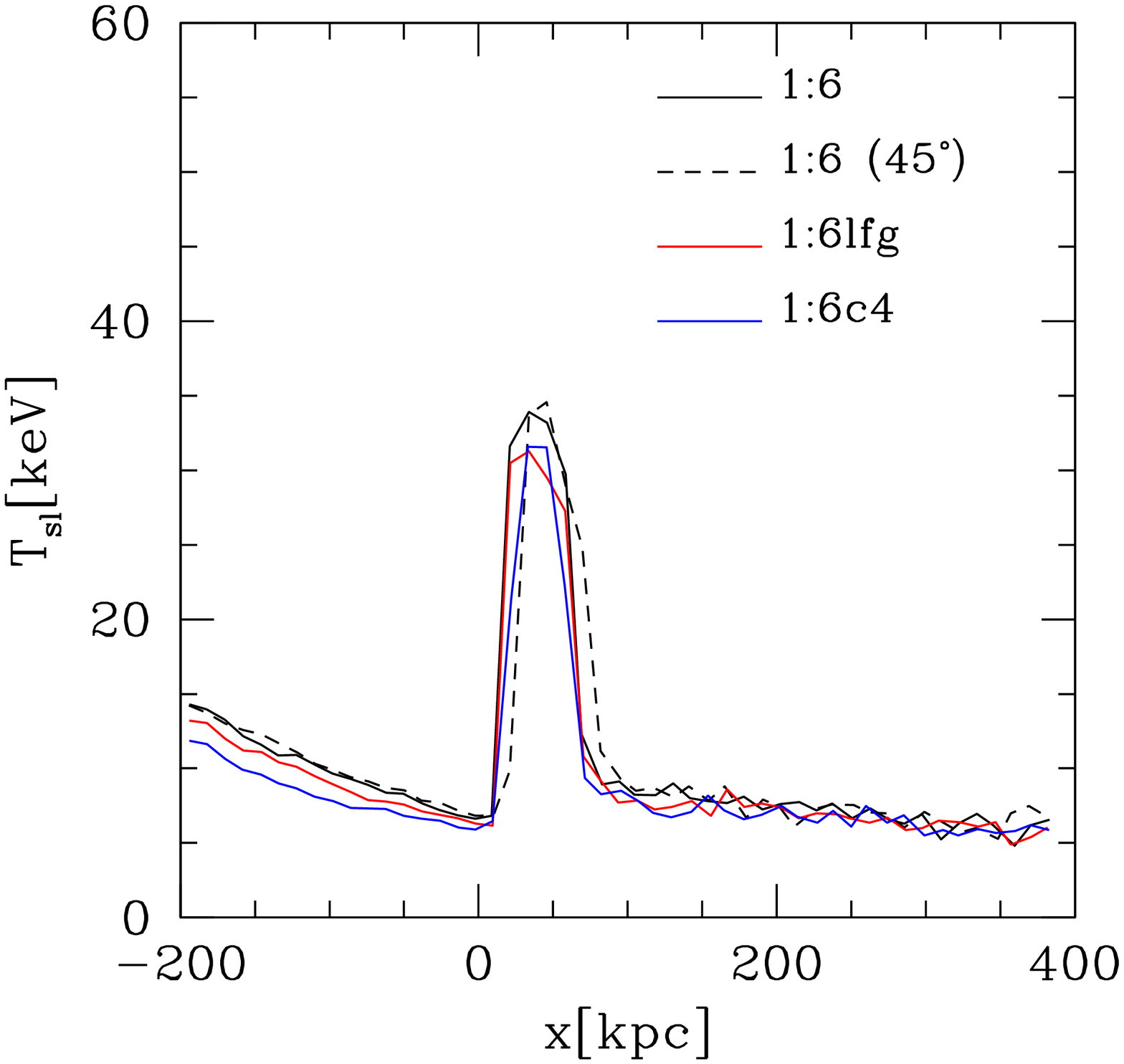}
\caption{Spectroscopic-like temperature profiles measured in a narrow slit (20 kpc) across the shock. The bullet is located at $x=0$ kpc with $x$ increasing towards the pre-shock region.
Upper left panel: 1:6 runs with different relative velocities are compared. Upper right: different mass ratios. Bottom left: comparison between adiabatic and cooling 1:6 and 1:3 encounters. Bottom right: effects of inclination, lower gas fraction and lower concentration of the main halo.}
\label{temppro1}
\end{figure*}
\citet{Mazzotta04} proposed a new definition of temperature, the spectroscopic-like temperature $T_{sl}$:
\begin{equation}
T_{sl} = \frac{{{\sum}_i}^{N_{gas}} m_i {\rho}_i  T_i ^{\alpha}/T_i^{1/2} }{{{\sum}_i}^{N_{gas}} m_i \rho_i T_i^{\alpha}/T_i^{3/2}}.
\end{equation}
 When applied to clusters hotter than 2-3 keV  this equation, with $\alpha = 0.75$, gives a good approximation (within few percent) of  the spectroscopic temperature obtained from data analysis of Chandra. 

The bullet cluster 1E0657-56 is characterized by the highest luminosity, temperature and the strongest bow shock of all known clusters \citep{Markevitch06} and  seems indeed to be the ideal candidate for adopting the spectroscopic-like temperature definition. 
Therefore in the remainder of this paper $T_{sl}$ will be analyzed and $T_{ew}$ only indicated for a comparison. 
Fig. \ref{temppro1} illustrates the projected spectroscopic-like temperature profiles across the shock for the different runs of Table \ref{runs}. All the values refer to the present time.
In particular, the upper left panel shows the temperature jumps associated with different relative velocities of the two clusters. Decreasing the initial relative velocity from $5000$ to $2000$ km s$^{-1}$ (in the system of reference where the main cluster is at rest) reduces the temperature peak by $\sim 7$ keV while the peak itself becomes broader since the thickness of the shock increases by almost a factor of two due to the lower pressure exercised by the pre-shock gas.
Both the 1:6v3000 and 1:6v2000 models seem to fit quite nicely the observed height ($\sim 27-30$ keV) and thickness (150-200 kpc) of the shock front \citep{Markevitch06} while the 1:6 run produces a peak which is too narrow ($\sim 100$ kpc) and pronounced ($\sim 35$ keV).
The upper right panel refers to different mass ratios. If the encounter is characterized by the same initial relative velocity and gas fraction the strongest shock is associated with the  most massive sub-cluster. A 1:3 adiabatic encounter produces a $\sim 45$ keV temperature peak with thickness $\sim 150$ kpc while the 1:8 temperature profile is not substantially different from the one of the 1:6 run. In the same plot we also show the shock created in the massive run 1:6v3000big. Clearly the maximum temperature is much higher than the observed one. 
Including radiative cooling (bottom left panel) has the effect of reducing the peak in temperature but does not influence the thickness of the shock region. The adoption of a simple cooling model is actually questionable. Indeed, although the estimate of the temperature jump is reasonable, the entire temperature profile drops by 5 keV due to the fact that once cooling is activated the main cluster gas component becomes thermally unstable in the early phases of the interaction and overcools in the central regions. 
If cooling is important, models with higher mass ratios and relative velocities (like the 1:3lfg run in plot) have still to be taken in account and can not be excluded as the high temperature peaks could actually cool significantly.   
The choice of a different line of sight (bottom right plot) does not affect significantly the temperature profile along the shock. Even decreasing the baryonic fraction in the clusters and assuming a much less concentrated main halo, the height and thickness of the temperature peak do not change.

\begin{figure}
\epsfxsize=8truecm \epsfbox{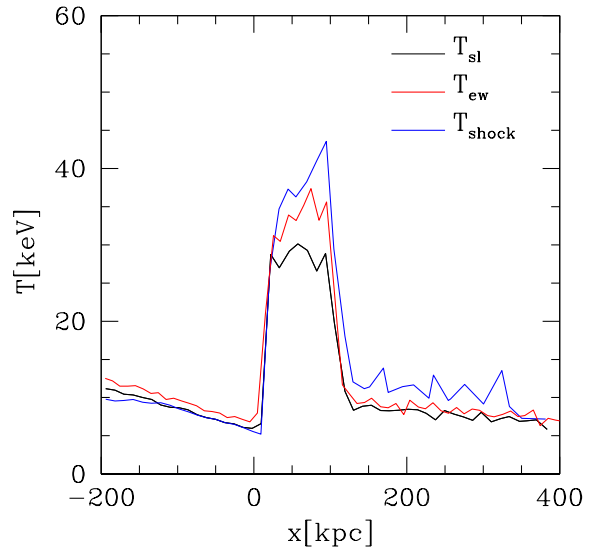} 
\caption{Run 1:6v3000: spectroscopic-like ($T_{sl}$), emission weighted ($T_{ew}$) and true ($T_{shock}$) temperature profiles across the shock.}
\label{temppro2}
\end{figure}

In Fig. \ref{temppro2} we compare the spectroscopic-like temperature profile across the shock region with the emission weighted one for our favourite run 1:6v3000. While the projected temperature profile calculated according to the two definitions is similar ($T_{ew}$ is only slightly higher than $T_{sl}$) in the pre and post-shock regions, the emission weighted temperature $T_{ew}$ in the region $0 \leq x \leq 150 $ kpc  is $\sim 20\%$ higher than $T_{sl}$.  $T_{shock}$ represents the actual temperature along the$x$-axis through the shock. Indeed the blue curve in Fig. \ref{temppro2}  gives the exact temperature jump across the shock (the ``true temperature'' of the shock) which is characterized by an even higher peak with respect to the projected ones. This deprojected temperature profile is actually the one shown by \citet{Springel07} in their Fig. 9 where they compare their model with observations which on the other hand refer to projected quantities \citep{Markevitch06}. As the calculated local temperatures of \citet{Springel07} fit the projected observed temperature profile very well we conclude that their projected temperature profiles are actually inconsistent with the observations by \citet{Markevitch06}.

\begin{figure*}
\includegraphics[%
  height=70mm]{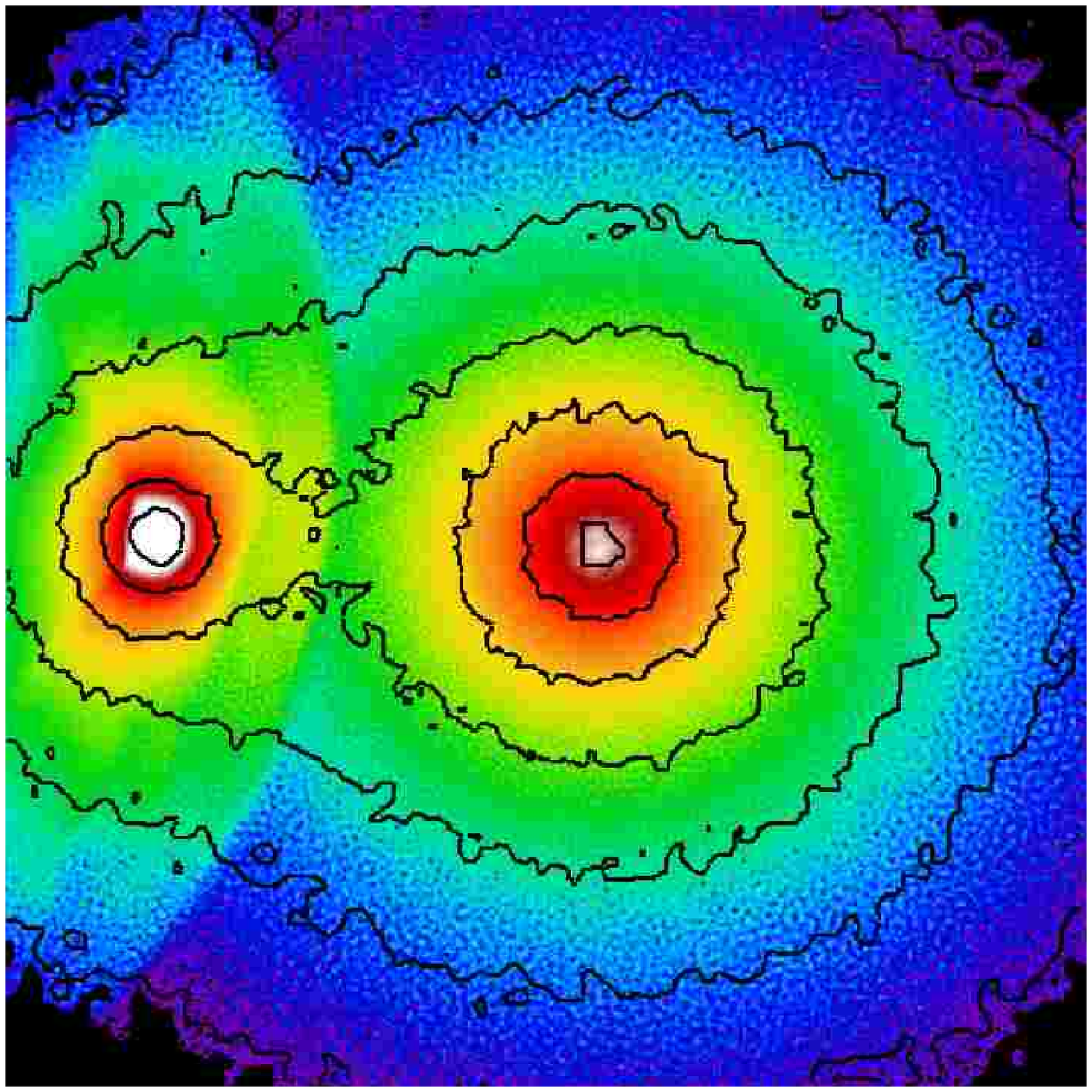}
\includegraphics[%
  height=70mm]{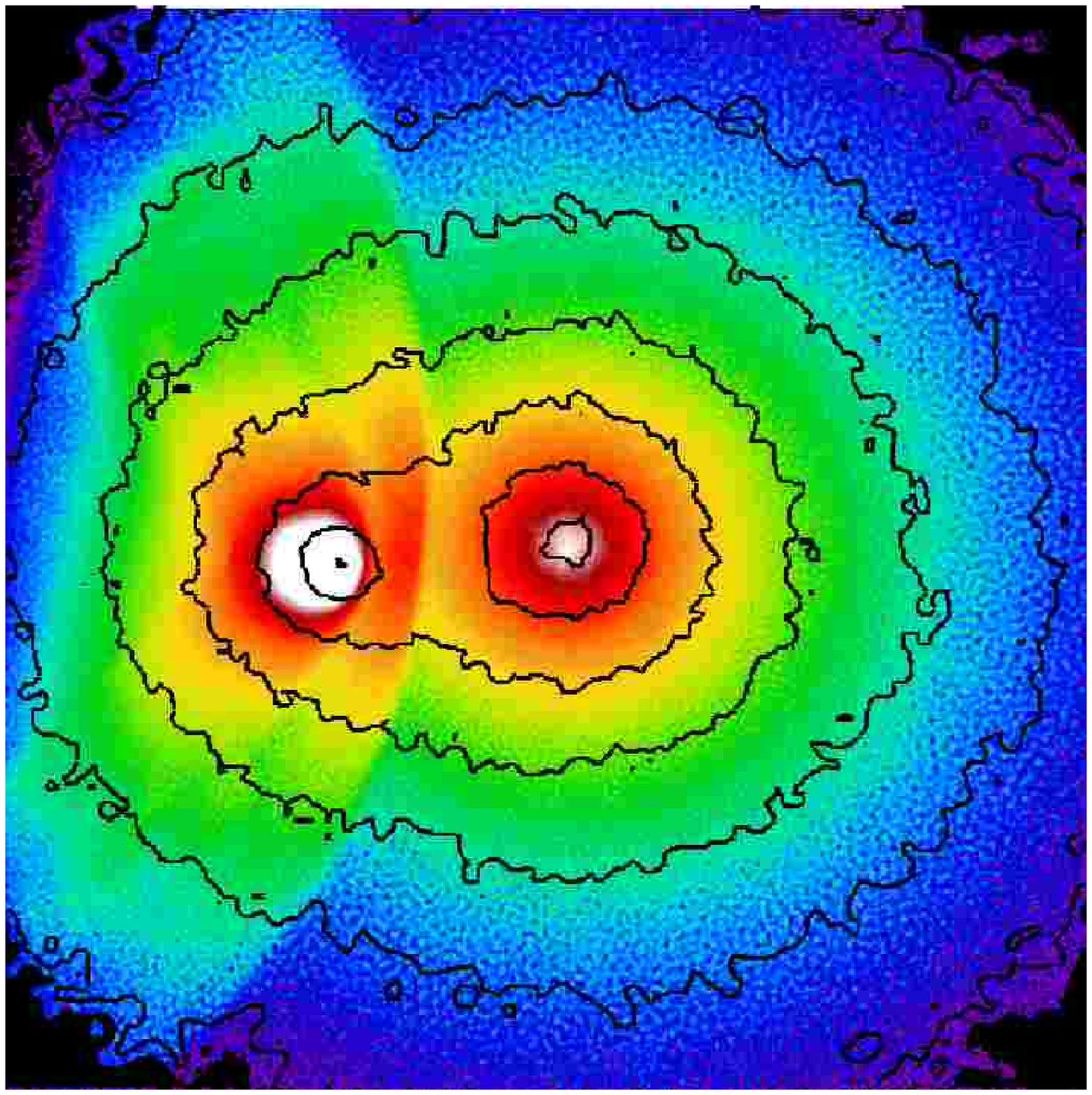}
\includegraphics[%
  height=70mm]{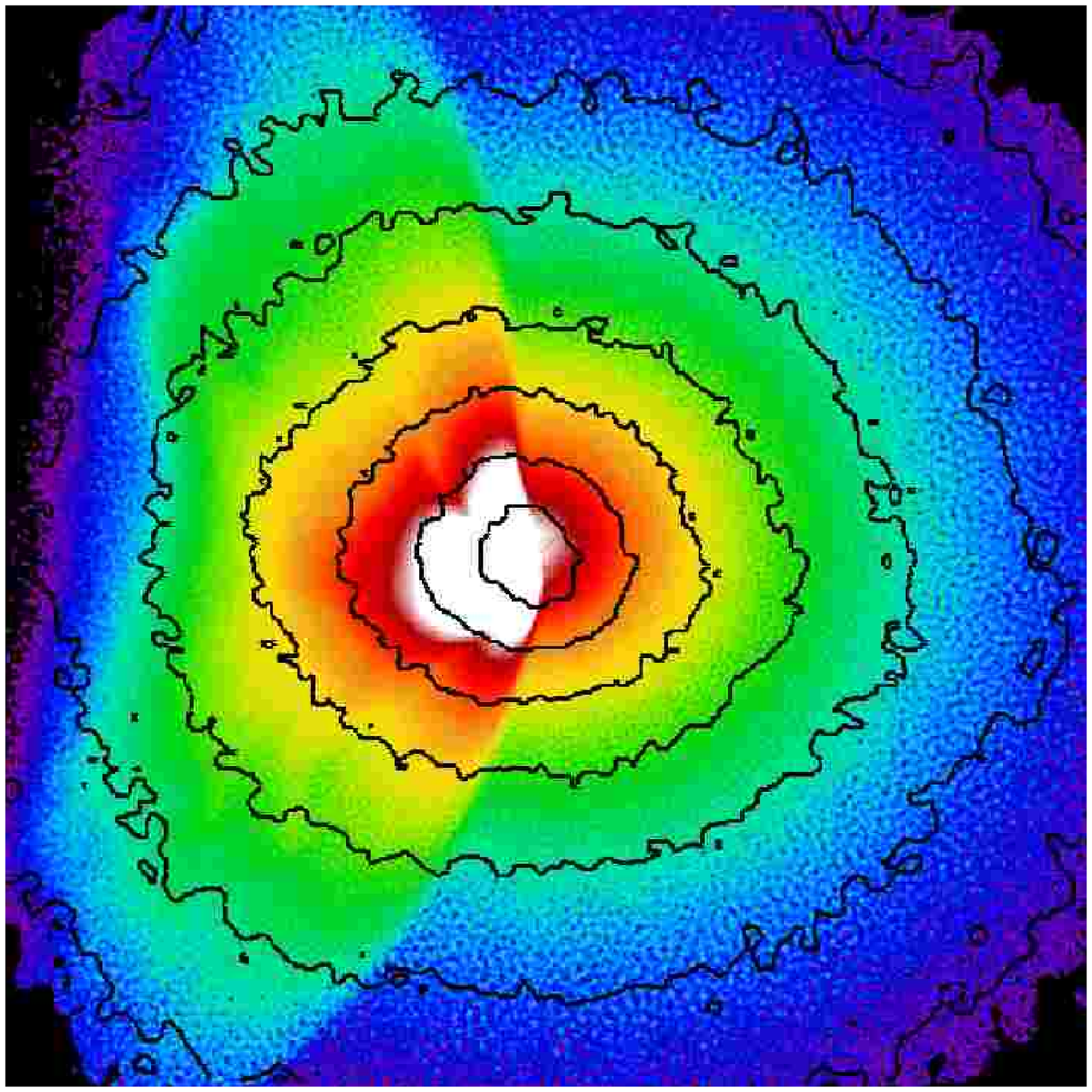}
\includegraphics[%
  height=70mm]{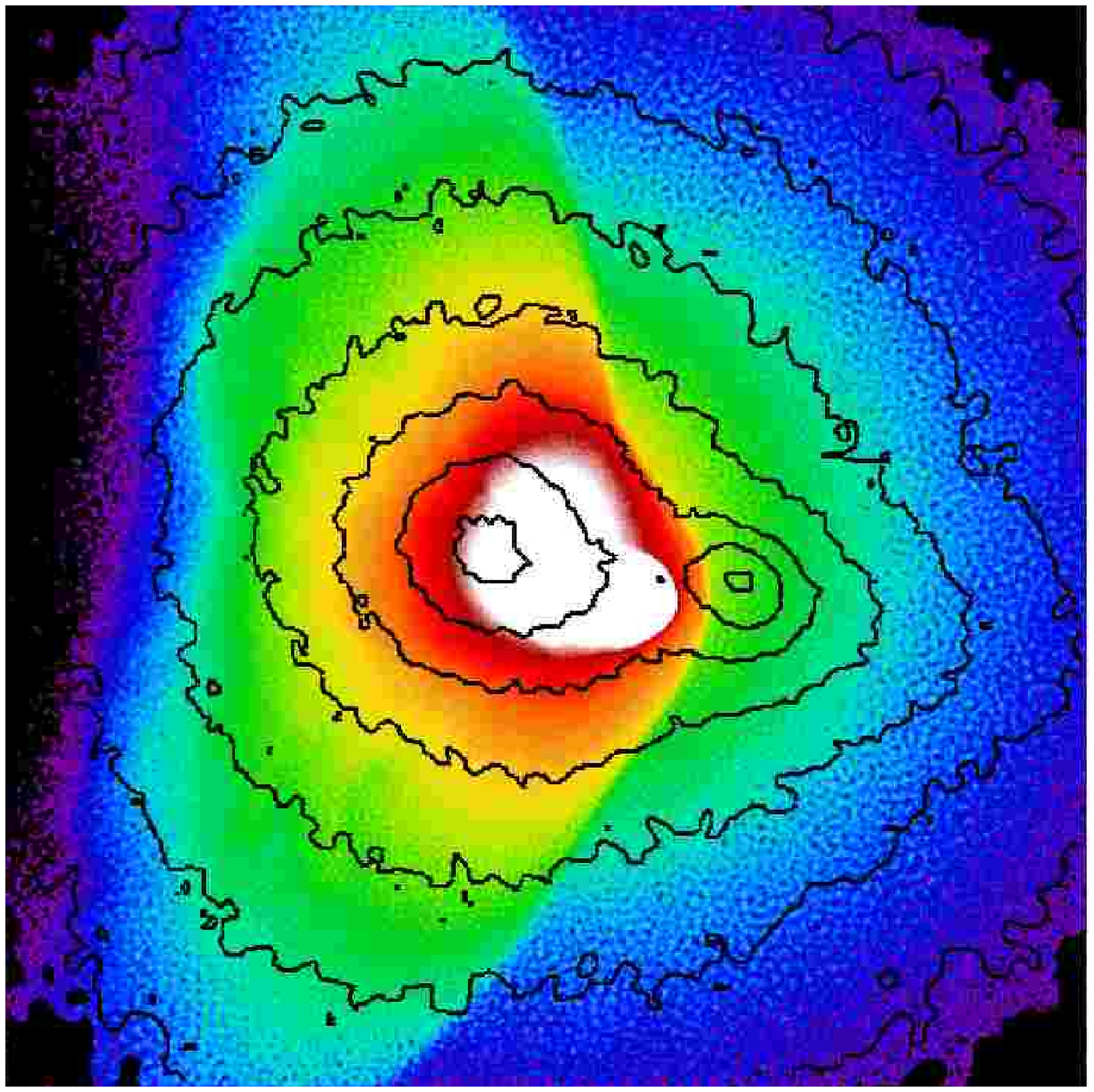}
\includegraphics[%
  height=70mm]{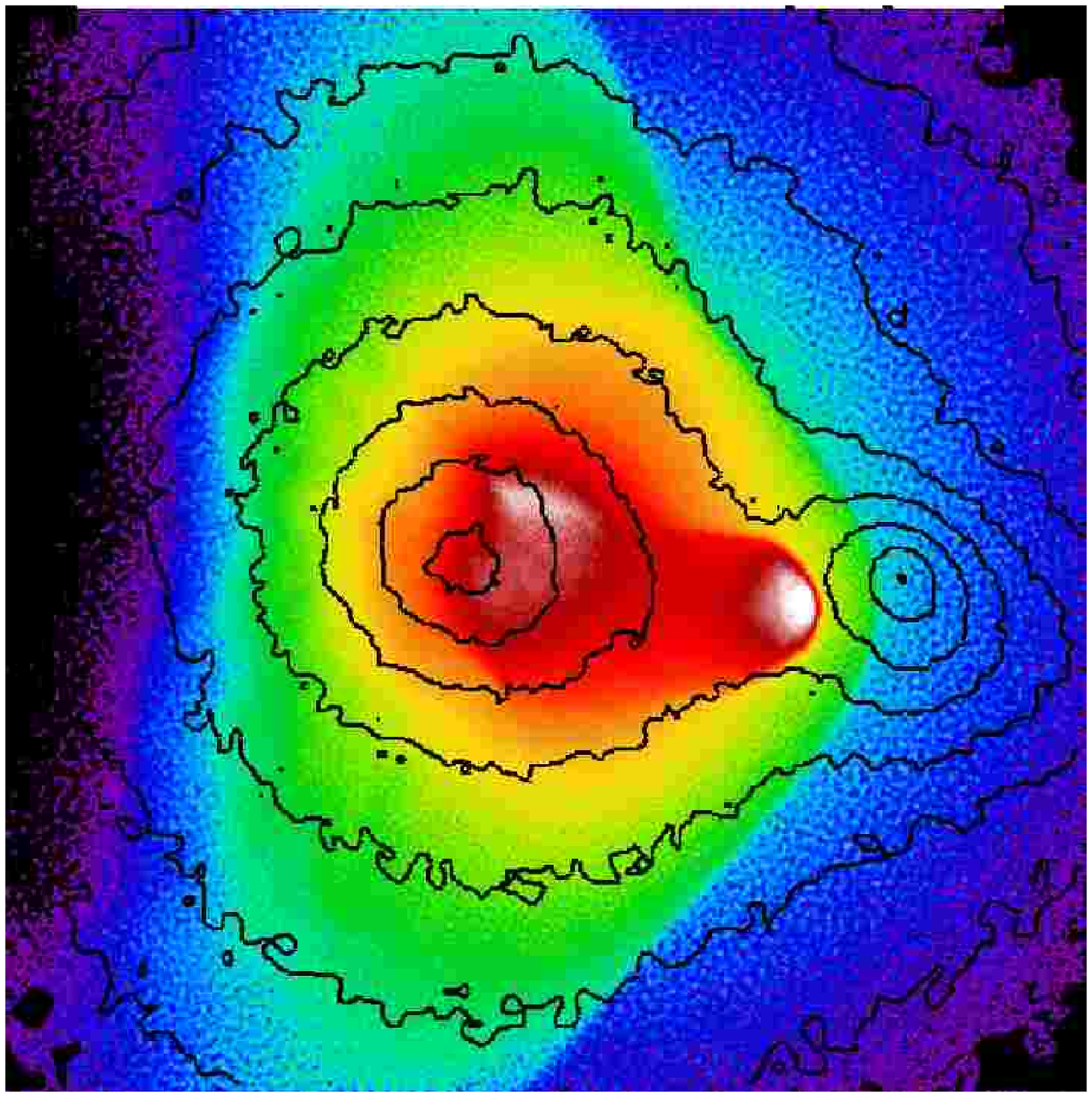}
\includegraphics[%
  height=70mm]{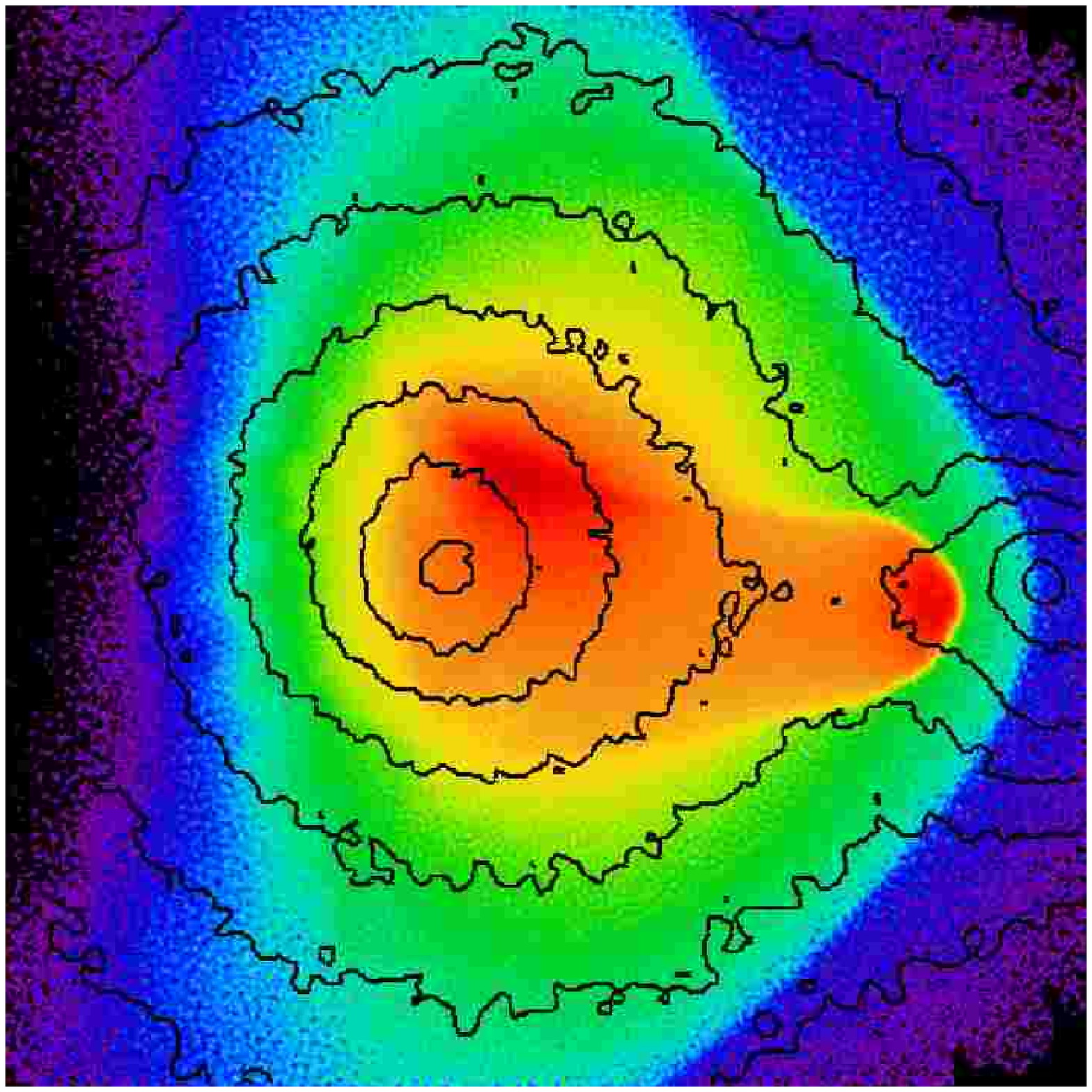}
\includegraphics[%
  height=5mm]{colorbar.ps}
\caption{Time evolution of run 1:6v3000. X-ray (0.8-4 keV) surface brightness maps. Logarithmic colour scaling is indicated by the key to the bottom of the figure, with violet corresponding to $10^{38}$ ergs$^{-1}$ kpc$^{-2}$ and white to $2.5 \times 10^{41}$ erg s$^{-1}$kpc$^{-2}$ . Projected isodensity contours of the total mass distribution are drawn on  top of temperature maps.}
\label{sequencexray}
\end{figure*}

\begin{figure*}
\includegraphics[%
  height=70mm]{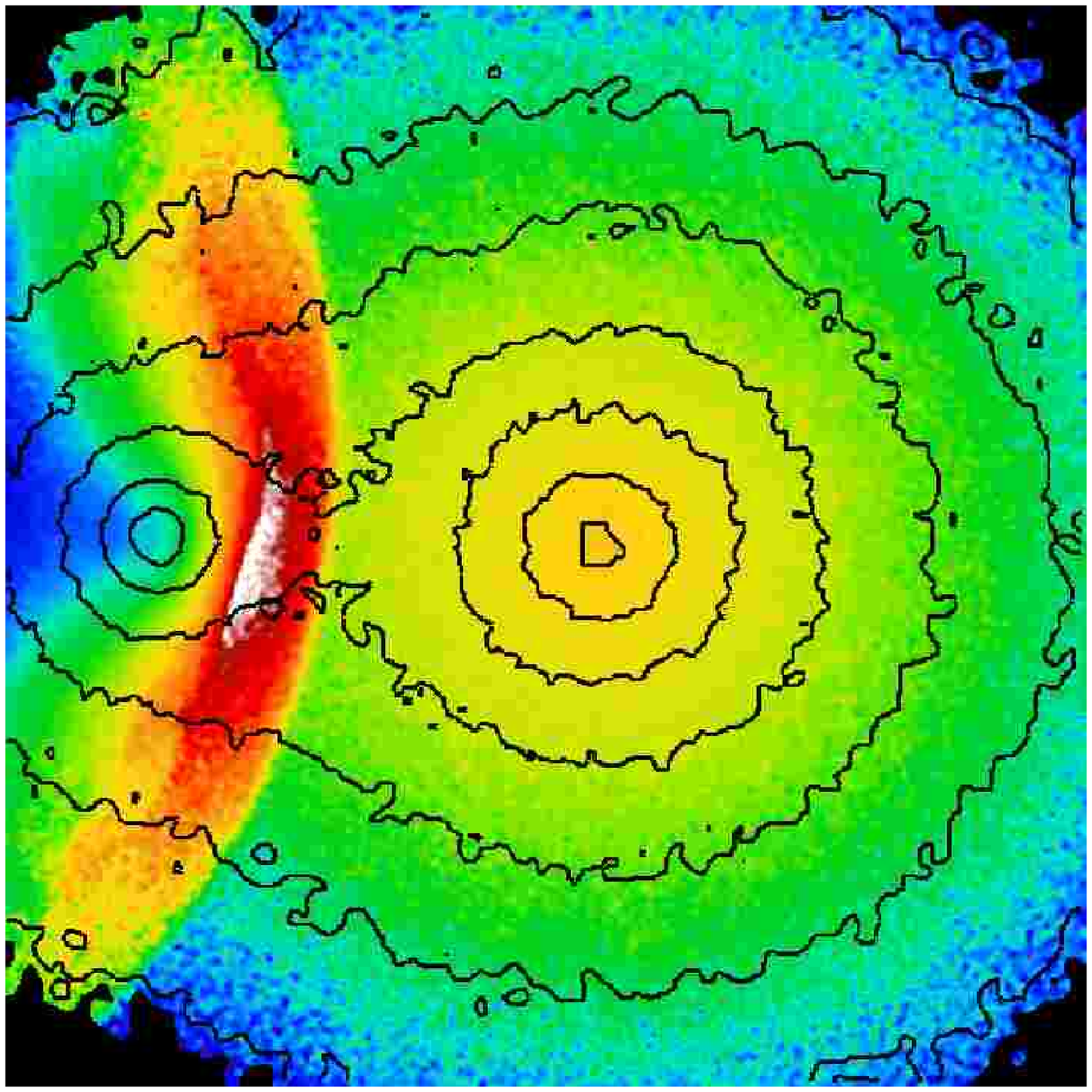}
\includegraphics[%
  height=70mm]{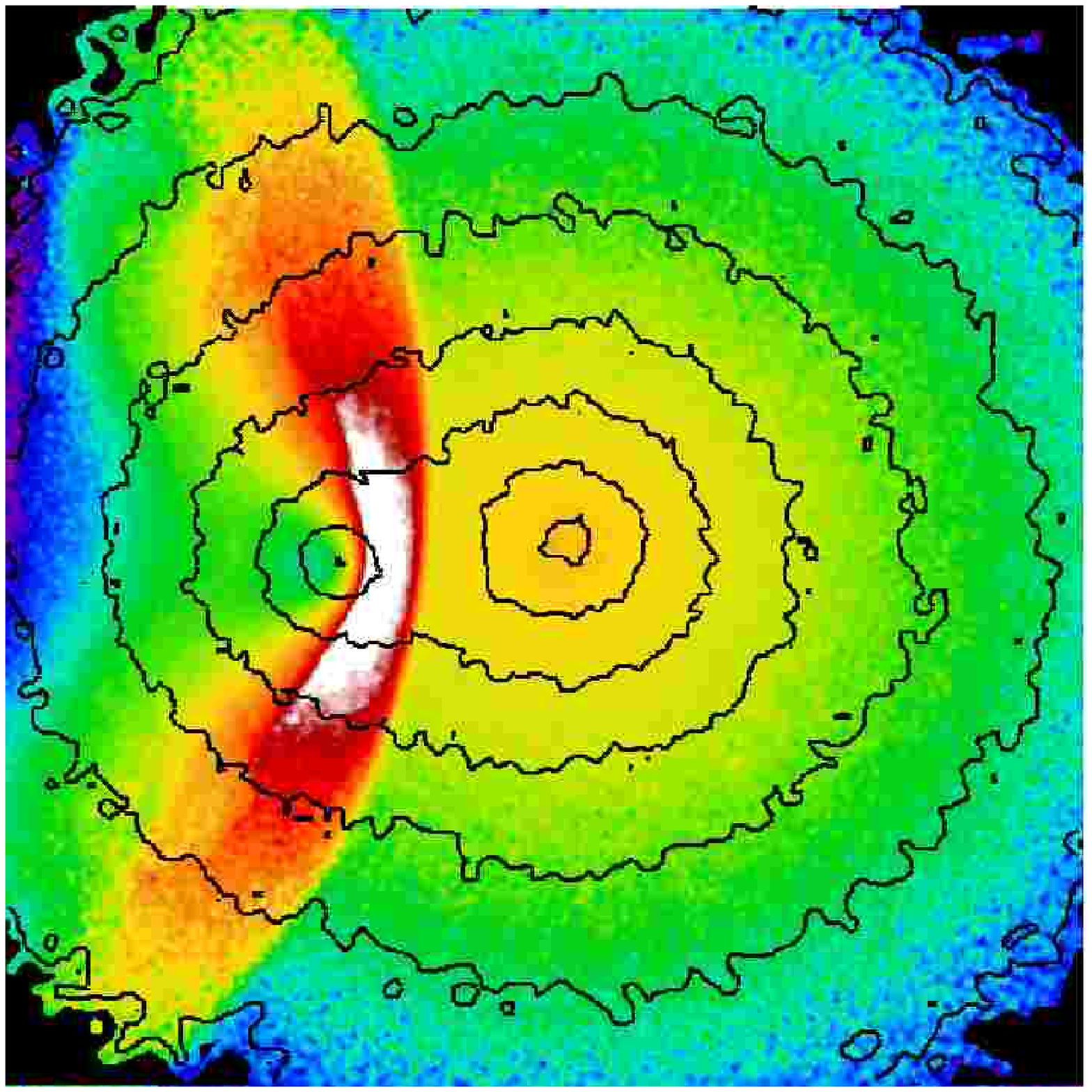}
\includegraphics[%
  height=70mm]{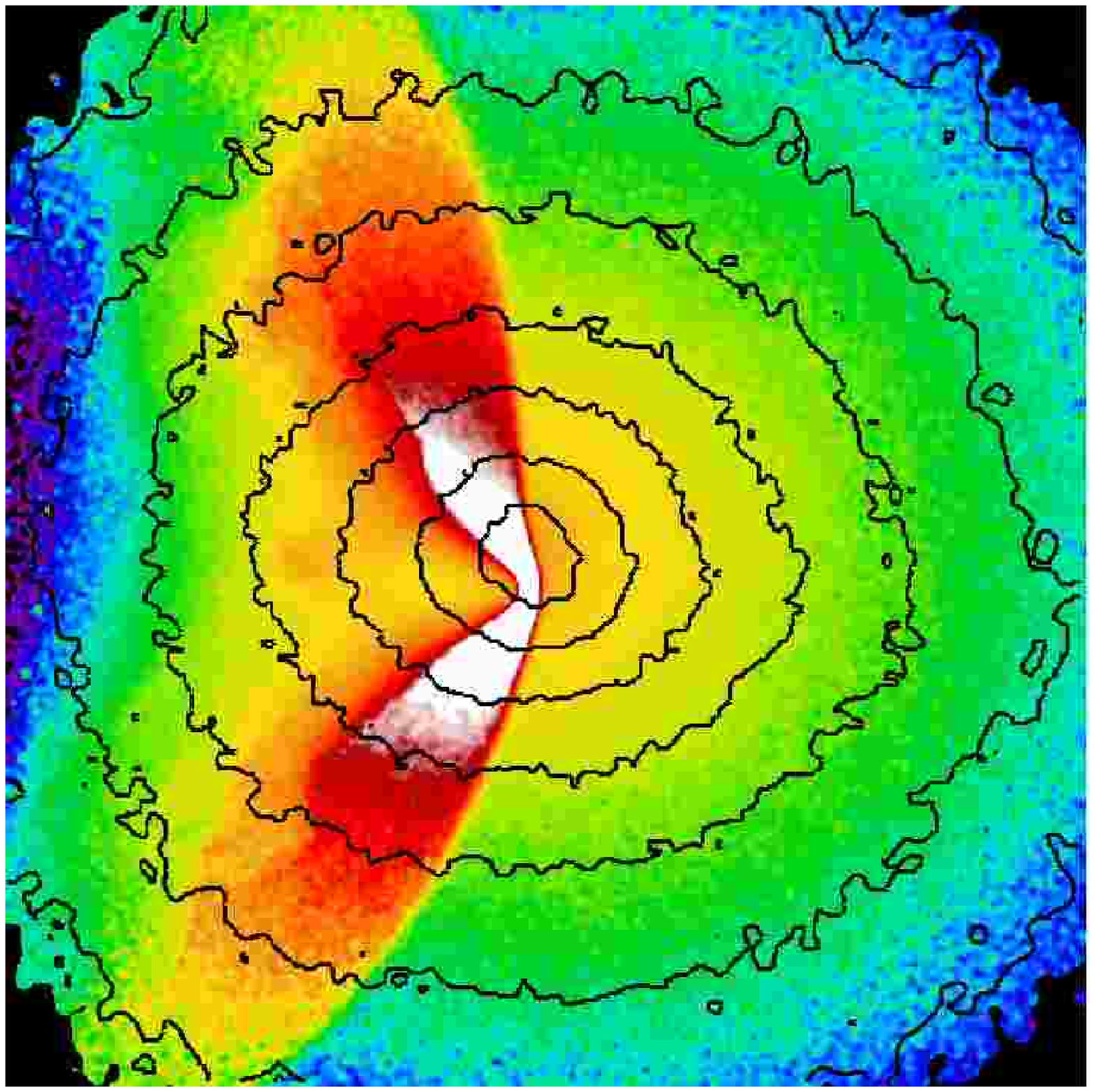}
\includegraphics[%
  height=70mm]{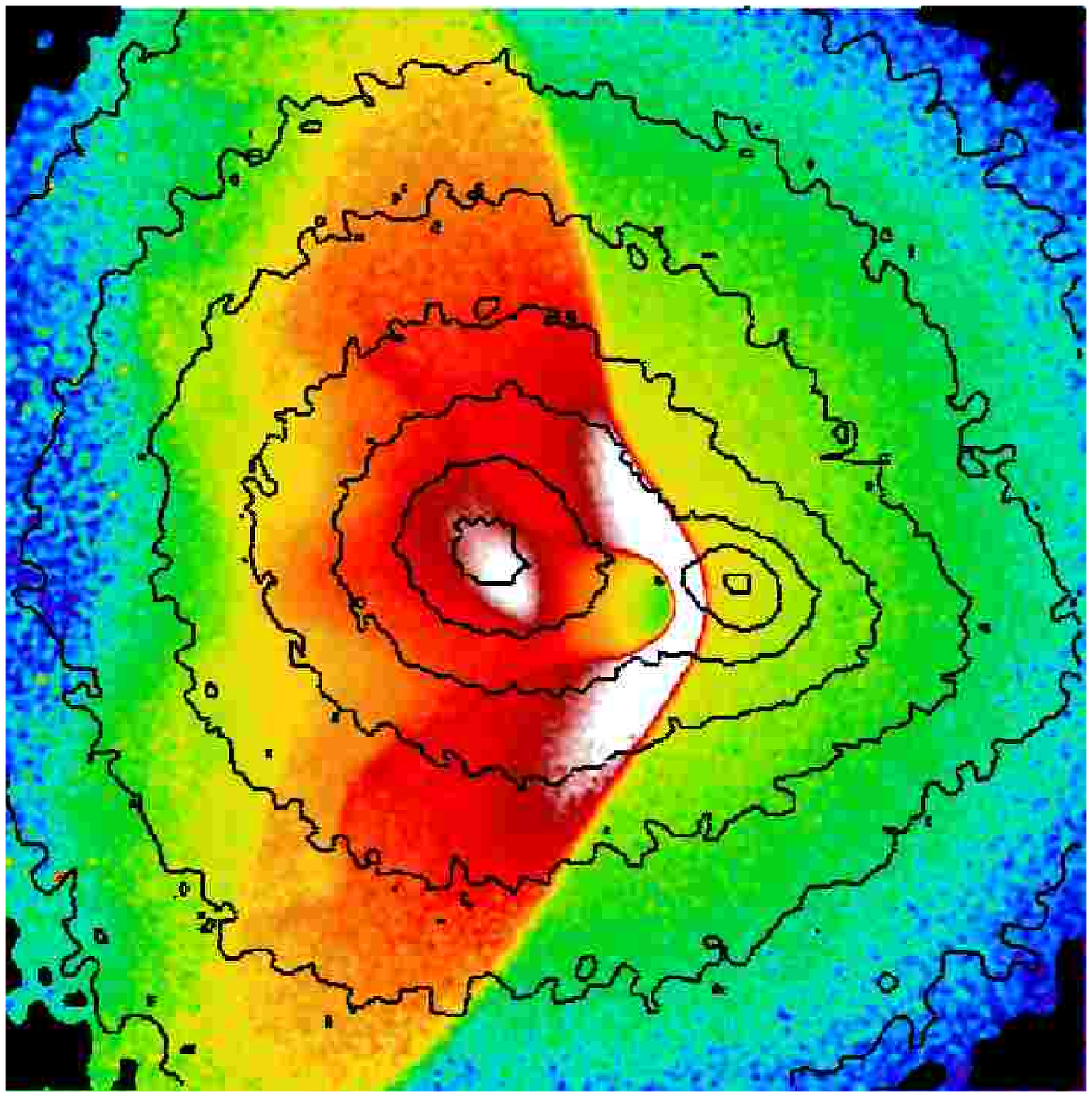}
\includegraphics[%
  height=70mm]{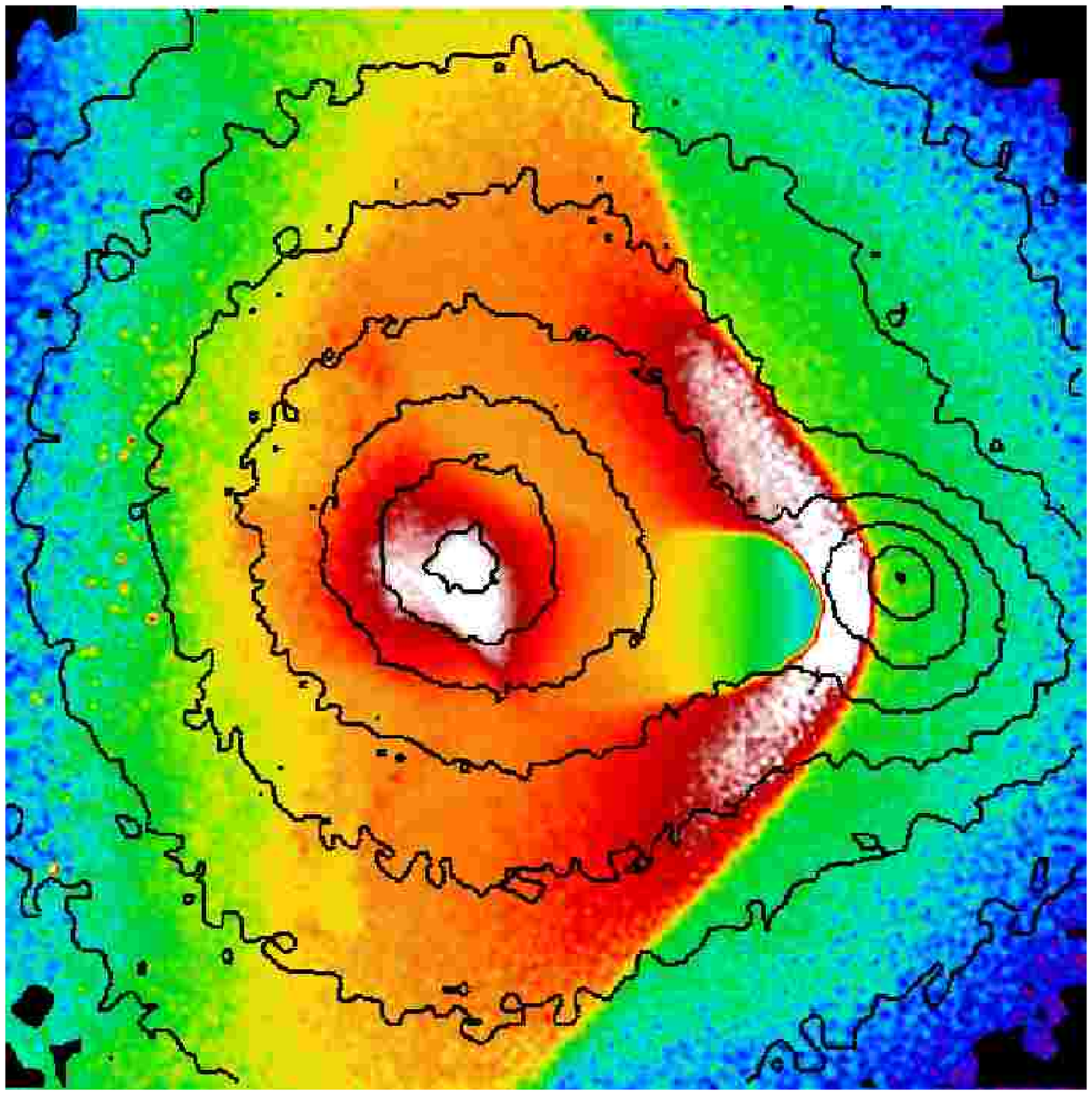}
\includegraphics[%
  height=70mm]{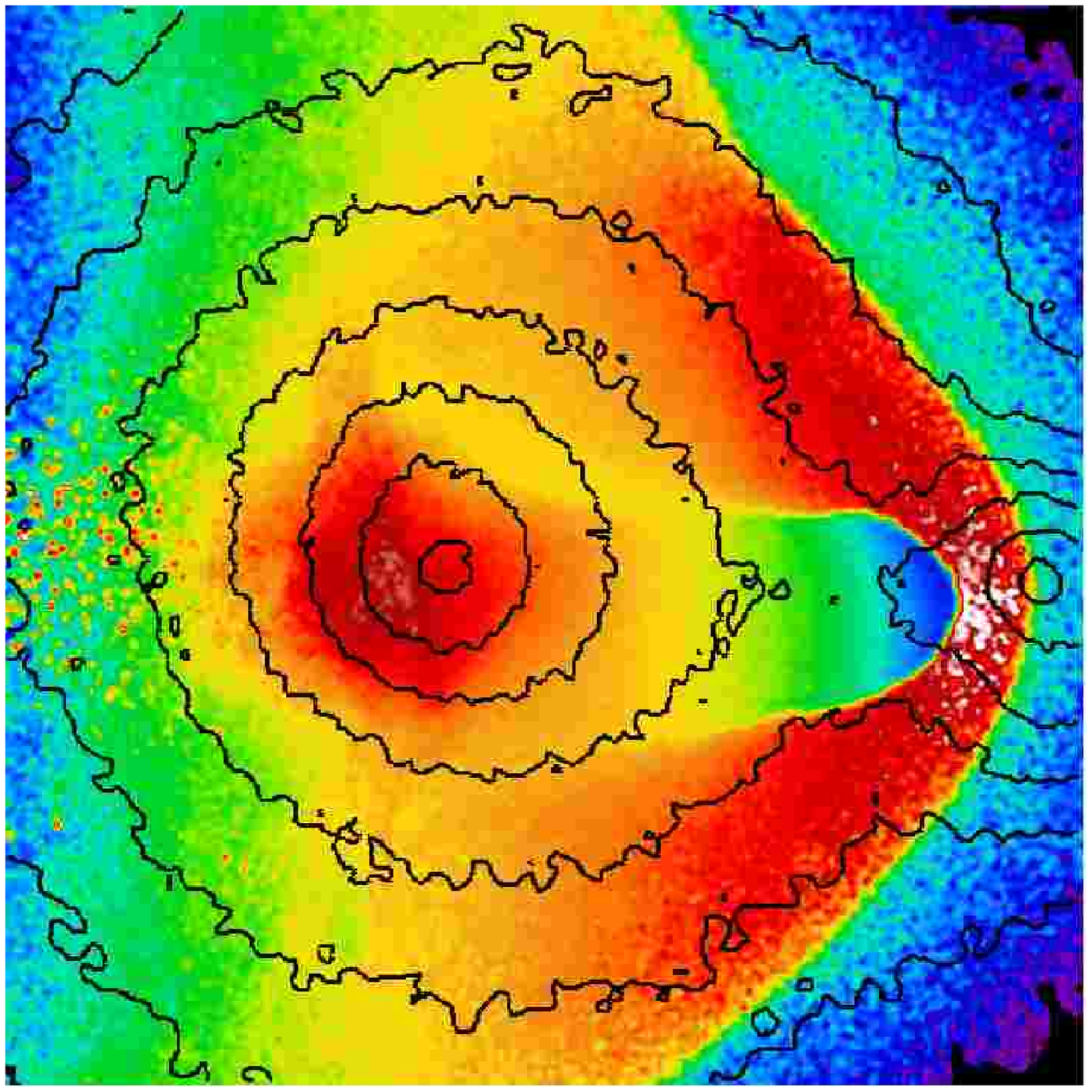}
\includegraphics[%
  height=5mm]{colorbar.ps}
\caption{Time evolution of run 1:6v3000. Spectroscopic-like X-ray temperature maps. Logarithmic colour scaling is indicated by the key to the bottom of the figure, with violet corresponding to 0.9 keV and white to 86 keV. Projected isodensity contours of the total mass distribution are drawn on  top of temperature maps.  }
\label{sequencetemp}
\end{figure*}

\begin{figure*}
\includegraphics[%
  height=70mm]{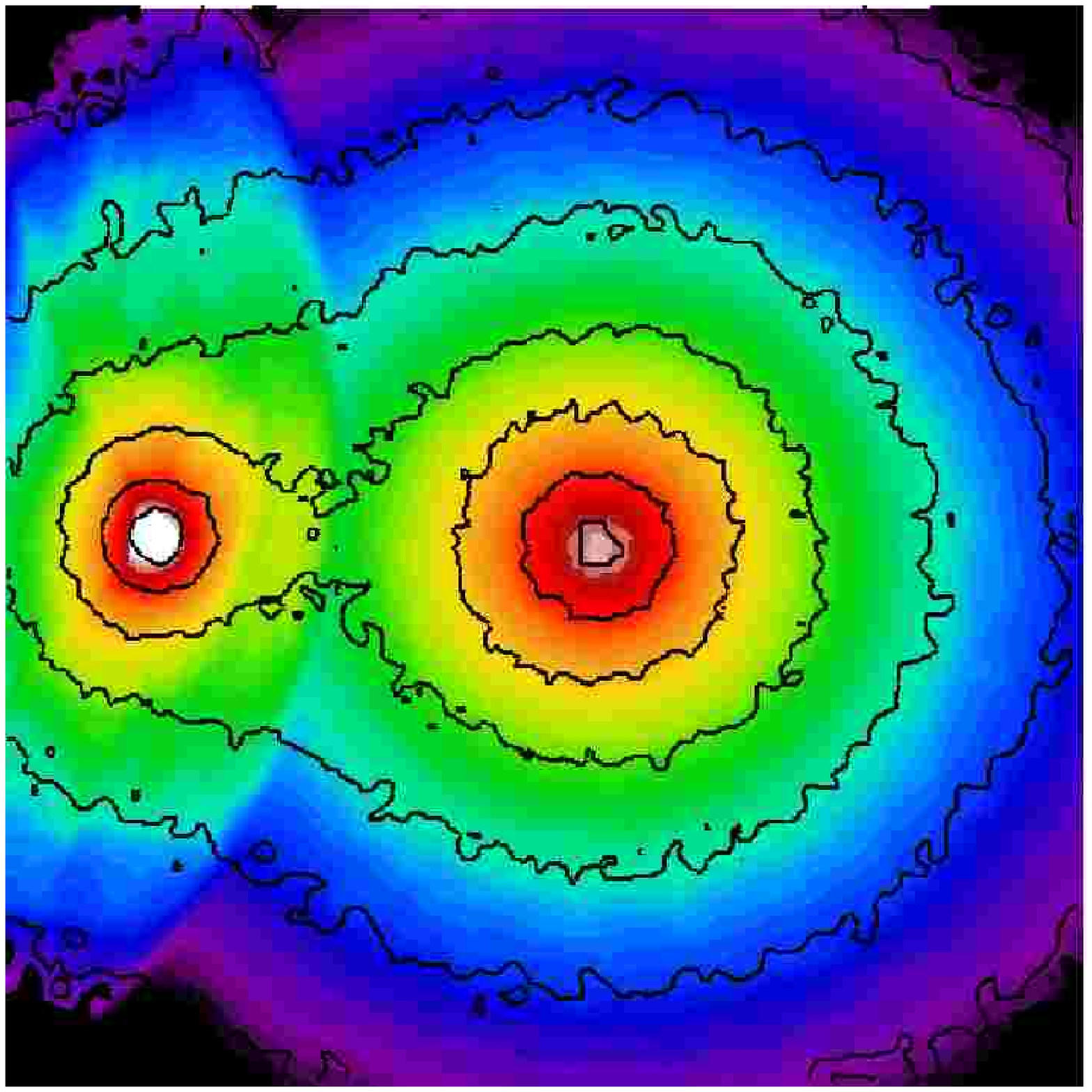}
\includegraphics[%
  height=70mm]{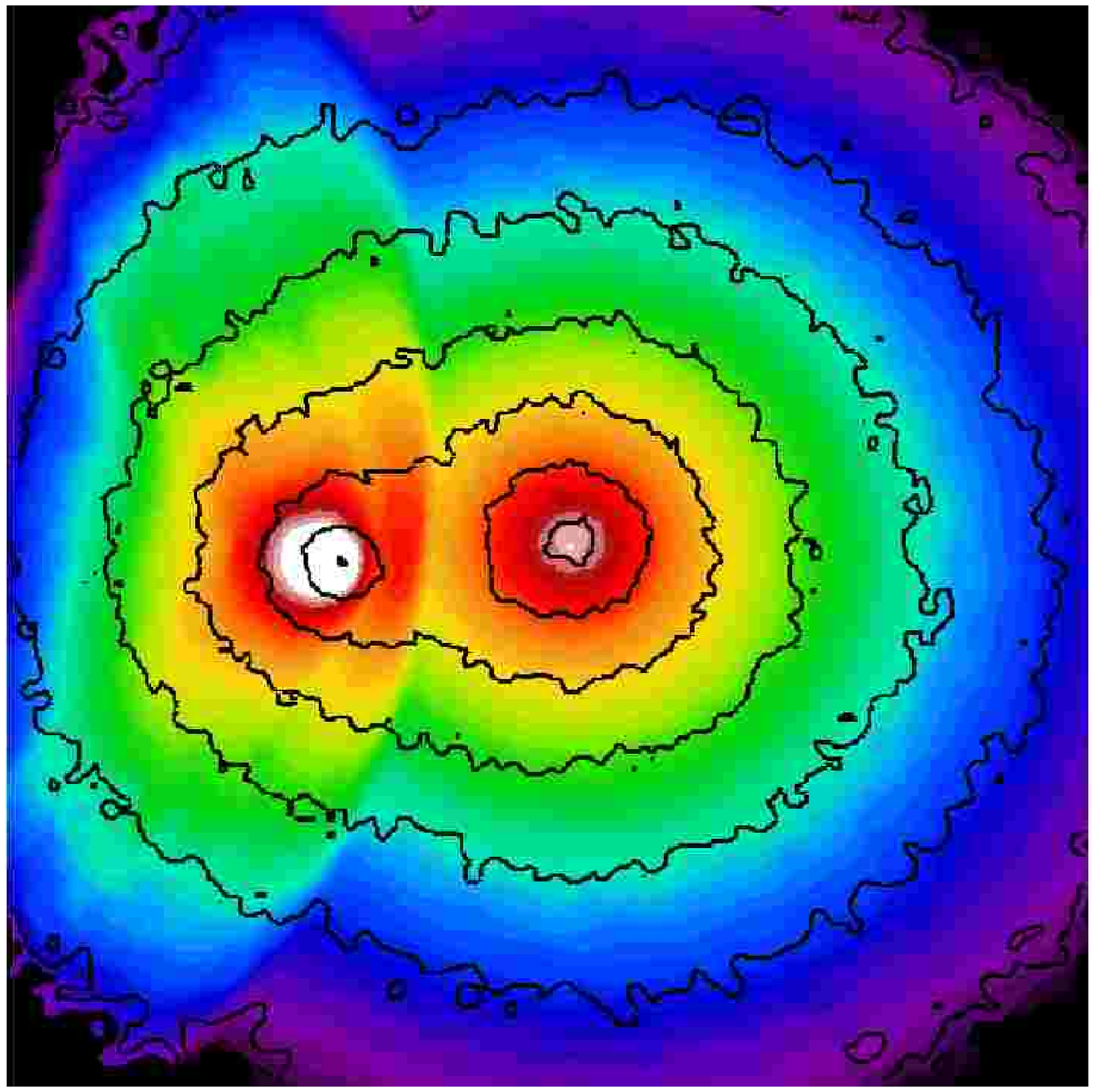}
\includegraphics[%
  height=70mm]{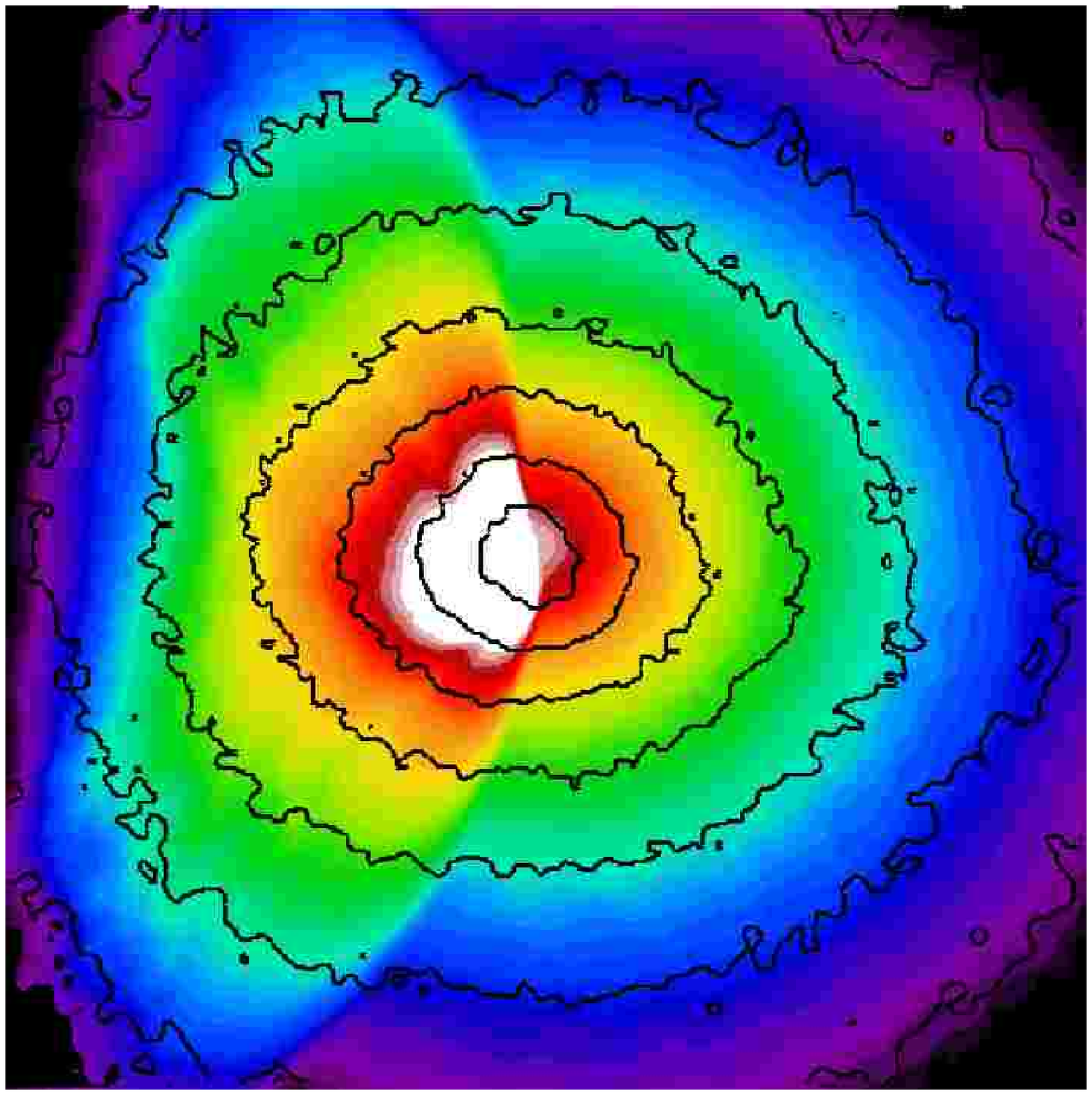}
\includegraphics[%
  height=70mm]{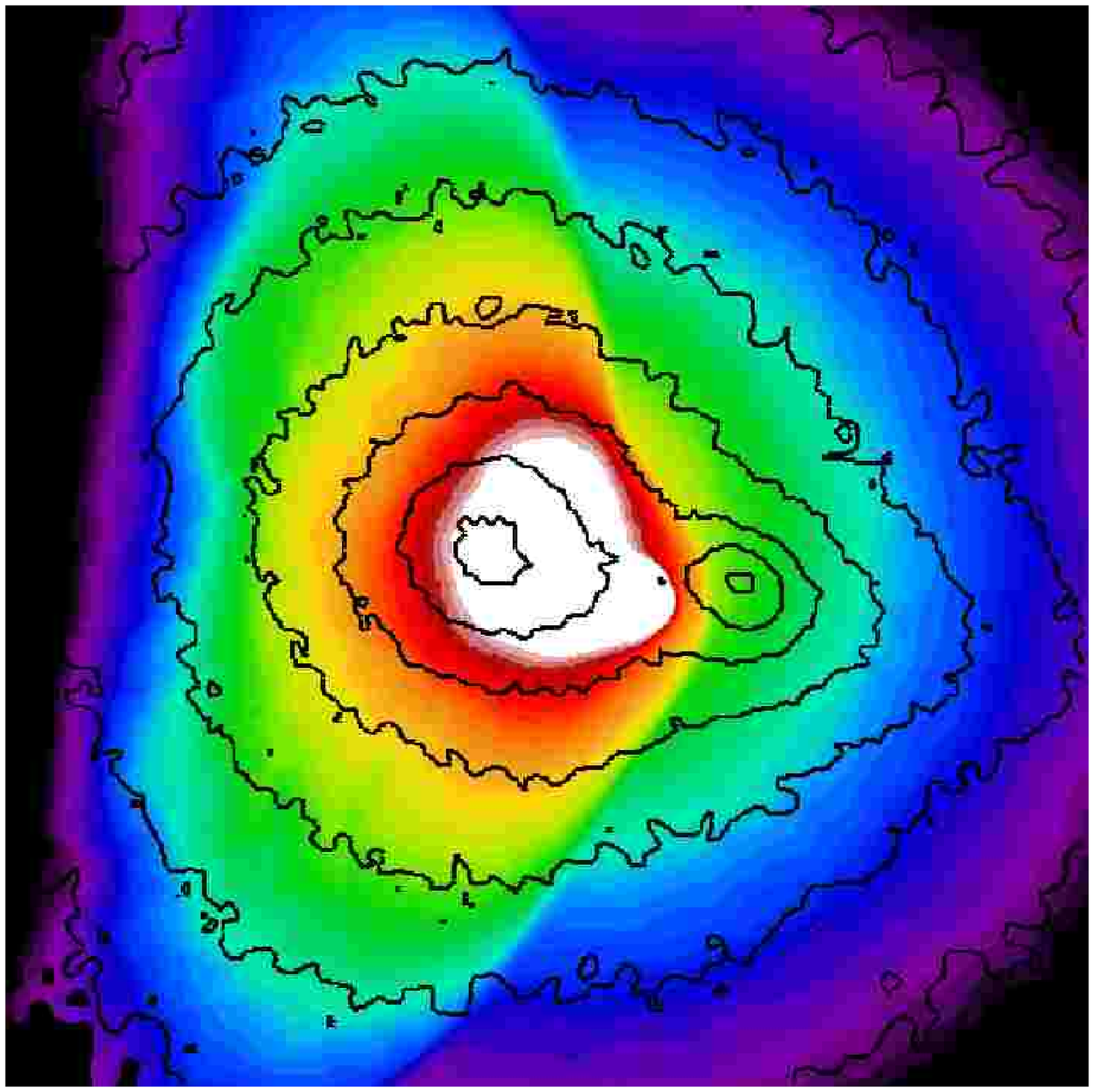}
\includegraphics[%
  height=70mm]{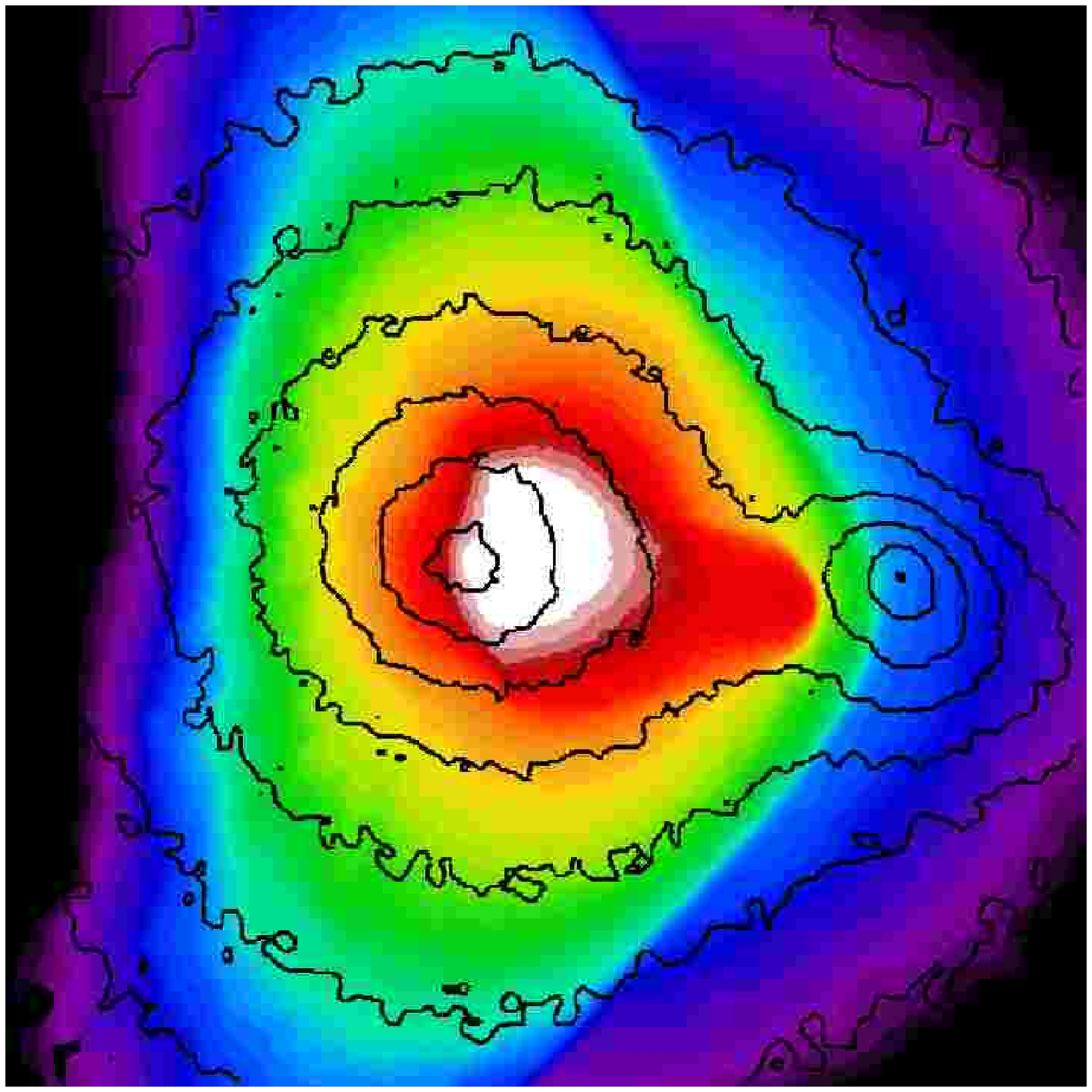}
\includegraphics[%
  height=70mm]{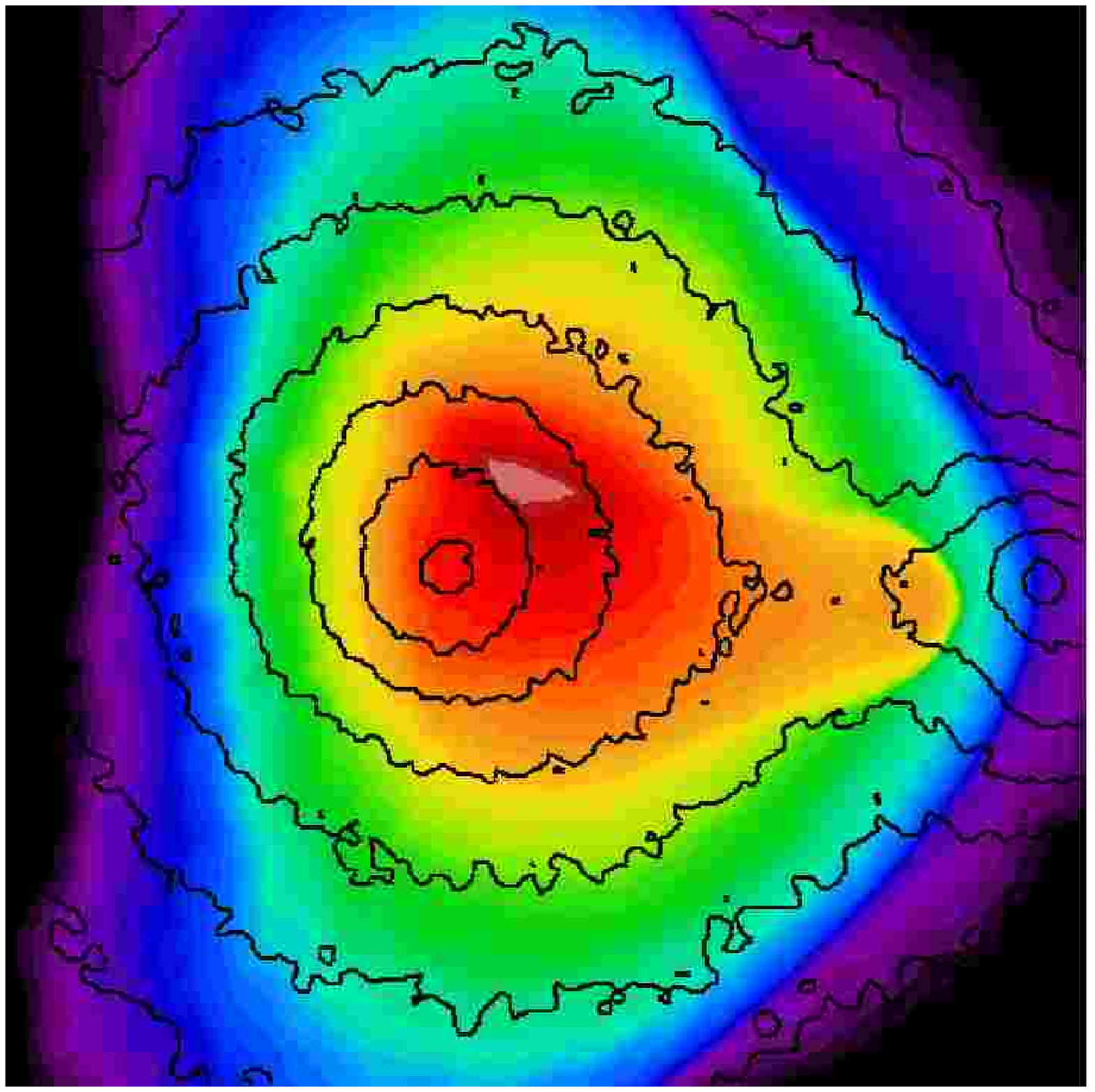}
\includegraphics[%
  height=5mm]{colorbar.ps}
\caption{Time evolution of run 1:6v3000. The colormap represents the gas surface density projected along an axis perpendicular to the plane of the encounter. Violet corresponds to a surface density of $2.3 \times 10^{2}$ M$_{\odot}$ kpc$^{-2}$ and white to $2.3 \times 10^{8}$ M$_{\odot}$ kpc$^{-2}$. Projected isodensity contours of the total mass distribution are drawn on  top of the image.    }
 \label{sequenceden}
\end{figure*}

In Fig. \ref{sequencexray}, \ref{sequencetemp}, \ref{sequenceden} we show for our selected adiabatic run 1:6v3000 the evolution with time of the $0.8-4 $ keV X-ray surface brightness, spectroscopic-like temperature and gas surface density during the central phases of the interaction. All the quantities are projected along an axis perpendicular to the plane of the collision. On top of the color maps we draw in black the projected isodensity contours of the total mass distribution, which is dominated by the dark matter component. 
Time increases from the left to the right and from the top to the bottom. The sequence of six panels covers an interval in time of 400 Myr, between 1.3 and 1.7 Gyr from the beginning of the simulation. The snapshot corresponding to the present time is the one on the bottom left.
The bullet approaches the main cluster from the left, with an initial velocity of 3000 km s$^{-1}$ in the main cluster rest-frame. 
The shock front has an arc-like shape \citep{Ricker01} and becomes progressively more asymmetric as the bullet moves closer to the core of the main cluster.  When the centers of the two clusters are less than 250 kpc separated, ram-pressure becomes effective in producing a displacement (visible in the projected density maps as well as in surface brightness) between the gaseous core of the bullet and the peak of its associated mass distribution. The offset in the main cluster is evident only when the bow shock passes through its core. 
In the central panels of Fig. \ref{sequencexray} and \ref{sequenceden} the X-ray luminosity and surface density saturate in order to distinguish features in the maps at later times.
Nevertheless, it is visible how the main cluster core gets compressed and displaced from the center of the potential towards the top right of the image and appears in the X-ray maps (last two panels) as an elongated structure, characterized by high surface brightness.

Temperature maps better describe the evolution of the shock region, which gets compressed and hotter during the core passage while in a later phase it cools down and becomes thicker due to lower pressure of the pre-shock gas.  
The bullet itself expands as it leaves the central regions of the main cluster.
As observed by \citet{Markevitch06} and previously noticed by \citet{Springel07}, 
despite of its strong X-ray emission (it is the brightest feature in the post core-core interaction X-ray maps) the bullet remains relatively cold. 
Even if radiative cooling is not activated, the core of the sub-cluster is heated only to a maximum temperature of $10^8$ K, while the shock front is much hotter ($5 \times 10^8$ K).
\begin{figure*}
\includegraphics[%
  scale=0.4]{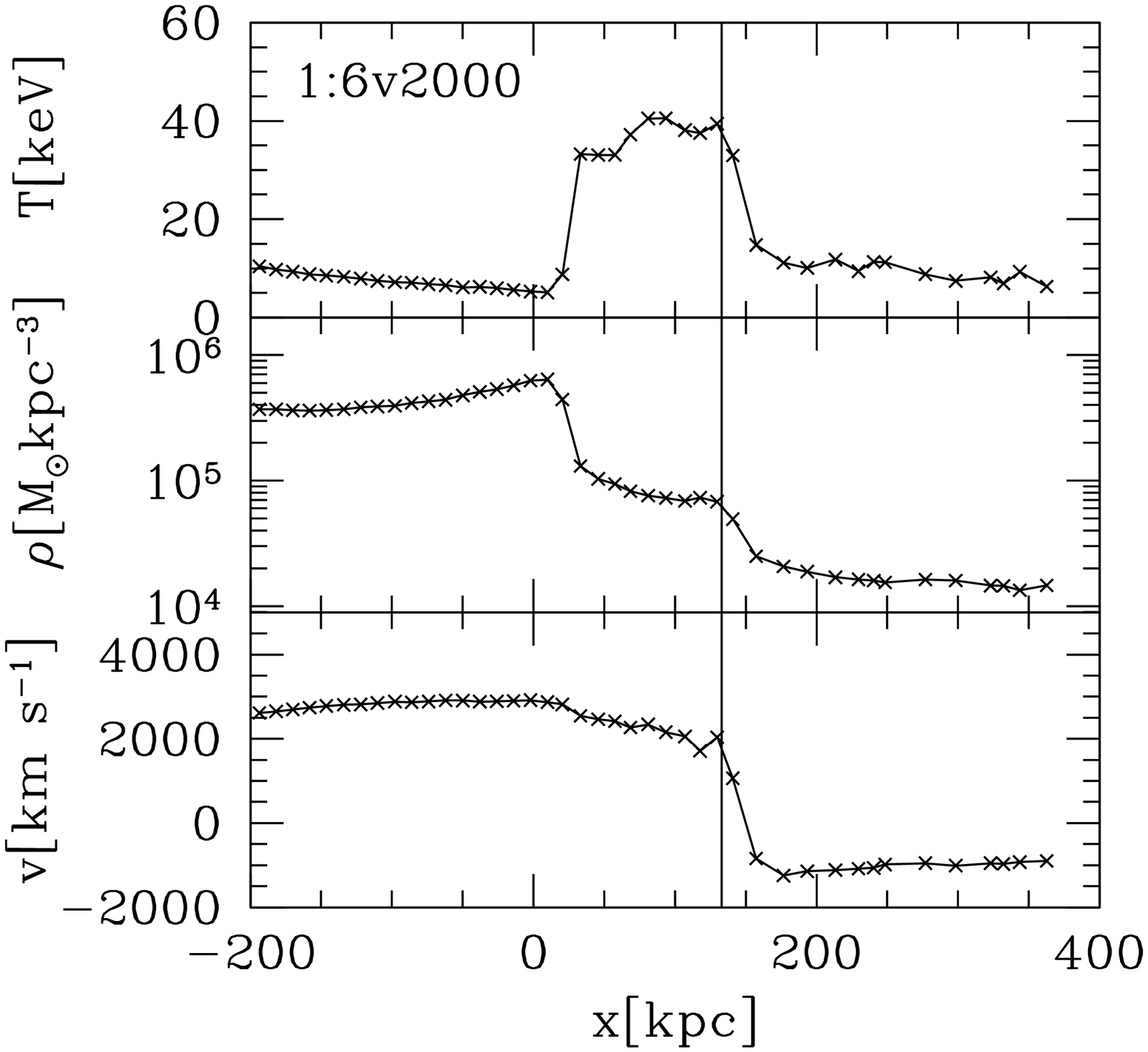}
\includegraphics[%
  scale=0.4]{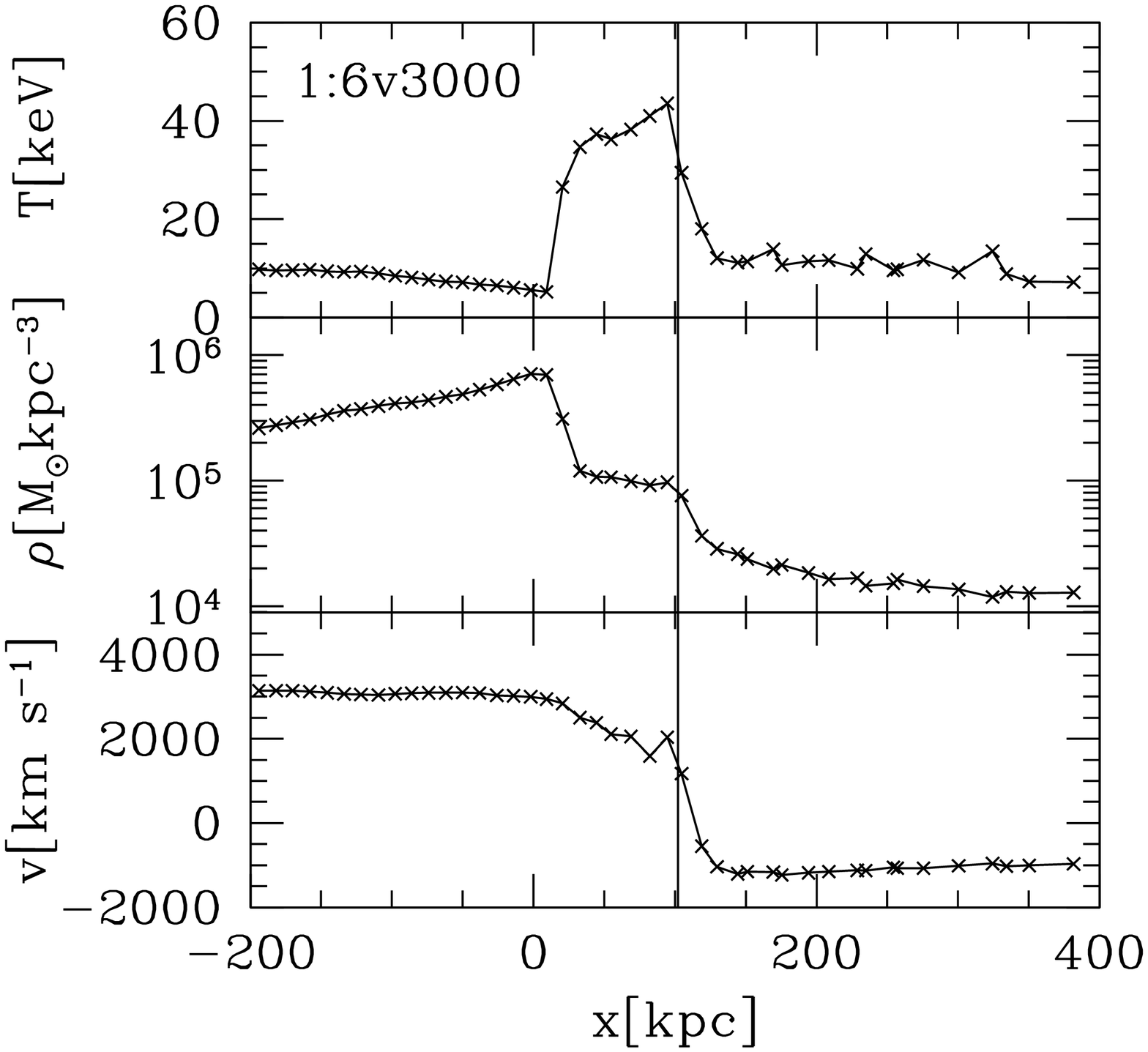}
\includegraphics[%
  scale=0.4]{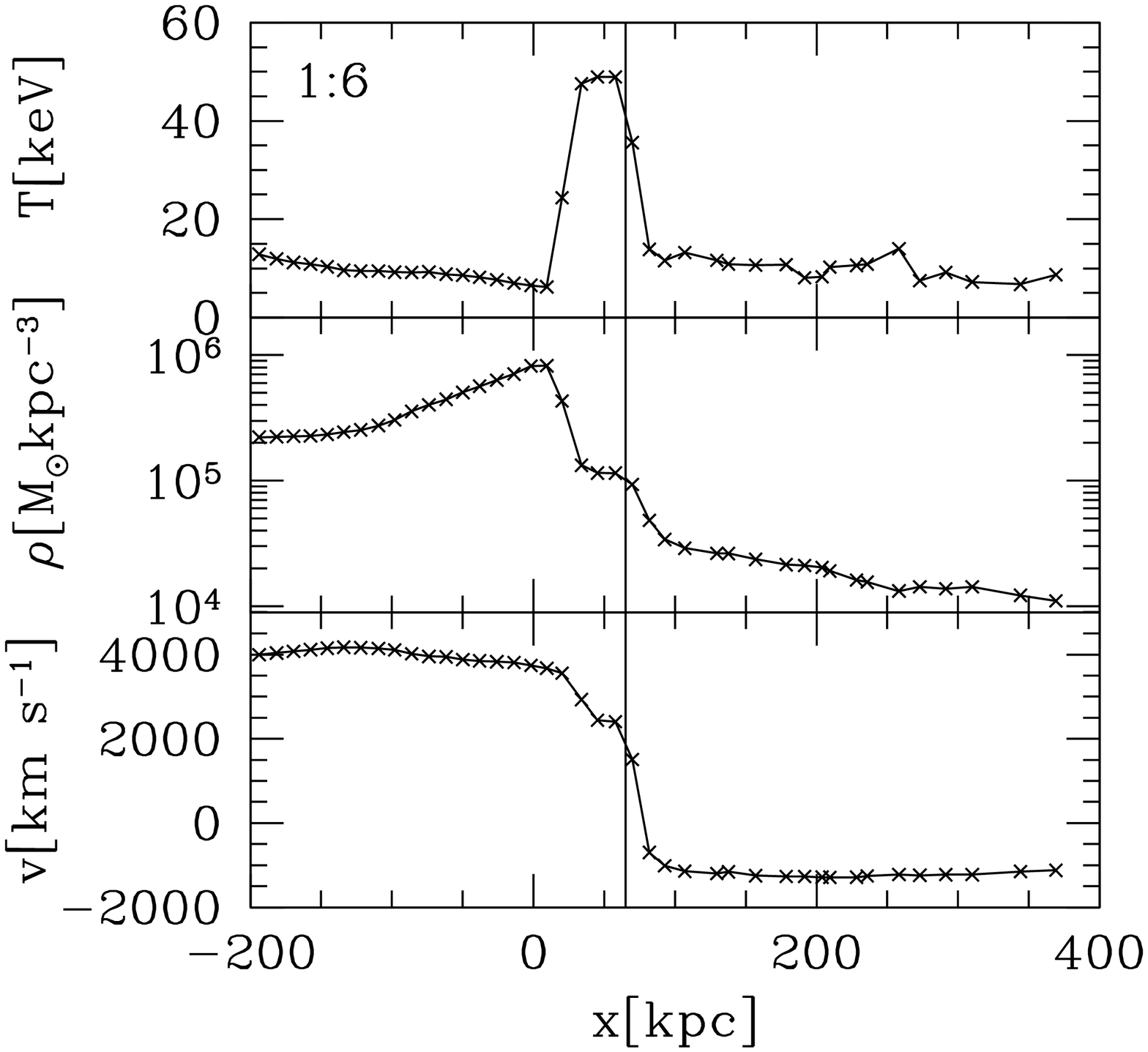}
\includegraphics[%
  scale=0.4]{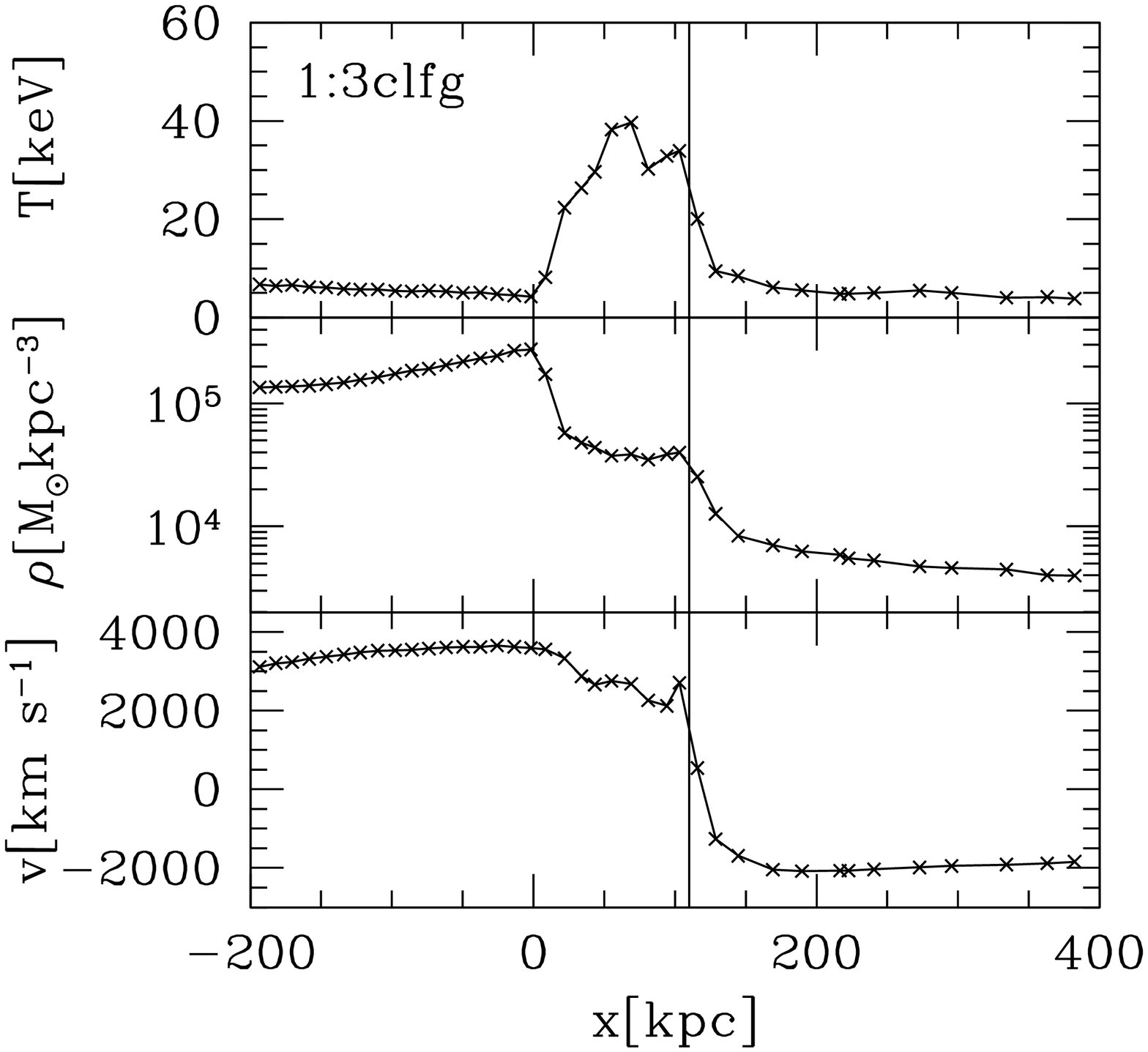}
\caption{Gas temperature, density and one dimensional velocity profiles across the shock discontinuity. The bullet is located at $x=0$ while the dashed vertical lines indicates the position of the bow shock.  }
 \label{shock}
\end{figure*}
 The projected temperature associated with the bullet is higher when it passes through the core of the main system (central panels of Fig. \ref{sequencetemp}) due to the line of sight overlap with the hot gas from the main cluster and the shock heated material stripped from the bullet itself  which surrounds it. As soon as the sub-cluster moves out into the cool external regions of the main cluster and loses part of the hot envelope of stripped gas, its projected temperature decreases to values comparable to the observed ones.   
Another peculiar feature in the temperature maps is the high temperature region next to the innermost total density contour of the main cluster and visible in the middle right and bottom left panels of Fig. \ref{sequencetemp} .
This area could be associated with the southeastern high temperature region observed by \citet{Markevitch02} (regions $f$, $i$ and $l$ in their Fig. 2) in the main cluster X-ray map and assumed to be coincident with the main merger site.
In our simulations the  high temperature region is filled with  hot gas stripped from the external regions of the sub-cluster and deposited within the core of the main system (compare with Fig. \ref{mixing}).
This high temperature material, combined with the main cluster gas which lies in the same projected region, produces the diffuse X-ray high emission feature visible at the present time in the main cluster below the primary peak (bottom left panel of Fig. \ref{sequencexray} or middle right panel of Fig. \ref{b>0} for a better colour contrast).  The eastern side of the high temperature region is less bright in X-ray emission and  lies beyond the luminosity peak associated with the main cluster.
As the sub-cluster moves to larger radii,  the high temperature region expands and cools.
Shortly ($\sim 70$ Myr) after the present time this gas is not anymore clearly distinguishable in the X-ray and density maps.

\section {3D analysis}

In this Section we will focus on a sample of runs from Table \ref{runs} -- namely 1:6, 1:6v3000, 1:6v2000 and 1:3clfg -- and investigate in details their three-dimensional characteristics. Three of these simulations are adiabatic models characterized by different initial relative velocities and will permit us to study the effects of the sub-cluster speed on the present and future state of the encounter. The remaining run 1:3clfg is the one which better reproduces -- together with 1:6v3000 and 1:6v2000 -- the observed jump in temperature across the shock front. 

Fig. \ref{shock} illustrates the behavior of hydrodynamical quantities across the shock discontinuity for a snapshot which corresponds to the present time. The horizontal axis is centered on the gaseous bullet and oriented perpendicular to the bow shock nose. The bow shock location is indicated by a vertical line while the edge of the bullet corresponds to the peak in density. The sub-cluster is moving outward from the main cluster core towards positive values of $x$. 
All the physical quantities are calculated and mass averaged on a one dimensional grid where each grid element has a volume of $15^3$ kpc$^3$.
$v$ is the component of the velocity perpendicular to the shock front and is calculated with respect to the system center of mass.
The Mach number ${\cal M}$ is determined from the temperature jump -- which shows a better defined discontinuity compared with the density jump -- using the Rankine-Hugoniot conditions.
We find values in good agreement with observations \citep{Markevitch06} for 1:6v3000 and the cooling run 1:3clfg (both with ${\cal M} \sim 3$). 1:6v2000 is characterized by a slightly lower value of  ${\cal M}$ ($\sim 2.9$) while the high velocity run 1:6 has a stronger shock, with ${\cal M} = 3.2$. 

As seen in the previous Section, the pre-shock temperature ($\sim $11-12 keV for the adiabatic runs) is  slightly higher than the one obtained by projection along the line of sight (Fig. \ref{temppro2} ). Assuming $T$=11 keV and ${\cal M} = 3$ we predict a pre-shock sound speed $c_S=1700$ km s$^{-1}$ and a shock velocity $v_S= {\cal M}c_S \sim 5100$ km s$^{-1}$. This value reduces to the observed shock velocity $v_S \sim 4700$ km s$^{-1}$ if we use the projected average pre-shock temperature ($T\sim 9$ keV) adopted by \citet{Markevitch06}. 

As previously noticed by other authors \citep{Springel07} the velocity jump shown in the last panel of each plot is much smaller than the theoretically inferred shock velocity.
Actually the pre-shock gas is not at rest but shows a negative velocity along the $x$-axis
 which however can only be partially explained by the fact that the center of mass of the system is moving in the positive direction of the $x$-axis following the bullet. Indeed the upstream velocity maintains a negative sign even with respect to the rest frame of the parent cluster, indicating a pre-shock infall  towards the bullet.
This effect is explained in \citet{Springel07} by studying the dynamical evolution of the system's global potential, which becomes deeper after the core-core interaction and induces infall of material from the region ahead of the shock. Nevertheless the infall velocity which characterizes our models is a significantly smaller than the sub-cluster velocity in contrast  to the values found by  \citet{Springel07} and ranges only between 500 and 900 km s$^{-1}$ in the different runs.
\begin{figure}
\epsfxsize=8truecm \epsfbox{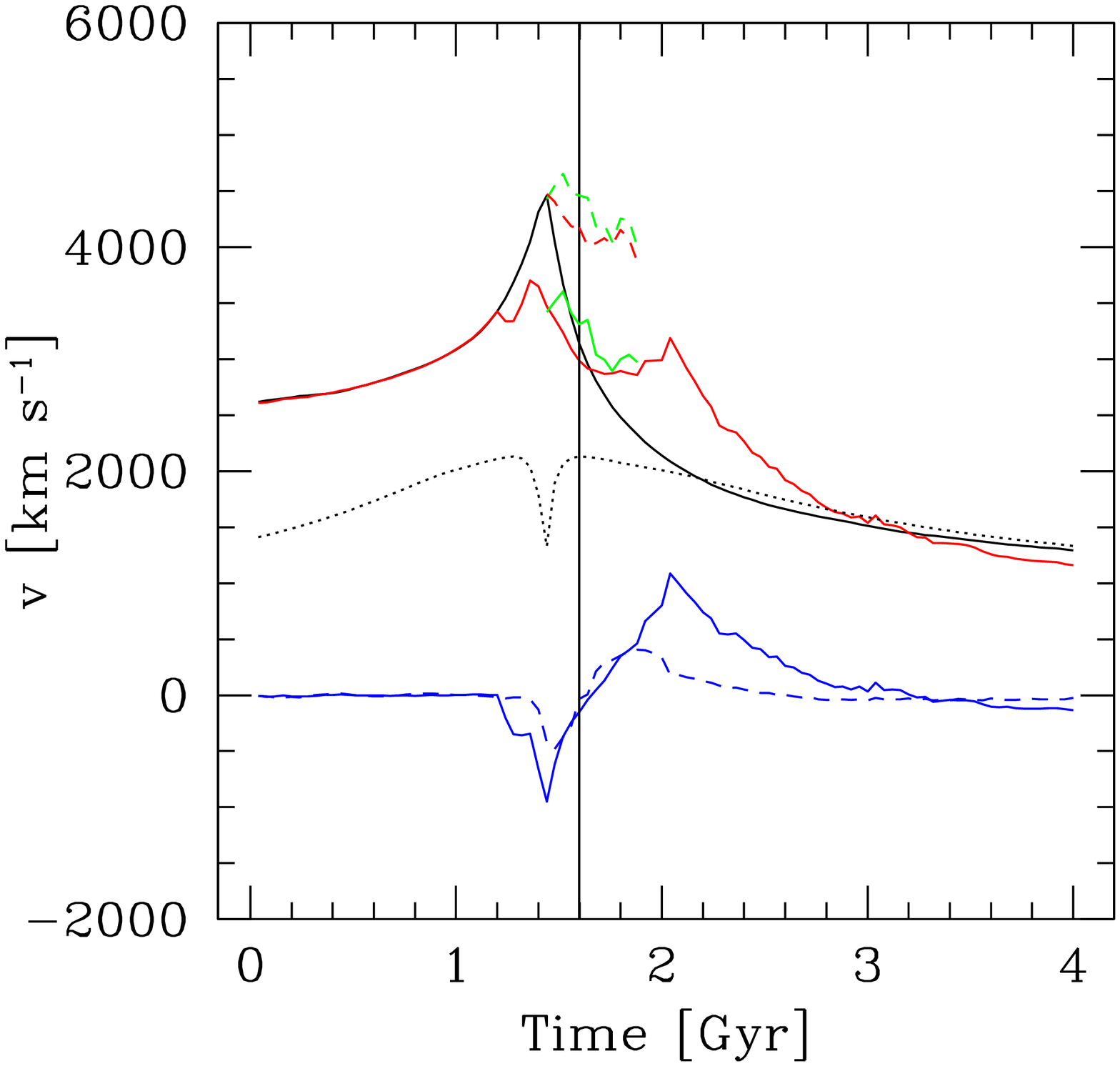} 
\caption{Run 1:6v3000. Characteristic bullet velocities plotted as a function of time. The black and red solid curves represent the velocity of the dark matter and gaseous component of the sub-cluster in the plane of sky, while the black dotted line indicates the escape velocity from the main cluster. The blue curves show the relative velocity of the two components of the sub-cluster in the two directions perpendicular to the line of sight.
The red solid and dashed curves show the velocity of the edge of the bullet in the center of mass and pre-shock gas rest-frame, respectively. Finally, the green curves represent the shock velocity obtained by differentiating the shock position.
The vertical line indicates the present time.  More details are given in the text.
   }
\label{vel3000}
\end{figure}

\begin{figure}
\epsfxsize=8truecm \epsfbox{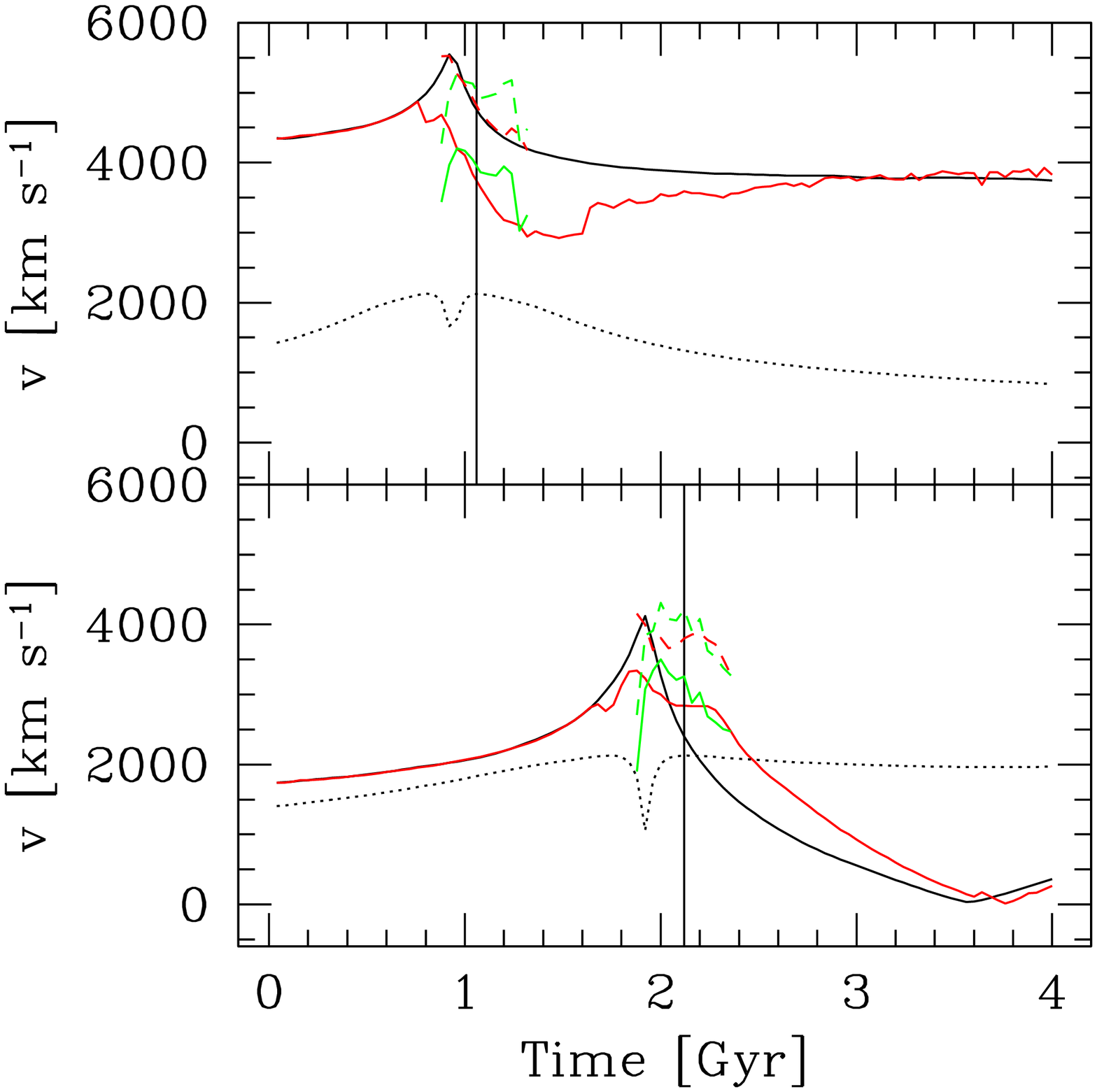} 
\caption{Run 1:6 (top panel) and 1:6v2000 (bottom panel). Same as in Fig. \ref{vel3000}
   }
\label{vel5000_2000}
\end{figure}
Figs. \ref{vel3000} and \ref{vel5000_2000} illustrate the characteristic velocities of the sub-cluster in the orbital plane for the three runs 1:6v3000, 1:6 and 1:6v2000. All the velocities are calculated in the center of mass rest frame and  the time corresponding to the present position is indicated by a vertical line.
The velocity of the dark matter component peaks at the moment of closest approach between the two cores and then decreases faster than for a ballistic orbit as a result of dynamical friction. The escape velocity at a given sub-cluster position is calculated assuming a spherical unperturbed host potential and is indicated by a black dotted curve. All the runs have initially unbound sub-clusters. Due to the effects of dynamical friction after the phase of core-core interaction the 1:6v2000 sub-cluster is actually bound to the main system, while sub-clusters with initial velocities $v=3000$ km s$^{-1}$ and 5000 km s$^{-1}$ have velocities slightly or much larger than the escape velocity from the main cluster.
The gaseous bullet initially follows its dark matter counterpart but before the point of closest approach it is slowed down by ram-pressure. At the same time the morphology of the sub-cluster gas distribution changes. The contact discontinuity assumes an arc-like shape (partially reducing the effect of ram-pressure) and the bow shock forms. It is interesting to notice that after the point of closest approach the evolution of the relative velocity between gas and dark matter in the sub-cluster strongly depends on the intensity of the ram-pressure force. In particular for relatively low ram-pressure values (1:6v2000 and 1:6v3000) the gaseous bullet is accelerated towards its dark matter counterpart as soon as it leaves the core of the host cluster where it experienced the largest external densities and ram-pressure. As a result the gravitational acceleration relative velocity between the gaseous and dark component of the sub-cluster (whose two components in the orbital plane are represented for 1:6v3000 by the blue solid and dashed curves in Fig. \ref{vel3000}) is larger than zero. At the time corresponding to the present position the two velocities look comparable in the case of 1:6v3000 while in 1:6v2000, where the ram-pressure acting on the bullet is lower, the acceleration starts earlier and the gaseous bullet is already $\sim 500$ km s$^{-1}$ faster than its dark counterpart.
For larger impact velocities -- as in the case of 1:6 -- ram-pressure is effective in slowing down the gaseous bullet even at large distances from the center of the main-cluster. The velocity of the gaseous bullet is therefore smaller than that of its dark counterpart until when the sub-cluster is well outside of the virial radius of the main system.

The green solid curve in each plot represents the velocity of the front shock obtained by differentiating the positions of the shock front at increasing times.
The shock velocity rapidly increases after the core-core interaction and at the present time it is always larger than the velocity of the gaseous bullet. 
In order to calculate the bullet and shock velocities in the system of reference of the pre-shock gas (red and green dashed curve, respectively) the infall velocity of the upstream gas is calculated at different times before and after the present one. 

The shock velocity $v_S$ at the present position is 4100, 4500 and 5100 km s$^{-1}$ in the case of 1:6v2000, 1:6v3000 and 1:6, respectively, consistently with the shock velocity $v_S= Mc_S $ inferred from the Rankine-Hugoniot jump conditions and with the value provided by observations.    
Only 1:6v2000 shows a two-dimensional shock velocity well below the observational uncertainties.
As found by \citet{Springel07}, after the point of closest interaction the shock velocity is always larger than the velocity  of the sub-cluster mass centroid, but the amount of the difference strongly depends on the model. In particular, in the case of the two low velocity runs 1:6v2000 and 1:6v3000  $v_S$ is $\sim 65\% $ and $ \sim 40\% $ larger than the velocity of the dark matter component, while in 1:6 the difference is almost negligible (only $6\%$).    \\

\begin{figure}
\epsfxsize=8truecm \epsfbox{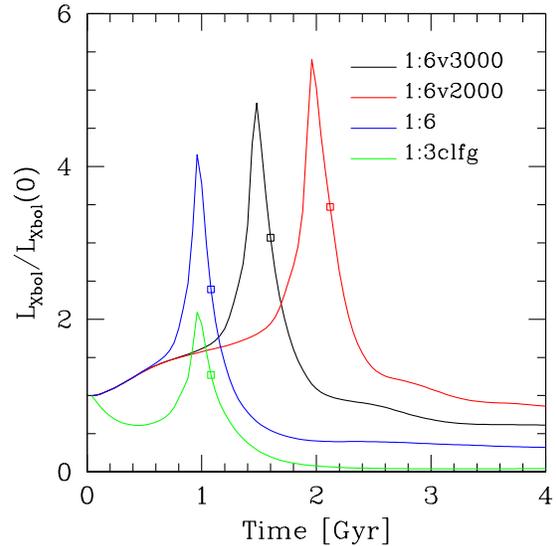} 
\caption{Bolometric X-ray luminosity $L_{bol}$ as function of time  for four selected runs. Luminosity is scaled to its initial value and calculated for the entire simulated volume. }
\label{Lvst}
\end{figure}

\begin{figure}
\epsfxsize=8truecm \epsfbox{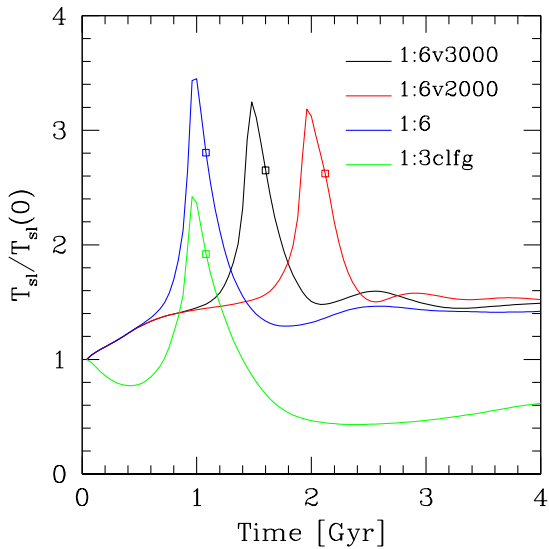} 
\caption{Average spectroscopic-like temperature $T_{sl}$ as function of time for four selected runs. Temperature is scaled to its initial value and averaged for the entire simulated volume. }
\label{Tvst}
\end{figure}

The cluster collision produces a drastic increase in luminosity and temperature. Figs. \ref{Lvst} and \ref{Tvst} show the bolometric X-ray luminosity $L_{Xbol}$ and average spectroscopic-like temperature $T_{sl}$ as functions of time. Both quantities are calculated for the entire simulated box and scaled to their initial values. 
The first phase of the interaction -- which involves only the external low density regions of the two clusters -- is characterized by an identical slow increment of luminosity and temperature in all the adiabatic runs. The jump is associated with the phase of core-core interaction and peaks right after the time of closest approach.  The present time is indicated with an empty squared and sits on the downturning curve.
The high velocity run 1:6 is associated with the largest increase in temperature ($T_{spec}/T_{spec}(0) \sim$ 3.5) and with the smallest jump in luminosity ($L_{bol}/L_{bol}(0) \sim $4). For decreasing sub-cluster velocities the peak in temperature becomes slightly less pronounced while the luminosity jump rises by a factor of 1.5.
The loss of baryonic material from the sub-cluster within the high density core of the main system is indeed larger for low velocity encounters and leads to higher luminosities  even if the increment in temperature is smaller with respect to the high velocity runs.
Excluding the bound run 1:6v2000, the amount of gas stripped from the bullet and lying within the virial radius of the main cluster is $1.6 \times 10^{13} M_\odot$ in 1:6 and $1.8 \times 10^{13} M_\odot$ in 1:6v3000, which in both cases corresponds to almost $\sim 60\%$ of the initial baryonic content of the sub-cluster.  
The difference becomes more pronounced if we consider only the core ($r< r_s$) of the main cluster, where the mass of sub-cluster gas is $1.4 \times 10^{12} M_\odot$ in the 1:6v3000 run and one order of magnitude less in the case of 1:6. 
At 4 Gyrs after the beginning of the simulation the total luminosity of the system is similar to the initial one in the case of the bound system 1:6v2000 where the center of mass of the bullet does not move out to distances beyond the virial radius of the main system. The luminosity drops to $50\%$ or even less of the initial luminosity for the runs 1:6v3000 and 1:6 respectively. This decrease in luminosity is motivated by the fact that at the final stages of the simulations a significant fraction of the gas is unbound and very extended. Due to its low density it does not contribute to the luminosity of the system despite the high temperature. In particular, for the same intracluster distance, the high velocity encounter 1:6 is associated to the highest fraction of unbound material, as will be shown later in this Section. On the other hand, the final temperature of the system is higher than the one associated to the initially isolated clusters and converges to a value $T_{sl}/T_{sl}(0) \sim 1.5$ almost independently of the sub-cluster velocity.
The cooling run 1:3clfg shows a somehow different behaviour: both, luminosity and temperature profiles have an initial decrement due to the cooling of the central regions of the two approaching clusters. Already during the early phases of the interaction the cooling run moves out of the equilibrium. The peaks in luminosity and temperature are much smaller than the corresponding adiabatic ones (not illustrated). The final luminosity approaches zero.

\begin{figure}
\epsfxsize=8truecm \epsfbox{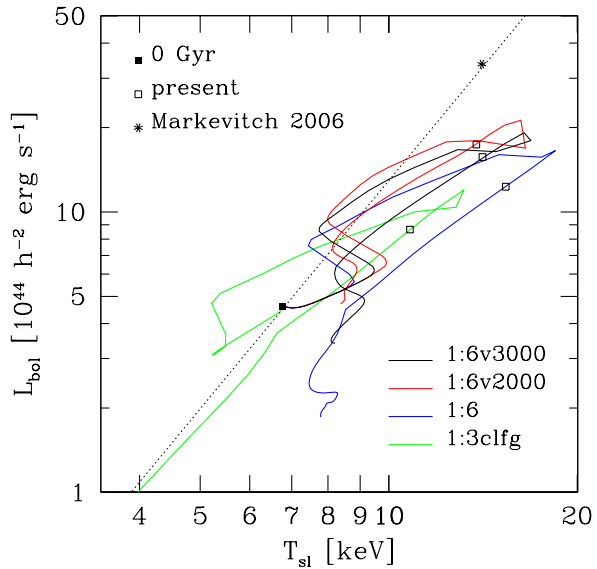} 
\caption{Evolution of the interacting system along the $L_{X}-T$ diagram. Here $L_{Xbol}$ and $T_{sl}$ are the bolometric  X-ray luminosity and the spectroscopic-like temperature within the virial radius of the main system. 
The dotted line represents the $L_{X}-T$ relation by \citet{Markevitch98} for local clusters.  The initial and present time of each simulation, as well as the observed position of the bullet cluster on the $L_{X}-T$ relation \citep{Markevitch06} are indicated.}
\label{TL}
\end{figure}
\begin{figure*}
\epsfxsize=15truecm \epsfbox{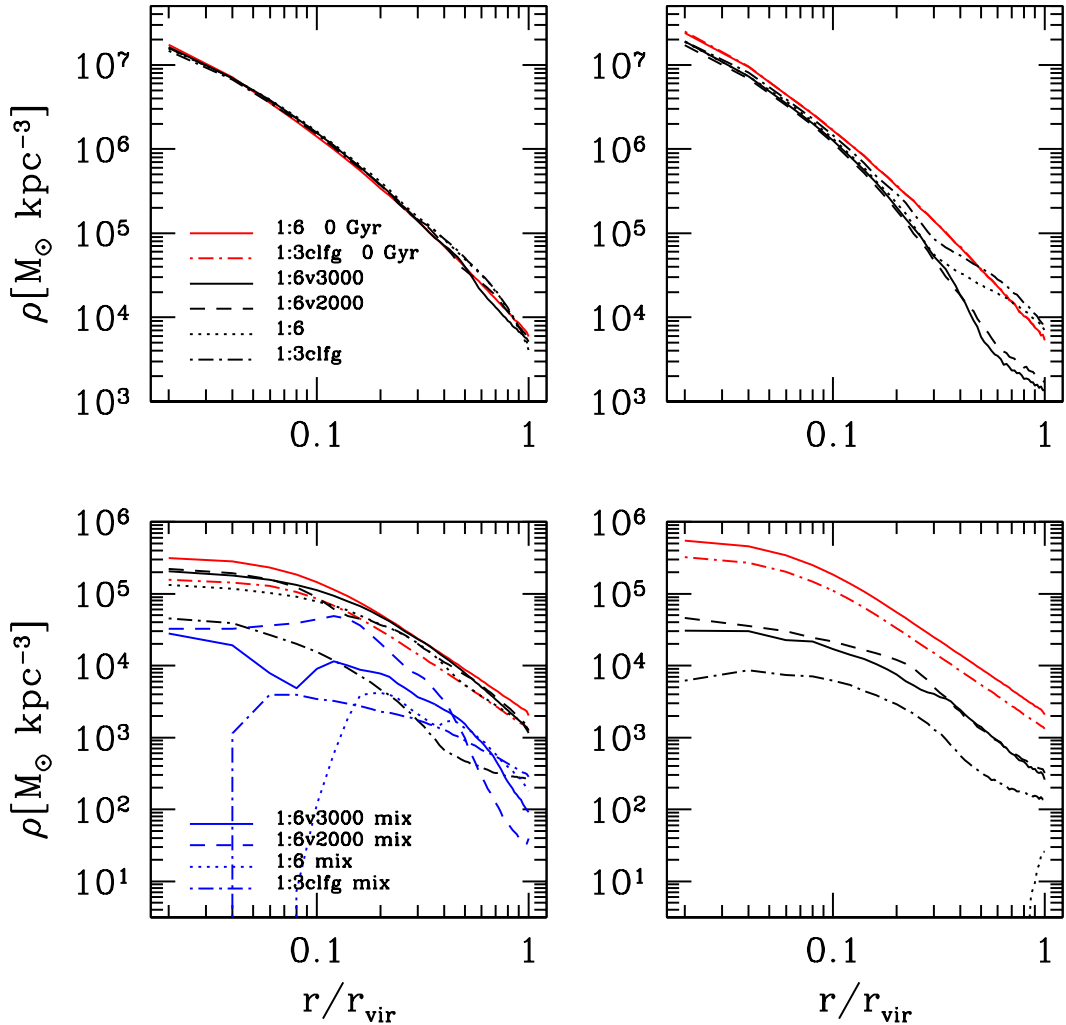} 
\caption{Dark matter (upper panels) and gas (bottom) density profiles of the main (left panels) and sub-cluster (right). Initial values and profiles at the final time (see text) are shown. Blue curves in the bottom left image refer to gas stripped from the sub-cluster and lying in the potential of the main system. Radius is scaled to the virial radius $r_{vir}$ of the dark matter distribution.}
\label{profiles}
\end{figure*}

\citet{Markevitch06} found that the bullet cluster lies exactly on the $L_{X}-T$ relation for nearby clusters \citep{Markevitch98} but its temperature is much higher than the one expected according to weak lensing mass estimates. 
Fig. \ref{TL} illustrates the drift of the simulated systems along the $L_{X}-T$ diagram. The position of 1E0657-560 is indicated by a star.  $L_{Xbol}$ and $T_{sl}$ are calculated in a cylindrical region centered on the center of mass of the main cluster and radius equal to its initial virial radius. The luminosity of each model is normalized in such a way that the initial main cluster lies on the $L_{X}-T$ relation for local clusters despite of the different initial  gas fractions. The starting time of the simulations is indicated with a black solid square. 
All the adiabatic runs present a similar evolution and move roughly parallel to the $L_{X}-T$ relation, as previously noticed by \citet{Rowley04} for major mergers in cosmological simulations. 
During the early stages of the encounter the cluster moves along a curve which is flatter than the observed $L_{X}-T$ relation: the compression of the low density gas at the outskirts of the cluster produces an increase in temperature which is only marginally accompanied by a luminosity growth.
The time when the core of the sub-cluster enters the virial radius of the main system represents an inversion point in the diagram: despite the formation of a bow-shock the temperature decreases due to the expansion of the main-cluster gas and the presence of low temperature baryons belonging to the sub-cluster within the virial radius of the main system. At the same time the luminosity rises as a result of the shock and the cluster moves perpendicularly towards the $L_{X}-T$ relation.
This phase is actually quite short (on average $\sim 0.12 $ Gyr) but it characterizes all the adiabatic runs.
Later on the cluster moves almost parallel to the $L_{X}-T$ relation toward larger values of $T_{sl}$, with the peak in temperature being reached at the point of closest approach.
Most of the runs show a small delay ($\sim 40$ Myr) between the time characterized by the highest temperature and the time with highest luminosity with the curve making a knot in the diagram.
The branch of the curve associated with the post core-core interaction is parallel to the increasing one but shifted  to smaller luminosities: during the strongest phase of the interaction some of the hot material is lost beyond the virial radius of the main cluster and indeed the largest shift in luminosity is observable in the high velocity run 1:6. Both, luminosity and temperature now decrease until they reach a second inversion point in the curve (in the case of 1:6 it is only a change in slope), associated with the egress of the bullet from the virial radius of the main system.
The cooling run 1:3clfg is characterized by a first  decrease in temperature which corresponds to the initial phase of thermal instability and central cooling of the main cluster. Later on it moves in the $L_{X}-T$ diagram similarly to the adiabatic runs although the peaks in luminosity and temperature are much less prominent.
Nevertheless in a pure cooling model the main cluster does not return to a state of thermal equilibrium at the end of the interaction since nothing prevents the central regions from cooling and the system moves toward extremely low  values of luminosity and temperature.

 \begin{figure}
\includegraphics[%
  height=75mm]{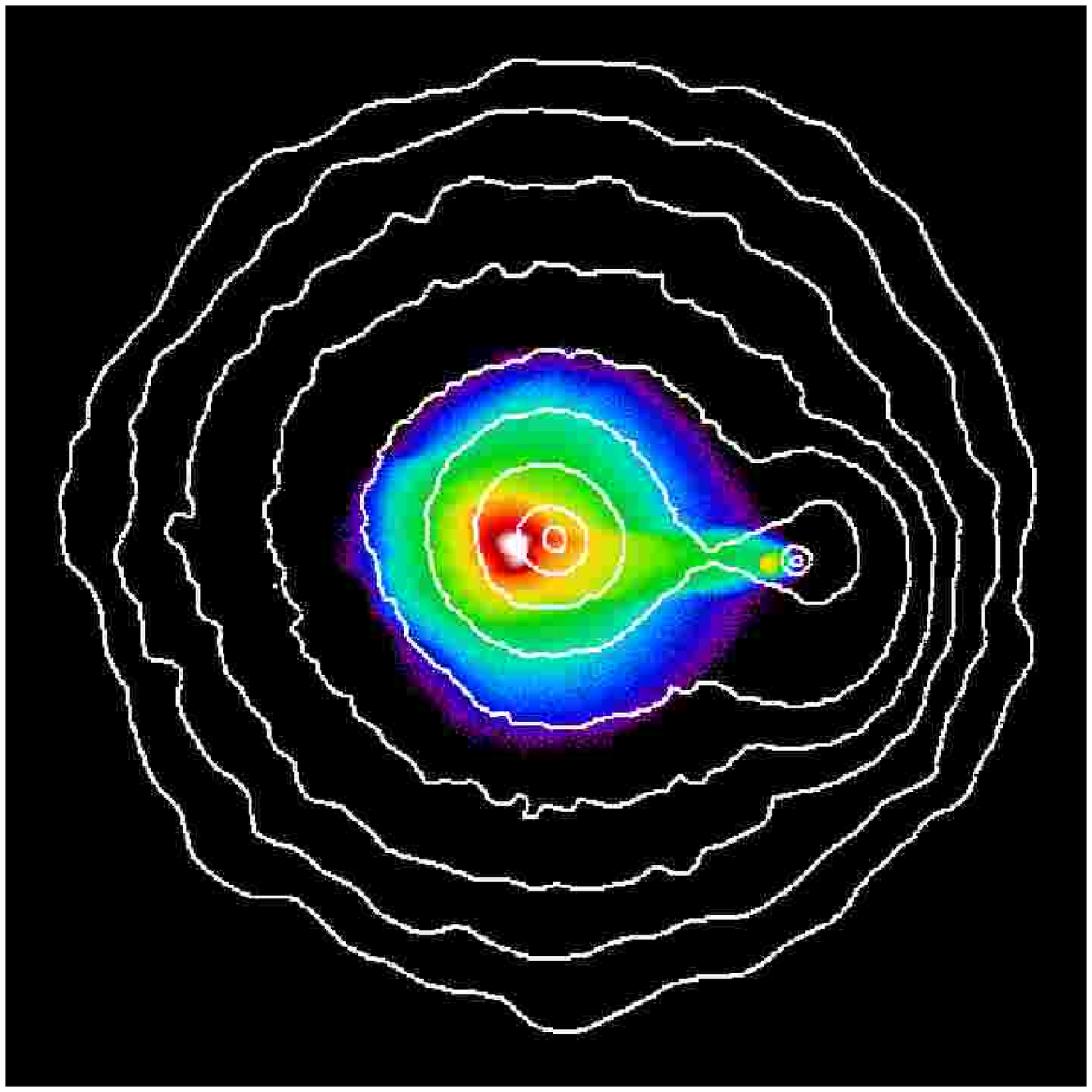}
\includegraphics[%
  height=75mm]{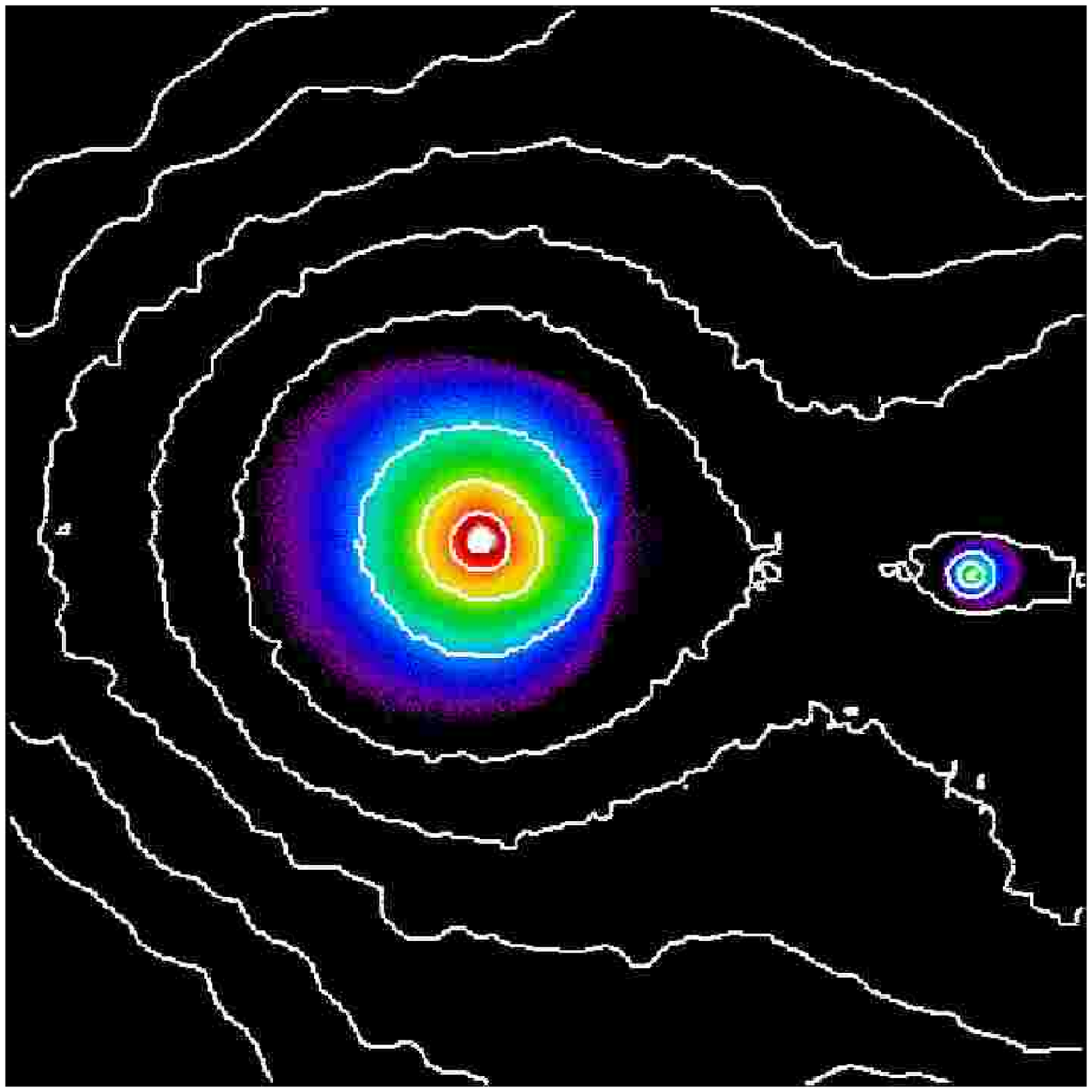}
\includegraphics[%
  height=4mm]{colorbar.ps}
\caption{X-ray surface brightness maps showing the late phases of the evolution of run 1:6v3000. Two successive times are considered, when the centers of the projected total mass distribution (white contours) are  $\sim 2500$ kpc (top panel, Time = 2.4 Gyr) and $\sim 5500$ kpc (bottom, Time = 4 Gyr) apart.   Logarithmic color scaling is indicated by the key to the bottom of the figure, with violet corresponding to 0.9 keV and white to 86 keV.  }
 \label{future}
\end{figure}

Fig. \ref{profiles} illustrates the final structure of the remnants. In the case of the two highest velocities the final time is chosen in such a way that the distance between the centers of the two systems is about $ 1000$ kpc larger than the sum of the virial radii at $Time=0$. This occurs at $Time=2.2$ and $4$ Gyr in the case of 1:6 and 1:6v3000, respectively. The low velocity run 1:6v2000 is analyzed at the time corresponding to the first apocenter when the core of the sub-cluster is close to the virial radius of the main cluster. 
The collisionless component of the main cluster (top left panel)  in all cases is not substantially affected by the interaction while the sub-cluster dark matter halo appears to be strongly perturbed and retains the original spherical symmetry only within its scale radius.
The sub-cluster central density profile (top right) is vertically shifted downward without a significant change of slope while for $r>0.2r_{vir}$ the loss of material becomes more significant and in the case of low velocity encounters ($v=2000$ and $3000$ km s$^{-1}$) with mass ratio 1:6 the profile shows a large jump of more than one order of magnitude between  $0.3r_{vir}$ and $0.5r_{vir}$. 
Beyond the scale radius the isodensity contours appear elongated and show a large plateau (see also Fig. \ref{future}) associated to tidally stripped material.
The evolution of gas density profiles is represented by the two panels on the bottom row of Fig. \ref{profiles}. 
In the 1:6 runs the interaction affects the central slope of the main cluster which becomes shallower while  the 1:3 encounter produces the largest deviation from the initial values, with the final density profile shifted down by $\sim 25\%$. 
As mentioned earlier, a not negligible part of the gas within the virial radius of the main cluster originally belonged to the sub-cluster and was subsequently stripped by ram-pressure during the central phases of the interaction. 
The density profiles of the stripped gas are drawn in blue for the different runs. 
In the case of 1:6v2000 only the gas outside the virial radius of the sub-cluster is considered.
Both 1:6v3000 and 1:6v2000 are characterized by a large fraction of sub-cluster gas lost to the core  of the main system, with a flat (1:6v2000) or even positive (1:6v3000) central slope, while the high velocity run 1:6, despite the larger ram-pressure values, has less time to deposit gas in the central regions and shows a clear cut-off for $r< 0.1 r_{vir}$. 
At larger radii, baryonic material is still accreting onto the remnant. Indeed, part of the main cluster gas has been pushed out by the bow shock and is now falling back into the cluster potential together with a fraction of the material lost by the sub-cluster.
Curves in the bottom right panel of  Fig. \ref{profiles} represent the sub-cluster gas density profiles. 
The low velocity sub-clusters 1:6v3000 and 1:6v2000 retain less than one tenth of their initial gas within $0.1 r_{vir}$ while the density profiles in the external regions drop by a factor $\geq 5$.
In general, the encounter flattens the gas density profile of the bullet core and this effect is evident in the cooling simulation as well.
The 1:6 bullet at $Time=2.2$ Gyr has lost almost all the gas within $r_{vir}$. Half of this baryonic material was stripped from its dark halo by ram-pressure during the central phases of the interaction and now accelerates towards its dark matter counterpart. It will be accreted  by  the sub-cluster halo at later times.

\begin{figure*}
\epsfxsize=15truecm \epsfbox{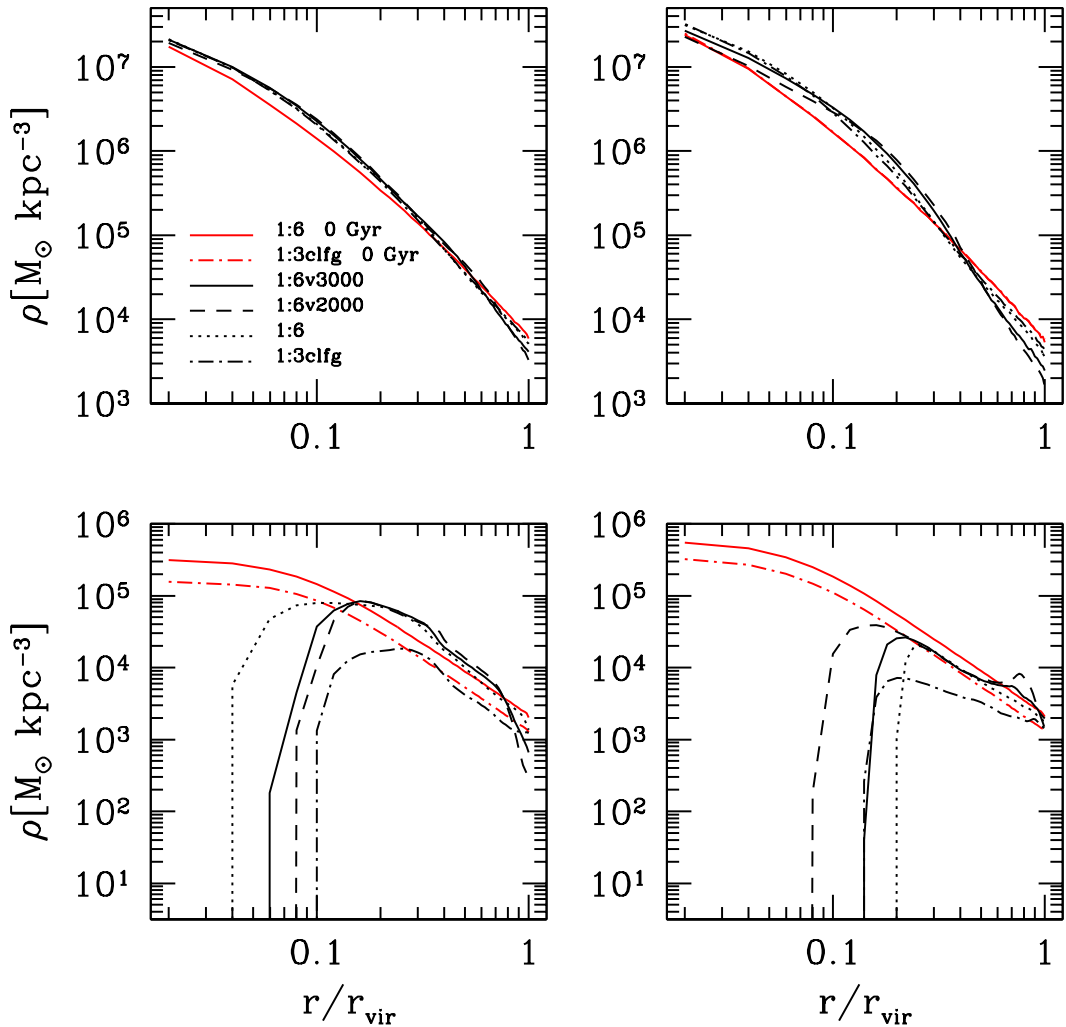} 
\caption{Dark matter (upper panels) and gas (bottom) density profiles of the main (left panels) and sub-cluster (right). Initial values and profiles at the present time are shown. Radius is scaled to the virial radius $r_{vir}$ of the dark matter distribution.}
\label{profiles2}
\end{figure*}

Fig. \ref{profiles}  compares dark matter and gas density profiles at times when the main and sub-cluster are close to a state of virial equilibrium. Despite the changes observable in the gas density profiles and the tidal stripping affecting the bullet beyond its core radius, the central slope of the collisionless component seems to be not strongly perturbed by the interaction \citep{Kazantzidis04, Kazantzidis06}. 
The situation changes  comparing the density profiles of the initial systems with those of the main and sub-cluster at the present time, when the bullet is located at almost one third of the host virial radius.  
As illustrated in the two upper panels of Fig. \ref{profiles2}, the density profile of the interacting dark matter halos increases in the inner regions and shows a decrement beyond $0.4 r_{vir}$. This behavior is similar for  different models and therefore independent of the orbital details, of the mass ratios and is seen for  both, the main and the sub-cluster.
The two-body relaxation time scale beyond the softening radius is longer than $10^{4}$ Gyr \citep{Boylan04, Arad05}, thus implying that dark matter density profiles are not affected by numerical relaxation. 
 As noticed in Section 3, fitting the projected masses around the centers of  main and sub-cluster at the present time we would get more concentrated or more massive systems with respect to the real ones.
The last two panels of Fig. \ref{profiles2}  represent gas  density profiles at the present time. The loss of gas in the central regions of the sub-cluster is expected as a result of ram-pressure stripping. But interestingly also the inner 200 kpc of the  main cluster are completely devoid of gas originally belonging to the main cluster.
Later on, after the displacement of the main cluster gaseous core, the baryonic material stripped from the bullet replenishes the central regions of the host system. 
As soon as the bullet moves toward the outskirts of the main system, the main cluster  gas has time to collapse again into the center of the potential (bottom left panel of Fig. \ref{profiles}). \\

According to \citet{Hayashi06} the probability to find a cosmological configuration where the most massive sub-halo has a velocity larger than $v_{sub}$ is well fitted by the function:
\begin{equation}
log f(>v_{sub}) =-\big(\frac{v_{sub}/v_{200}}{1.55}\big)^{3.3},
\end{equation}
where $v_{200}$ is the virial velocity of the main cluster.
Adopting for $v_{sub}$ the present time velocity of the sub-cluster dark matter component (in Table \ref{runs}) we find  a probability of $5 \times 10^{-14}$, $4.3 \times 10^{-4}$ and $0.036$ for runs 1:6, 1:6v3000 and 1:6v2000 respectively. 

\section{Conclusions}
We used high resolution N-body/SPH simulations to perform an extensive parameter study of the ``bullet'' cluster system 1E0657-56.
Collisions of two NFW halos with hot isothermal gas components in hydrostatic equilibrium were studied, adopting initial  relative velocities of 5000, 3000 and 2000 km s$^{-1}$, which is 3.8, 2.3 and 1.5 times the main cluster's virial velocity, respectively. We varied masses, mass ratios, impact parameter, baryonic fraction and concentrations. Most of the runs are adiabatic. Radiative cooling is included in two cases, adopting a standard cooling function for primordial gas. We analyzed the projected properties of the system at a time where the distance between the centers of the mass distribution associated with the main and sub-cluster is comparable to the value provided by lensing observations.

We have shown that:
\begin{itemize}
\item Most of the main features in the observed  X-ray maps are not well reproduced by encounters with zero impact parameter. Indeed, depending on the relative velocity of the bullet, a perfectly head-on encounter either destroys the X-ray peak associated with the main cluster or does not produce a significant displacement between gas and dark matter in the main system. Moreover, in a pure head-on collision, the gaseous core of the main cluster is displaced along the line which connects the centers of the two total mass distributions, contrary to observations. An impact parameter  corresponding to the core radius of the main cluster gas distribution provides enough ram-pressure to produce a displacement comparable to observations and introduces asymmetries in the main cluster emissivity map, similar to those detected in X-ray. 
\item A low concentrated ($c=4$) main cluster  does not survive  the collision  and its X-ray emissivity peak is destroyed. 
\item Encounters with mass ratios as large as 1:3 do not match the observed X-ray morphology and the size of the projected temperature jump across the shock discontinuity. Introducing cooling in the simulations, the temperature peak is cooled to a value comparable with the observed one. On the other hand, lensing mass reconstructions seem to suggest even larger mass ratios. This remains an open question. Gravitational lensing analysis of the simulations presented in this paper will be part of a follow-up paper (Mastropietro, Maccio` \& Burkert in prep). 
\item A pure cooling model neglecting energetic feedback still does not give a realistic description of the interacting system. What we are witnessing is overcooling in the central regions  following a thermal instability in the early phases of the interaction.
Nevertheless, cooling simulations provide a significant reduction of the temperature peak across the shock indicating that a more realistic treatment of the gas physics make models with higher mass ratios more realistic.
\item The choice of a different gas fraction does not affect significantly the results.
 \item   The morphology of  X-ray maps is best simulated by adiabatic 1:6 encounters. In particular the run 1:6v3000, with initial velocity $v=5000$ km s$^{-1}$ ($v \sim 2570$ km s$^{-1}$ in the center of mass system of reference) and present time dark matter velocity $v \sim 3100 $ km s$^{-1}$ (again in the center of mass rest-frame), reproduces most of the main X-ray features: the peculiar morphology of the X-ray emission associated with the main cluster, the relative surface brightness between the main and the sub-cluster, the shape of the bow shock and of the contact discontinuity. Decreasing the relative velocity (1:6v2000 has a present time dark matter velocity which is almost 300 km s$^{-1}$ smaller than that used by Springel \& Farrar 2007) the bullet becomes much less bright with respect to the center of the host system in disagreement  with observations, although the displacement in the two clusters is better reproduced. The high velocity run  1:6 produces a contact discontinuity much broader than observed.
\item A significant fraction of the X-ray emission next to the center of the main cluster is associated with gas stripped from the external regions of the bullet. In the 1:6 runs the main cluster X-ray peak presents two distinct components: a compact strongly emitting region associated with the displaced core of the host system and a more diffuse component  spatially  coincident with gas stripped from the sub-cluster during the central phases of the interaction.
The relative luminosity of this secondary component increases with decreasing bullet velocities.
\item The observed difference in line of sight velocity is compatible with an inclination of about 10 degrees of the plane of the encounter. A bottom-down inclination of our simulations keeps both the clusters close to the plane of the sky and does not produce essential changes in the the X-ray maps. But  it affects the morphology of the main cluster peak and the luminosity of the secondary peak with respect to that associated with the main cluster core. 
\item The projected temperature jump across the shock discontinuity gives important indications about the nature of the encounter. Indeed, both, the height and the thickness of the peak change, with the peak becoming broader for decreasing bullet velocities due to  the lower pressure, exercised by the pre-shock gas after the central phases of the encounter. Assuming an adiabatic equation of state, both runs 1:6v3000 and 1:6v2000, match quite well the observed temperature jump, while for larger velocities (1:6) the peak becomes too narrow and pronounced compared to observations. 
\item Temperature maps reveal some interesting features. In particular, as already found by \citet{Springel07}, the bullet remains relatively cold despite of its dominant X-ray emission.  For a short time, a high temperature region appears  next to the center of the main cluster mass distribution immediately after the central phases of the interaction. This feature -- still visible  at the present time -- is not spatially coincident with the main cluster gaseous core. It is associated with hot gas stripped from the sub-cluster within the core of the host system. Its location partially corresponds to the X-ray diffuse peak, while its eastern component could be related to the high temperature region observed by  \citet{Markevitch02} southeast of the main-cluster peak. 
\end{itemize}

For a selected sub-sample of runs we performed a detailed three dimensional analysis following their past, present and future evolution for four Gyr.
\begin {itemize}
\item Only the low velocity bullet 1:6v2000 is actually bound to the main cluster at the end of the simulation while in 1:6v3000 and 1:6 the sub-cluster has a velocity larger than the escape velocity  from the host system.
\item The velocity of the gaseous component of the bullet starts diverging from that of its dark matter counterpart before the point of closest approach between the two clusters, when the gas is  slowed down by ram-pressure.
As predicted by \citet{Milosavljevic07}, due to the drop in ram-pressure after the central phases of the interaction and gravitational acceleration by its dark halo counterpart, at the present time the gaseous bullet is not necessarily slower than the dark matter halo at the present time. We explored this question in greater details and found that the behavior of the sub-cluster gas after the point of closest approach strongly depends on the initial velocity of the bullet.
For relatively low velocities (and ram-pressure values, run 1:6v3000 and 1:6v2000)  the gaseous bullet is accelerated towards its dark matter counterpart as it leaves the core of the main cluster. At the present time the gaseous bullet moves as fast as its dark matter halo in 1:6v3000 and even 500 km s$^{-1}$ faster in 1:6v2000. 
For larger encounter velocities (1:6) ram-pressure is more effective in slowing down the bullet even beyond the core radius of host cluster and the gaseous bullet is always slower than dark matter. 
\item It has been already noticed  by \citet{Milosavljevic07} and \citet{Springel07} that  the sub-cluster velocities do not coincide with the shock velocity $v_s$ as measured by observers.
In the case of the three runs  (1:6v3000, 1:6v2000 and 1:6clfg) which best reproduce the projected temperature jump, the  Mach number and shock velocity determined using the Rankine-Hugoniot conditions across the shock  discontinuity show a good agreement with the values provided by \citet{Markevitch06}. 
The two dimensional shock velocity calculated by tracking the shock position as function of time is consistent with these values. We find that, although after the point of closest approach $v_s$ is always larger than the velocity of the sub-cluster dark matter halo, the difference depends on the model and becomes less significant for higher velocities. In particular in the high velocity run 1:6 the present time shock velocity is only $6\%$ higher than the dark matter bullet.  
\item The collision produces a drastic increase in luminosity and temperature. The highest velocity impacts are associated with the largest increase in temperature (and shocks) and the smallest peaks in luminosity due to the fact that in high velocity encounters a significant fraction of the baryons stripped from the bullet are deposited  at large radii within the main cluster.
\citet{Markevitch02} found that the bullet cluster lies on the $L_X-T$  relation for local clusters \citep{Markevitch98} at very high temperature.
We followed the evolution of the main cluster in the $L_X-T$ diagram during the different phases of the interaction and find that it moves roughly parallel to the $L_X-T$ relation for nearby clusters. The maximum temperature is reached at the point of closest approach and the peak in luminosity immediately afterwards, with a small delay which produces a knot in the diagram.  The branch of the curve associated with the post core-core interaction is  still parallel to the  $L_X-T$ relation but shifted towards smaller luminosities with respect to the early increasing branch, due to the loss of gas beyond the virial radius of the main cluster. 
\item After the encounter, as soon as the bullet is close or beyond the virial radius of its host system, the dark matter density profile of the main cluster does not deviate anymore significantly from the original one, while the gas profile becomes shallower in the central regions, with a large decrement in density observed only in run 1:3clfg. 
The situation changes drastically if we compare the density profiles of the initial systems with those at the present time, when the bullet is still located well within the virial radius of the main cluster. In this case the interacting systems are not in virial equilibrium and the dark matter densities of both, the main and sub-cluster, increase in the inner regions and show a decrement beyond 0.4 $r_{vir}$.
Comparing these present time profiles with NFW halos would give wrong estimates of the halo parameters.
At a time corresponding to the present configuration the center of the host system is completely devoid of main cluster gas as a result of the displacement of the main cluster gaseous core by the bullet. On the other hand, the baryons stripped from the bullet have replenished the central regions of the host. 
This is only a temporary situation however since a  few Gyrs later the main  cluster gas falls back and becomes again the dominant component in the central regions, although a significant fraction of gas stripped from the bullet is still present.
The final sub-cluster dark matter density profiles seem significantly affected by the interaction beyond their scale radius, where the isodensity contours are elongated and show a plateau.
\end{itemize}

\section{Acknowledgements}

We would like to thank R. Piffaretti, S. Borgani, A. Biviano, T. Naab, A. Maccio' and M. Girardi  for useful discussions. We acknowledge T. Quinn for support with the TIPSY X-ray package and D. Clowe for providing the upper panel of Fig. 1.
The work was partly supported by the DFG Sonderforschungsbereich 375 ``Astro-Teilchenphysik''.
The numerical simulations were performed on a local SGI-Altix 3700 Bx2 which was partly funded by the cluster of excellence ``Origin and Structure of the Universe''.

\label{lastpage}

\end{document}